\RequirePackage{fix-cm}
\documentclass[natbib]{svjour3}       
\usepackage{graphicx}
\usepackage{enumitem}
\usepackage{graphicx}
 \usepackage{mathptmx}      
\usepackage{latexsym}
\usepackage{amsfonts}
\usepackage{color}
 \journalname{Space Science Reviews}
\def\sech{\mathop{\rm sech}\nolimits}

\def\lapprox{\mathrel{\hbox{\rlap{\hbox{\lower4pt\hbox{$\sim$}}}\hbox{$<$}}}}
\def\gapprox{\mathrel{\hbox{\rlap{\hbox{\lower4pt\hbox{$\sim$}}}\hbox{$>$}}}}

\newcommand{\be}{\begin{equation}}
\newcommand{\ee}{\end{equation}}

\newcommand {\mb} {\mathbf}
\newcommand {\mbb} {\mathbb}
\newcommand {\bea} {\begin{eqnarray}}
\newcommand {\eea} {\end{eqnarray}}

\newcommand{\Bvec}{\mathbf{B}}

\newcommand{\bln} {}
\newcommand{\eln} {}

\newcommand{\asr}{{Adv.\ Space Res.}}
\newcommand{\ag}{{Ann.\ Geophys.}}

\newcommand{\af}{{Ark.\ Fys.}}
\newcommand{\aap}{{Astron.\ Astrophys.}}
\newcommand{\apj}{{Astrophys.\ J.}}
\newcommand{\apjl}{{Astrophys.\ J.\ Lett.}}
\newcommand{\apjs}{{Astrophys.\ J.\ Suppl.\ Ser.}}
\newcommand{\ar}{{Astron.\ Rep.}}
\newcommand{\aspcs}{{Astron.\ Soc.\ Pac.\ Conf.\ Ser.}}

\newcommand{\baas}{{Bull.\ Am.\ Astron.\ Soc.}}
\newcommand{\grl}{{Geophys.\ Res.\ Lett.}}
\newcommand{\jastp}{{J.\ Atmos.\ Solar-Terr.\ Phys.}}
\newcommand{\jatp}{{J.\ Atmos.\ Terr.\ Phys.}}
\newcommand{\jgr}{{J.\ Geophys.\ Res.}}
\newcommand{\lrsp}{{Living Rev.\ Sol.\ Phys.}}
\newcommand{\mnras}{{Mon.\ Not.\ R.\ Astron.\ Soc.}}
\newcommand{\nat}{{Nat.}}
\newcommand{\pasj}{{Pub.\ Astron.\ Soc.\ Jpn.}}
\newcommand{\plms}{{Proc.\ Lond.\ Math.\ Soc.}}
\newcommand{\pof}{{Phys.\ Fluids}}
\newcommand{\pop}{{Phys.\ Plasmas}}
\newcommand{\prl}{{Phys.\ Rev.\ Lett.}}
\newcommand{\prsla}{{Proc.\ R.\ Soc.\ (Lond.) A}}
\newcommand{\qjrms}{{Q.\ J.\ R.\ Meteorol.\ Soc.}}
\newcommand{\solphys}{{Sol.\ Phys.}}
\newcommand{\planss}{{Planet.\ Space Sci.}}

\newcommand{\ssfcmp}{{Soc.\ Sci.\ Fenn.\ Comment.\ Math.\ Phys.}}
\newcommand{\ssr}{{Space Sci.\ Rev.}}
\newcommand{\pt}[1]{\frac{\partial}{\partial t}}
\newcommand{\pr}[1]{\frac{\partial}{\partial r}}
\newcommand{\pz}[1]{\frac{\partial}{\partial z}}
\begin{document}

\title{Ionized Plasma and Neutral Gas Coupling in the Sun's Chromosphere and Earth's Ionosphere/Thermosphere
}

\titlerunning{Plasma-Neutral Coupling: Chromosphere and I/T}        

\author{Leake, J.E. [1] \and DeVore, C.R. [2,3] \and Thayer, J.P. [4] \and Burns, A.G. [5] \and Crowley, G. [6] \and 
Gilbert, H.R. [3] \and Huba, J.D. [2] \and Krall, J. [2] \and Linton, M.G. [2] \and Lukin, V.S. [2] 
\and Wang, W. [5]}

\institute{[1] George Mason University \textit{email: jleake@gmu.edu} \and [2] Naval Research Laboratory \and [3] 
NASA Goddard Space Flight Center \and [4] University of Colorado \and [5] National Center for Atmospheric Research 
(NCAR) \and [6] Atmospheric and Space Technology Research Associates} 

\authorrunning{Leake, DeVore, Thayer, et al.} 


\date{Received: date / Accepted: date}

\maketitle

\begin{abstract}
We review our understanding of ionized plasma and neutral gas 
coupling in the weakly ionized, stratified, electromagnetically-permeated 
regions of the Sun's chromosphere and Earth's ionosphere/thermosphere.
Using representative models for each environment we derive fundamental descriptions of the
coupling of the constituent parts to each other and to the electric and magnetic fields, and we examine the variation in 
magnetization of the ionized component. Using these descriptions
we compare related phenomena in the two environments, and discuss electric currents, energy transfer and dissipation. 
We present a coupled theoretical and numerical study of plasma instabilities in the two environments that serves as
an example of how the chromospheric and ionospheric communities can further collaborate. We also suggest future collaborative studies that will help improve our understanding of these two different atmospheres which share many similarities, but have large disparities in key quantities.

\end{abstract}

\section{Introduction}
\label{introduction}

In a universe where partially ionized gases abound, interactions between ionized plasma and neutral gas play a critical role in planetary and stellar atmospheres, including those of the Earth and the Sun. They also are important at the heliopause, where the solar wind meets the interstellar medium, and in other astrophysical contexts. Plasma-neutral interactions modulate momentum and energy exchange among the neutral gas, electrons, and ions, and between the ionized plasma and electromagnetic fields. The physics of plasma-neutral coupling adds another layer of complexity to problems that previously have been addressed by assuming a fully ionized plasma or some other single-fluid approximation. The importance of these transitional layers in the inner heliosphere -- in particular, the solar chromosphere and terrestrial ionosphere/thermosphere -- lies in their impact on space weather processes that can profoundly affect Earth and society. This motivates our attention to the underlying physics of plasma-neutral coupling.

The solar chromosphere is the highly dynamic, complex region above the relatively cool visible surface of the Sun and beneath the very hot corona. It is characterized by several transitions that occur with increasing altitude: from predominantly neutral to ionized hydrogen; from essentially unmagnetized to strongly magnetized charged particles; from collisional to collision-less behavior; and from gas-dominated to magnetic field-dominated dynamics. These transitions vary in space and time as the chromosphere is driven continually from below by convective motions and magnetic evolution. The chromosphere is the source of the solar wind and also modulates the flow of mass and energy into the corona. The chromosphere's complexity is increased further by its state of thermodynamic and ionization non-equilibrium, which makes understanding its observed emission and absorption spectra very challenging.

The ionosphere/thermosphere (hereafter I/T) is a similarly transitional region in the Earth's upper atmosphere, in which the gas is ionized to varying degrees by incident solar radiation. It encompasses the same physical transitions as those occurring in the chromosphere.  In this paper we use the term I/T to denote that region of the Earth's upper atmosphere between about 80-600 km altitudes.  The word ``thermosphere" technically denotes a distinct region based on the temperature profile of the neutral component of the upper atmosphere.  The word ``ionosphere" refers to the ionized component of the gas in the upper atmosphere, usually in the same 80-600 km altitude region.  Thus, I/T includes both the neutral and ionized constituent parts of the weakly ionized mixture. The I/T is bounded below by the mesosphere, and above by the magnetosphere, and as a result it is driven continually from above and below.  
Understanding how the many various forcing mechanisms interact to cause variability in the I/T system remains a major challenge.

Previous authors have discussed similarities and differences between the Sun's chromosphere and the Earth's ionosphere \citep{Haerendel_2006,Fuller_2009}.  These authors emphasized the strong collisional coupling of the minority plasma constituents to the majority neutral species in both atmospheres; a collisional coupling of the neutrals to the plasma that is very substantial in the solar chromosphere but relatively weak in the terrestrial I/T; and the impact of these interactions on the highly anisotropic electrical conductivities along and across ambient magnetic fields in the two environments. \citet{Haerendel_2006} further presented analogies between density fluctuations in solar spicules and sporadic E layers, atmospheric heating by Alfv\'en waves in chromospheric plages and auroral arcs, and plasma erosion driven by currents aligned with the magnetic fields in solar flares and auroral ion outflows. \citet{Fuller_2009} provided a detailed summary of processes occurring in the ionosphere, followed by a survey of phenomena in the Sun's chromosphere and comparisons between the two. They discuss the similar variability of the charged particle magnetizations in the two atmospheres and the role of convective overshoot from the photosphere/thermosphere below the chromosphere/ionosphere, and highlight major contrasts in the dynamic/static character of the magnetic fields and the resultant dominance of magnetohydrodynamics/electrodynamics in describing macroscopically the evolution of those fields.

{\color{black} One of the goals of this review is to highlight the commonalities and differences between the chromosphere and I/T in order to develop  cross-disciplinary collaboration between the two communities, which typically use different approaches to the same fundamental physics. In doing so, we hope to identify important questions concerning the transition from weakly ionized dense mixtures to fully ionized tenuous plasmas linked by electromagnetic fields, and present methods by which we can enhance our physical understanding of such systems using
improved analytical and numerical modeling of plasma-neutral coupling in the chromosphere and I/T. }

The paper is structured as follows. In \S \ref{numbers}, we examine representative static models of the chromosphere and I/T, and compare the two environments in terms of their fundamental neutral, plasma, and magnetic properties and some key dimensionless ratios. In \S \ref{equations}, we present the governing equations for a weakly ionized reacting plasma-neutral mixture. {\color{black} In \S \ref{electrodynamics} 
we investigate in detail the physics and the equations that govern the coupling of the ionized plasma and neutral gas to each other and to the electromagnetic field. We compare the magnetization and mobility of the ionized component of the two environments and relate them to the evolution of electric currents. 
In \S \ref{mhd_ed_processes} we discuss processes which are examples of such coupling, and consider the contrasting approaches that the I/T and chromosphere communities use to describe essentially the same phenomena. 
In \S \ref{EM}, we address the transfer and dissipation of energy, first focusing on the state of the field's knowledge. Then we discuss the importance of the conversion of electromagnetic energy into thermal and kinetic energy, and look at the efficiency of plasma-neutral coupling in this energy transfer. }In \S \ref{rayleightaylor}, we present an illustrative analytical and numerical case study of the Rayleigh-Taylor instability, which is common to the chromosphere and I/T yet also highlights the contrasting conceptual and mathematical approaches employed by the two communities. We conclude in \S \ref{conclusions} with some parting thoughts about current challenges to our understanding of plasma-neutral coupling on the Sun and at the Earth.

\section{Basic Properties}
\label{numbers}

In this section,  basic properties of the Sun's chromosphere and Earth's I/T are presented 
and compared. As we will show quantitatively, there are both significant similarities and substantial 
differences between these environments. {\color{black} From fundamental principles, we deduce qualitative 
implications about how the majority neutral and minority plasma constituents couple hydrodynamically in both 
atmospheres. Then in \S \ref{electrodynamics} we will discuss how they couple principally magnetohydrodynamically in the chromosphere (where the fluctuating 
and ambient magnetic fields frequently are of the same magnitude) but electrodynamically in the I/T (where the 
fluctuating fields are far smaller than the ambient field, in general).}
 Several of the following general introductory considerations are elaborated on in 
more detail in the later sections of the paper, which deal with the governing multi-fluid equations, 
{\color{black} generalized Ohm's law and its low-frequency limit}, the mobility of plasma and electric currents, electromagnetic energy transfer, and 
the Rayleigh-Taylor instability.

\subsection{Models for the Chromosphere and Ionosphere/Thermosphere}
\label{models}

The chromosphere {\color{black} may be best} represented by the semi-empirical quiet-Sun model ``C7'' developed and tabulated by 
\citet{2008ApJS..175..229A}, hereafter referred to as the ALC7 model \citep[see also][]{1981ApJS...45..635V,
1993ApJ...406..319F,2006ApJ...639..441F}. In this model, simulated  line and spectral emissions are matched 
to the observed spectra to obtain estimates of the total density, ionization level and temperature in the 
chromosphere. The transition region above is modeled by assuming an energy balance between the downward 
total energy flow from the overlying hot corona and local radiative loss rates.
{\color{black} Downward diffusion through the background neutral gas of the predominant hydrogen and helium ions, enhanced by the ambipolar electric field generated by the more freely diffusing electrons, contributes significantly to the energy transport in the lower chromosphere, while electron thermal conduction dominates in the upper chromosphere.
Although this one-dimensional model certainly does not capture all variations, either temporally or in 
three dimensions, it is useful for characterizing the generic structure of the chromosphere.}

The I/T {\color{black} may be best} represented by the NCAR Thermosphere-Ionosphere-Mesosphere-Electrodynamic General 
Circulation Model (NCAR TIMEGCM). TIMEGCM is a time-dependent, three-dimensional model that solves 
the fully coupled, nonlinear, hydrodynamic, thermodynamic, and continuity equations of the neutral gas {\color{black} along}
 with the ion energy, momentum, and continuity equations from the upper stratosphere 
through the ionosphere and thermosphere \citep{1988GeoRL..15.1325R,1992GeoRL..19..601R,1994GeoRL..21..417R}.
TIMEGCM predicts global neutral winds, neutral temperatures, major and minor species composition, 
electron and ion densities and temperatures, and the ionospheric dynamo electric field. The input 
parameters are solar EUV and UV spectral fluxes, parameterized by the F10.7 cm index, plus auroral 
particle precipitation, an imposed magnetospheric electric field, and the amplitudes and phases of 
tides from the lower atmosphere specified by the Global Scale Wave Model \citep{1999JGR...104.6813H}.
{\color{black} Many features of the model, such as increased electron temperature in the E and F layers, have been validated against observations
\citep[e.g.,][]{Lei_2007}. } For the atmospheric profiles used in this paper, TIMEGCM was run under equinox conditions with a F10.7 value of 150, a Kp index of 2, and tidal 
forcing. The vertical profile was taken from 47.5$^\circ$ N latitude at 12:00 local time. 
Thus like the model chromospheric profile described above, we represent the I/T structure by a single 1-D, time-independent profile extracted from the TIMEGCM.

In displays of quantities provided by, or derived from, these models for the chromosphere and I/T, 
we use as the primary (left) ordinate axis the normalized total gas pressure ($P/P_{0}$), where $P_{0}$ 
is defined to be the pressure at a selected reference height in the domain. The approximate corresponding altitude is shown as the 
secondary (right) ordinate axis. For the chromosphere, we chose the Sun's visible surface, the photosphere, 
as the reference height. The pressure there is 1.23$\times$10$^{4}$ Pa in the ALC7 model. 
For the Earth's atmosphere, we selected a reference height of 30 km, even though it is in the stratosphere and outside the I/T region.  {\color{black} This
choice of lower boundary is to allow comparison of electrodynamics between the two atmospheres later in this review.}
We set the top of the chromosphere where the pressure has decreased from its 
base value by six orders of magnitude; this occurs at an altitude of 1989 km, above which the temperature rises steeply in the transition region. Ten orders of magnitude of pressure reduction were 
included in the I/T, which extends up to about 640 km in altitude.

\subsection{Neutral Gas, Plasma, and Magnetic Field}
\label{profiles}

\begin{figure}[h]
\includegraphics[width=\textwidth]{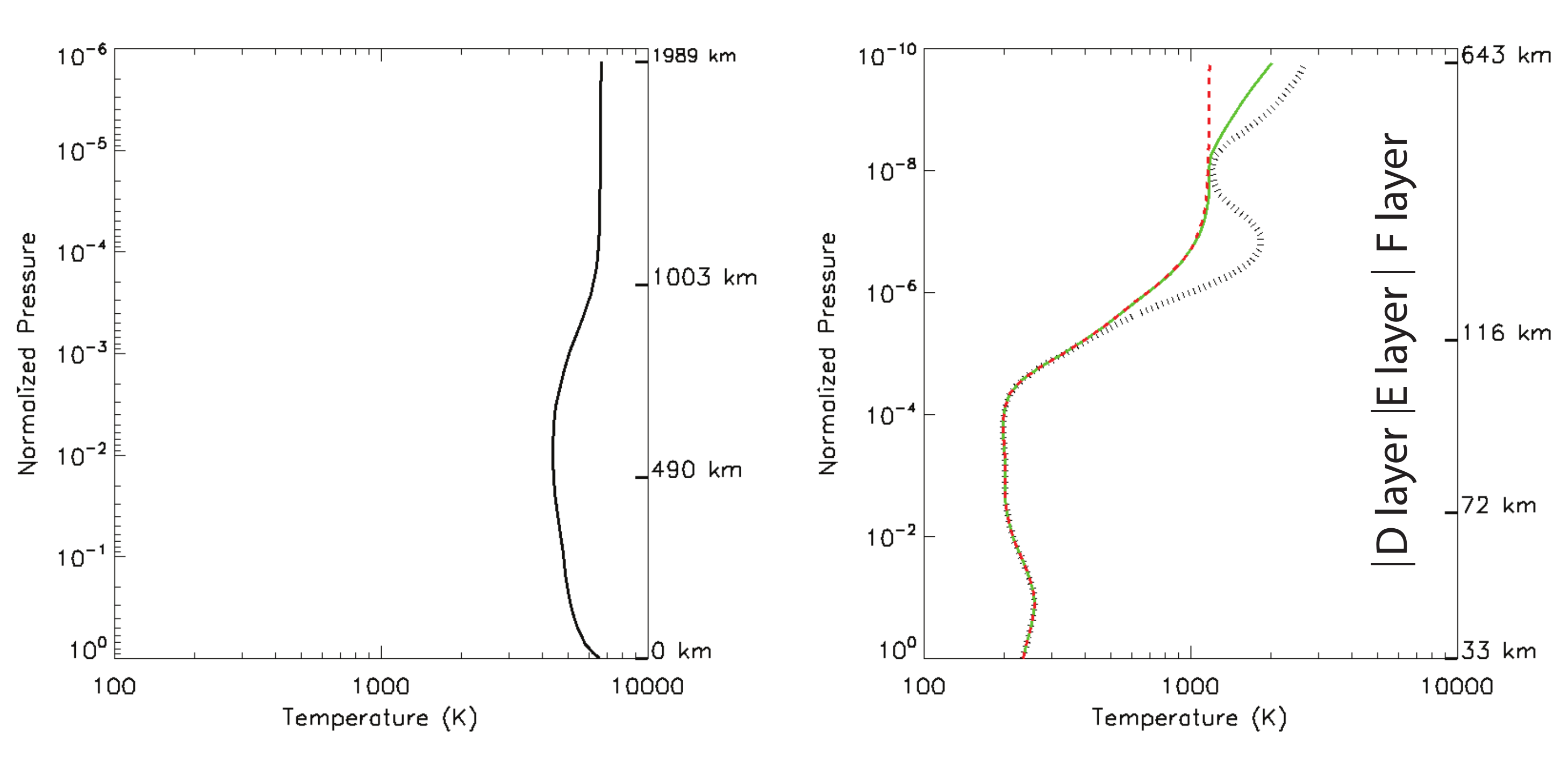}
\caption{Single-fluid temperature (K) in the ALC7 chromosphere (left) and neutral (red dashed line), ion (green solid line), and electron (black dotted line)
temperatures in the TIMEGCM I/T (right). 
\label{fig:temperatures}}      
\end{figure}

Figure \ref{fig:temperatures} depicts the temperature profiles in the ALC7 chromosphere and the TIMEGCM I/T. 
Overall, the chromosphere is about one order of magnitude hotter (4,400--6,700 K) than the I/T (200--2,800 
K). All species temperatures are assumed to be equal in the model chromosphere (left), whereas the neutral, electron and ion temperatures are allowed to differ in the model I/T (right).  We also note that the I/T values are plotted down to the reference pressure level near 30 km (stratosphere), for comparison with the chromospheric profile, even though the I/T altitudes are at 80-600 km.  Both the solar and Earth profiles show a decline with altitude in the lower atmosphere 
to a minimum value, beyond which the temperature begins to rise in the upper atmosphere. In the chromosphere, the temperature 
increase is due to both local heating, the nature of which is not well understood, and downward heat conduction from the 
overlying, much hotter solar corona. The increasing ionization fraction of the chromosphere  and transition region above with height is a direct 
consequence of this temperature increase. In the I/T, on the other hand, the neutral gas is photo-ionized by 
incident UV radiation from the Sun. The excess kinetic energy of the liberated photoelectrons is thermalized by 
electron-electron collisions and is transferred to the ions and neutrals by collisional thermal equilibration, 
accompanied by additional electron liberation due to impact ionization. Additional heating processes contribute 
at high latitudes, including Joule heating and energetic particle precipitation in the polar caps. 
Because the equilibration rate is much higher between ions and neutrals than with electrons, the ion and neutral 
temperatures are essentially equal to each other and equal to (at low altitudes) or below (at mid altitudes)
the temperature of the electrons. Above about 300 km altitude in the thermosphere, Coulomb coupling between electrons and ions becomes 
increasingly important, and the three temperatures increasingly separate, with electrons the hottest and neutrals 
the coolest. We speculate that a fully comprehensive model of the chromosphere would show similar qualitative 
trends in the coupling of the neutral and plasma temperatures, but the details of the temperature profiles of 
the species would depend sensitively on how the unknown heating mechanisms partition thermal energy among the 
particles.

The approximate altitude ranges of the ionospheric D, E, and F-layers are shown in  the right-hand panel of Figure \ref{fig:temperatures}.
The lowest, D, layer of the ionosphere extends 
from about 60 to 90 km in altitude. Its dominant neutral is molecular nitrogen (N$_2$), while its dominant 
ion is nitric oxide (NO$^+$) photo-ionized by penetrating Ly $\alpha$ radiation at $\lambda$ 121.5 nm. 
Water cluster ions can also be significant in the D-layer. The 
middle, E, layer extends upward from 90 km to about 150 km altitude. N$_2$ remains the dominant neutral species, 
but at this height solar soft X-ray and far UV radiation, together with chemical reactions, add molecular oxygen 
ions (O$_2^+$) to the plasma. 
The highest, F, layer ranges from about 180 km to well over 500 km in altitude. At these heights, due to molecular 
dissociation at the elevated temperatures and ionization by extreme ultraviolet radiation, atomic oxygen is dominant 
in both its neutral (O) and ionized (O$^+$) states. 

In contrast to the rich compositional structure of the Earth's 
I/T, the composition of the chromosphere is relatively uniform with altitude, consisting primarily of hydrogen 
(H, H$^+$), secondarily of helium {\color{black} (He, He$^+$) up to 10\% that of Hydrogen}, and thereafter a smattering of minority neutrals and ions 
up to iron at much smaller concentrations. However, the radiation in the chromosphere is dominated by spectral lines 
in some of these minority species, such as calcium, magnesium and iron. Plasma-neutral coupling in the cool material 
of solar prominences (see \S \ref{rayleightaylor}) has been shown to lead to species separation and preferential 
draining of He relative to H \citep{2002ApJ...577..464G,2007ApJ...671..978G}. This occurs due to the very strong 
charge-exchange collisional coupling of H to H$^+$, which retains the majority hydrogen atoms while the minority 
helium atoms leak out much more freely. In the solar chromosphere, the constantly churning convection driven from 
below maintains the roughly uniform composition through turbulent mixing. 
In contrast, in the Earth's atmosphere, above about 100 km (the turbopause) turbulent mixing is too weak to homogenize the atmosphere and maintain a uniform composition, so the atmospheric composition becomes stratified under gravitational attraction according to the species molecular weights.


\begin{figure}[h]
\includegraphics[width=\textwidth]{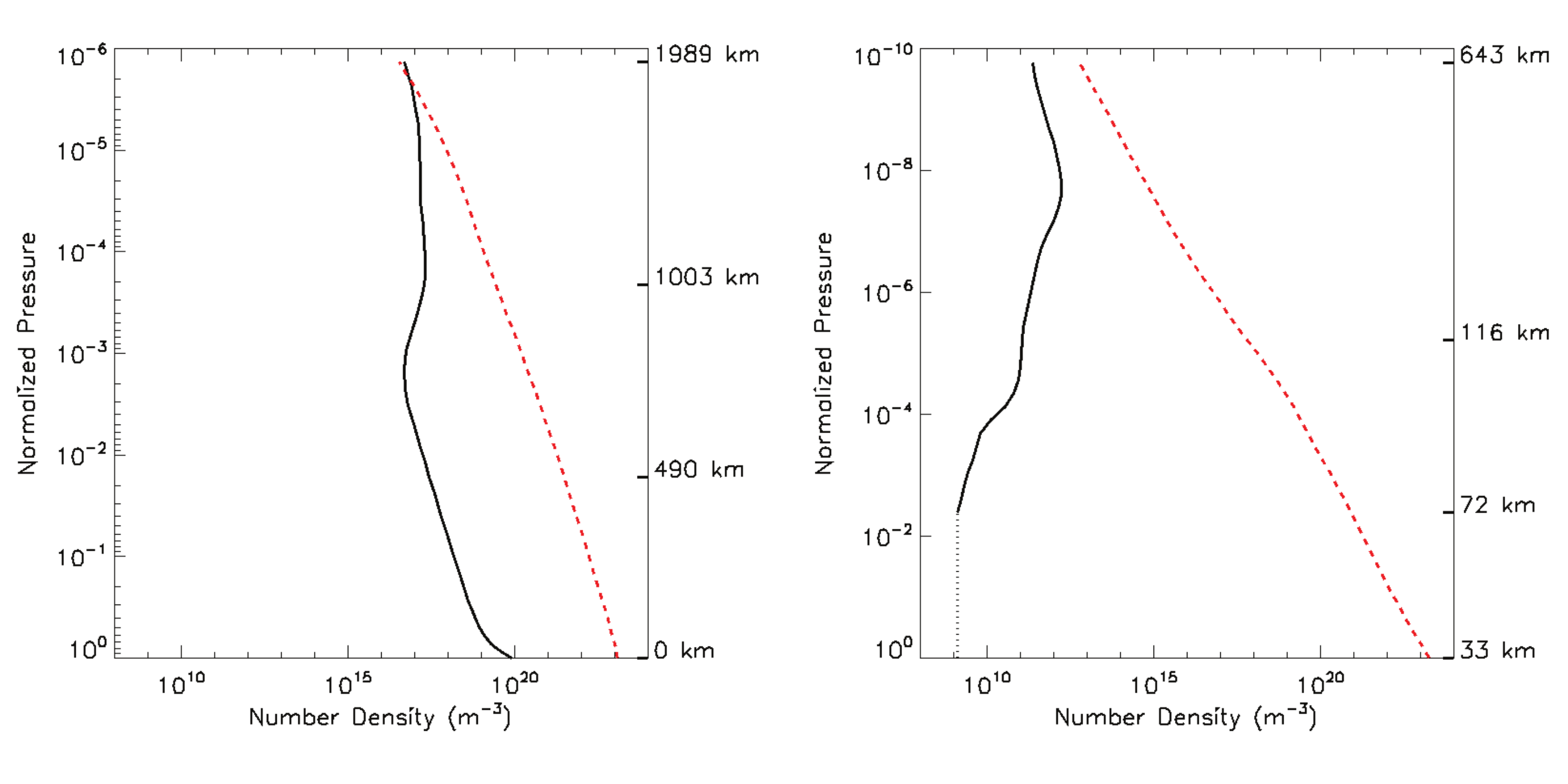}
\caption{Neutral (red dashed line) and plasma (black solid line) number densities ($\textrm{m}^{-3}$) vs.\ normalized pressure in the ALC7 chromosphere 
(left) and the TIMEGCM I/T (right).
\label{fig:densities}}    
\end{figure}

The number densities of the neutral and plasma constituents of the atmospheres are shown in Figure \ref{fig:densities}. The 
neutral densities are nearly equal in the ALC7 chromosphere (left) and TIMEGCM I/T (right) over their common range of 
normalized pressures, falling from a value of about $10^{23}$ m$^{-3}$ where the base pressure was chosen in each 
atmosphere. In contrast, the plasma densities differ by several orders of magnitude between the chromosphere and the I/T, 
due to the combination of their order-of-magnitude temperature difference and to the disparate ionization processes that dominate 
in the two atmospheres. Thus, the neutral and plasma densities become nearly equal at the top of the Sun's chromosphere, 
whereas in the Earth's I/T region, the plasma density is much smaller than the neutral density throughout the displayed altitude range.
Equality of the plasma and neutral densities in the I/T occurs only at much higher altitudes than those shown in our graphs. 
The dotted line below 70 km in Figure \ref{fig:densities} denotes a region below the ionosphere proper, where the plasma density is small and poorly characterized in TIMEGCM so we have elected to hold it fixed at its value at 70 km altitude.Ó


\begin{figure}[h]
\includegraphics[width=\textwidth]{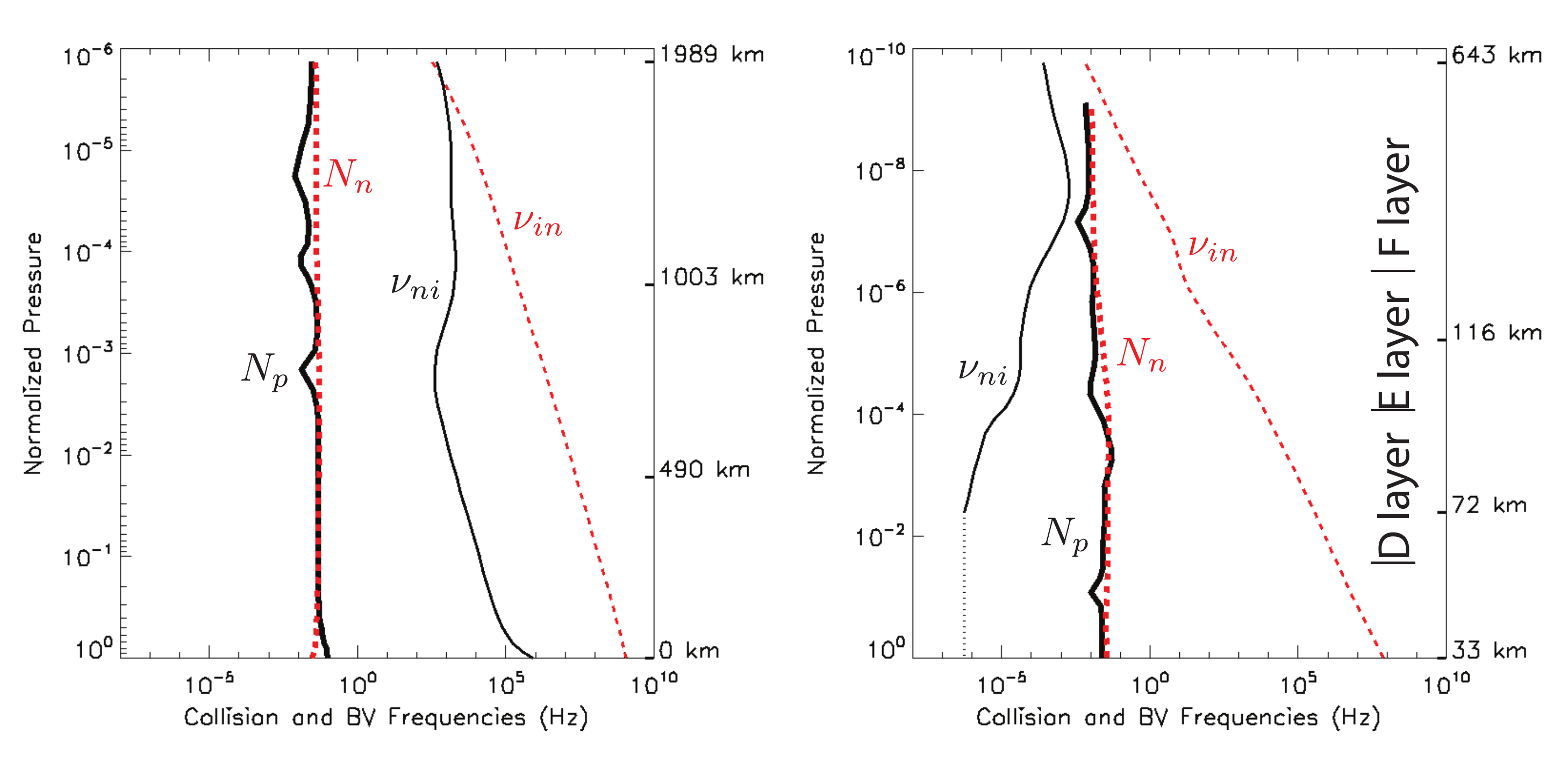}
\caption{ {\color{black} Frequencies in the ALC7 chromosphere (left panel) and the TIMEGCM I/T (right panel).
The thin lines show collision frequency for ions on neutrals ($\nu_{in}$ - red dashed line) and neutrals on ions ($\nu_{ni}$ - black solid line). The 
thick lines show the Brunt-V\"ais\"al\"a frequency for neutrals ($N_{n}$ - red dashed lines) and plasma ($N_{p}$ - black solid line). Note that 
$N_{p}/\nu_{in} < 1$ in both atmospheres, but that $N_{n}/\nu_{ni} < 1$ in the chromosphere while $N_{n}/\nu_{ni} > 1$ in the I/T.}
\label{fig:incollisions} }    
\end{figure}

These very similar neutral densities, but radically different plasma densities, in the chromosphere and I/T have important 
consequences for the roles of plasma-neutral coupling in the dynamics of the mixture. The collision frequencies of ions on 
neutrals ($\nu_{in}$) and of neutrals on ions ($\nu_{ni}$) are shown in Figure \ref{fig:incollisions}. Clearly, the profile 
shapes mostly reflect those of the neutral and plasma densities, respectively. A dependence on the thermal speed of the 
colliding particles introduces variations directly proportional to the square root of the temperature, and inversely 
proportional to the square root of the average particle mass; the ion-neutral collision frequency therefore is about 
an order of magnitude higher in the hydrogen-dominated chromosphere than in the nitrogen and oxygen-dominated I/T. 
The 
frequency $\nu_{in}$ reaches 1 GHz at the base of the ALC7 chromosphere (left) and about 1 MHz at the base of the 
TIMEGCM D layer at an altitude of about 70 km (right). Therefore the ions respond very strongly to the neutrals at the base of both the chromosphere and ionospheric D-layer and respond well above. In contrast, 
the neutral-ion collision frequency $\nu_{ni}$ is relatively high ($\sim$ 1 MHz) at the base of the chromosphere, but 
is miniscule ($\sim$ 1 $\mu$Hz) at the base of the D layer; thus, the response of the neutrals to the ions is strong 
in the chromosphere but extremely weak in the I/T. As a result, it is generally a reasonable approximation to treat the 
neutral density and wind velocity as given in studies of the lower I/T, while feedback on the neutral gas from the ensuing 
plasma dynamics is ignored. However this is a poor approximation in Earth's polar regions during geomagnetic storms, where the ions can strongly influence the neutral gas, and it is not 
well justified in the chromosphere, particularly for disturbances occurring at long scale lengths and low frequencies.

An explicit example of how these considerations apply to important chromospheric and ionospheric phenomena can be found in 
incompressible motions of the atmospheres. As will be shown by a linear analysis of the multifluid equations in \S 
\ref{rayleightaylor} below, the motions are characterized by the Brunt-V\"ais\"al\"a frequency $N$ 
\citep{1927QJRMS..53...30B,1925SSFCMP...2...19V},
\begin{equation}
  N^2 \equiv \frac{g}{L},
  \label{EqNum2}
\end{equation}
where $g$ is the gravitational acceleration and $L$ is the local scale height of the particle density, neutral or plasma. If the 
associated density is stably stratified (i.e., decreasing with height), then the motions are purely oscillatory. If, on the other 
hand, the density is unstably stratified (increasing with height), then the motions consist of one exponentially damped and 
one exponentially growing mode, the Rayleigh-Taylor instability \citep{Rayleigh_1882,Taylor_1950}. A glance at Figure 
\ref{fig:densities} indicates that the I/T must be susceptible to Rayleigh-Taylor instability, due to the high altitude peak 
in the plasma number density, in the $F$ layer. The plasma number density also increases with altitude for a region of the 
model chromosphere.


Figure \ref{fig:incollisions} shows the Brunt-V\"ais\"al\"a frequencies for the neutral gas and for the plasma, calculated from 
their density profiles in the ALC7 chromosphere (left) and the TIMEGCM I/T (right). The frequencies for the plasma are similar in magnitude to, but much more variable than, those of the neutral gas. The variability is  due to changes in the slopes of the density profiles with height. The frequencies for the neutral gas turn out to be roughly equal in the chromosphere and I/T (between 0.01 and 0.05 Hz).
 This occurs because the 
Sun's much stronger gravity is compensated for by the much higher thermal speed of its neutrals: the neutral frequency $N_n$ is 
essentially the ratio of those quantities, after using the pressure scale height to approximate the neutral-density scale height 
$L_{n}$. {\color{black} In both environments, the frequency $N_p$ set by $L_{p}$ is smaller than the ion-neutral collision frequency,
 $N_p / \nu_{in} < 1$, so that the Brunt-V\"ais\"al\"a oscillations (or Rayleigh-Taylor 
instabilities) of the plasma are affected by coupling to the neutrals.  In the chromosphere, the oscillations and instabilities 
of the neutrals are similarly affected by coupling to the plasma, since there we have $N_n / \nu_{ni} < 1$. However, 
the opposite is true in the I/T, where $N_n / \nu_{ni} > 1$: the oscillations of the stably stratified neutral gas are 
unaffected by the plasma, and the neutral motion is essentially undisturbed by the evolution of the unstably stratified plasma.}
These contrasting consequences of the plasma-neutral coupling will be borne out by the analysis and numerical simulations shown 
below in \S \ref{rayleightaylor}.

{\color{black}
Both the Sun's chromosphere and the Earth's I/T are permeated by magnetic fields. The Lorentz force on charged particles 
acts in the direction perpendicular to the field, so that when the field is sufficiently strong, the properties of the plasma 
become highly anisotropic {\color{black} although, as will be shown later, anisotropy also depends on magnetization, the ratio of gyrofrequency to collision frequency, and so also depends on plasma temperatures}. Some consequences of this anisotropy will be discussed in subsequent sections of the paper. For the 
general considerations presented here, we will assume that the magnetic field of the I/T is locally uniform with a field strength of
5.15$\times$10$^{-5}$ T, which is a good approximation as
the I/T magnetic field is basically a dipole. Combined with the 1D snapshot taken from the TIMEGCM model described above, this 1D, time-independent model 
for the I/T is simple but reasonable. The chromosphere's magnetic field, by contrast,  
has a temporally and spatially varying magnetic field. For example, the magnetic field is known to decay with height, but the field value at the surface is different above quiet Sun
regions compared to active regions. To capture some of this variance we adopt the approach of \citet{Goodman2000,Goodman2004a} and 
use a height-dependent 1D magnetic field model:
\be
B(z) = B_{0}\exp{\left(-\frac{z}{2L_{T}}\right)}
\label{eqn:chromB}
\ee
where $L_{T}=\frac{k_{B}T}{m_{p}g}$ is the local scale height. Furthermore we choose three different values for $B_{0} ~ : ~10,100,100 ~ \textrm{G}$}.


\begin{figure}[h]
\includegraphics[width=\textwidth]{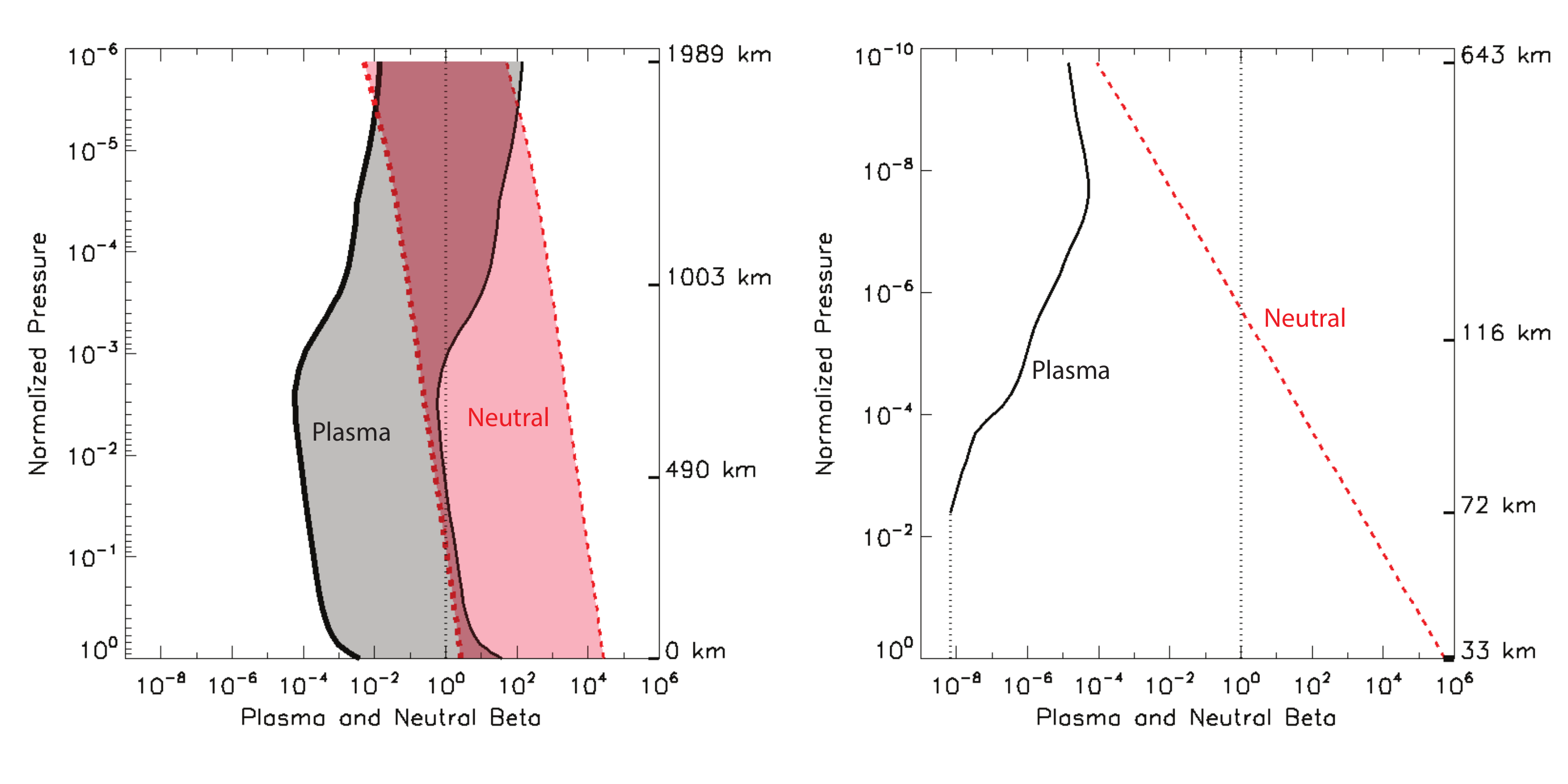}
\caption{ {\color{black} Neutral (red dashed lines) and plasma (black solid lines) $\beta$ in the ALC7 
chromosphere (left) and the TIMEGCM I/T (right). The I/T magnetic field is assumed to be a single value of 5.15$\times$10$^{-5}$ T.  For the chromosphere,  a magnetic field height-dependence (Equation \ref{eqn:chromB}) is assumed, for a choice of three surface field values $B_{0}$. 
The lines for the two extreme values of $B_{0}=10$ G (thin line) and $B_{0}=1000 $ G (thick line) are shown, with a shaded area between them.}
\label{fig:betas}   }    
\end{figure}

A key parameter governing the coupling of the fluid to the magnetic field is the so-called $\beta$, the ratio of the thermal 
pressure ($P$) to the magnetic pressure ($B^2/2\mu_0$) appearing in the equations of motion,
\begin{equation}
  \beta \equiv \frac{2 \mu_0 P}{B^2},
  \label{EqNum10}
\end{equation}
where $\mu_0$ is the magnetic permeability. Values of $\beta$ derived from both the plasma and neutral pressures are displayed 
in Figure \ref{fig:betas} for the ALC7 chromosphere (left) and the TIMEGCM I/T (right).  This dimensionless number 
measures the relative strength of the pressure and magnetic forces exerted on the fluid. It also measures the local amplification 
of the ambient magnetic field that can be accomplished by stagnation-point flows that compress the field in the perpendicular 
direction and evacuate the thermal pressure in the parallel direction.  Total pressure balance between the pre- ($B_0$, $P_0$) 
and post-compression ($B = B_0 + \delta B$, $P = 0$) states implies a fractional amplification of the field strength given by 
\begin{equation}
  \frac{\delta B}{B_0} = \left( 1 + \beta \right)^{1/2} - 1.
  \label{EqNum11}
\end{equation}
Direct dynamical compression of the field by plasma motions not coupled to the neutrals can occur only up to strengths set 
by the beta ($\beta_p$) associated with the plasma pressure ($P_p$).  {\color{black} For the chromosphere, $\beta_{p}$ has an approximate maximum of 100 (Figure \ref{fig:betas}) and the resulting maximum amplification is $\sim 9$, while in the I/T $\beta_{p}$ has a maximum of $10^{-4}$ and so the maximum amplification is $\sim 5\times 10^{-5}$. In principle, plasma motions strongly coupled by collisions 
to flows of the neutral gas could compress the field up to strengths set by the beta ($\beta_n$) associated with the neutral pressure ($P_n$). Using maximum values of $\beta_{n}$ from Figure \ref{fig:betas} of $2\times 10^{4}$ for the chromosphere and $2\times 10^{5}$ for the I/T, would allow amplification factors as large as 140 and 450, respectively.} However, this process 
requires the neutral gas to couple sufficiently strongly to the magnetic field to maintain total force balance through the 
intermediary of fast neutral-ion collisions. This coupling in the I/T is far too weak to maintain such a balance; 
therefore, such strong terrestrial magnetic-field fluctuations are never observed. The coupling in the chromosphere is much 
stronger, and likely is responsible for at least some of the much larger magnetic-field fluctuations observed in the lower 
solar atmosphere. As already noted, the strongest chromospheric magnetic fields originate above sunspots, which form in the 
higher-pressure photosphere and convection zone below the chromosphere. {\color{black} Typical neutral (and plasma) 
flows can range from km/s in convective cells to as large as 1000 of km/s in chromospheric jets. In the lower I/T typical neutral flows are 100 m/s, but at higher latitudes during storms these can increase to 500 m/s.}

\subsection{Summary}
\label{numsum}

The basic atmospheric profiles, and quantities derived from them, that have been discussed in this section clearly show both 
quantitative similarities and differences between the solar chromosphere and the terrestrial I/T. The neutral number 
densities, neutral beta, ion-neutral collision frequencies and Brunt-V\"ais\"al\"a frequencies of both neutrals and plasma 
track reasonably closely over the common range of normalized pressures in the two atmospheres.  On the other hand, the plasma 
number densities, plasma beta, and neutral-ion collision frequencies are very different, due to the much higher ionization fraction on the Sun vs.\ the Earth. 
The plasma is driven strongly by the neutral gas in both atmospheres, as a result, while the neutrals are driven by the ions fairly strongly in the chromosphere but relatively weakly in the I/T.
Relative magnetic 
field fluctuations of order unity are ubiquitous on the Sun, but the fluctuations are far smaller in magnitude on the Earth. 
Both atmospheres exhibit regions of instability driven by a combination of gravity and convection where upward-increasing 
particle number densities occur, modulated by the frequency of collisions between the ions and neutrals. In the remainder 
of the paper, further implications of this rich combination of similarities and differences between the Sun's chromosphere 
and Earth's I/T will be elucidated.

\section{Governing Equations}
\label{equations}

{\color{black} In the partially ionized, collisional mixture of the Sun's chromosphere and Earth's I/T, each species can be treated as a fluid, i.e. each species is collision dominated. There are phenomena where particle distribution functions that deviate from Maxwellian are important, such as during flares in the chromosphere and auroras in the ionosphere, but we consider only here the fluid description of the chromosphere and I/T.

Even though the species are collision dominated, for transient or high frequency (i.e., larger than the collision frequency between species) phenomena the coupling of the individual species to each other by collisions may not be sufficient to allow a description which disregards the difference in the inertial terms in the momentum equation for each species. We define such a description here as a single-fluid model. Here we present a multi-fluid model adopted for both the chromosphere and I/T. Multi-fluid here refers to a model which solves the continuity, momentum and energy equation for all three components, ions, electrons and neutrals. However, we subsequently combine the ion and electron equations to create a two-fluid model which solves the continuity, momentum and energy equation for the neutral fluid and the ionized fluid (electron-ion drift is captured by the Hall term in the generalized Ohm's law). This model is especially relevant  when magnetic fields are present, as they directly affect the ionized component of the mixture but not the neutral component. The details of the model's application in the chromosphere can be found in \citet{Meier12b}, \citet{Leake2012}, and \citet{Leake2013}, where it is shown that the ionized and neutral fluids can decouple as current sheets form and thin in the chromosphere, and hence a multi-fluid model is vital. There are occurrences other than magnetic reconnection sites when multi-fluid models are required. Examples include when high frequency waves from flares propagate down into the chromosphere \citep[e.g.,][]{2002SoPh..206..285V,2010PASJ...62..993K,2011AGUFMSH31A1990E,Russell_2013}, and when waves interact non-linearly to create flows and currents on smaller and smaller scales \citep[see review by][]{1990SSRv...54..377N}.
Due to a low ionization level in the I/T compared to the chromosphere, neutral-ion collisions can be much less frequent than in the chromosphere, and phenomena that occur on timescales of minutes require a multi-fluid model \citep{1988GeoRL..15.1325R,1992GeoRL..19..601R,1994GeoRL..21..417R,Fuller-Rowell_1996,Millward_1996}. 
}

In the multi-fluid model, the three fluids, ions (\textit{i}), electrons (\textit{e}), and neutrals (\textit{n}), can undergo  
recombination and ionization interactions. The ions are assumed to be singly ionized which is a good approximation for the dominant species in both atmospheres. 
The rate of loss of ions/electrons (or gain of neutrals) due to recombination is $\Gamma^{rec}$, and the rate of gain of ions/electrons (or loss of neutrals) due to ionization is $\Gamma^{ion}$.
\newline

\subsection{Continuity}
Assuming charge quasi-neutrality ($n_{i}=n_{e}=n$, where $n_{\alpha}$ is the number density of species $\alpha$), the ion, electron, and neutral continuity equations are:
\bln
\bea
 \frac{\partial n}{\partial t} + \nabla \cdot (n \mb{V}_i) & =
& \Gamma^{ion} - \Gamma^{rec}, \label{eq:cont_i}\\
\frac{\partial n}{\partial t} + \nabla \cdot (n \mb{V}_e) & =
& \Gamma^{ion} - \Gamma^{rec}, \label{eq:cont_e}\\
\frac{\partial n_n}{\partial t} + \nabla \cdot (n_n \mb{V}_n) & 
= & -\Gamma^{ion} + \Gamma^{rec}.\label{eq:cont_n} \eea \eln
Subtracting Equation \ (\ref{eq:cont_e}) from (\ref{eq:cont_i}) then yields
\bln
\be
\nabla \cdot \left( n \left[ \mb{V}_i - \mb{V}_e \right] \right) = 0,\label{eq:cont_diff}
\ee
\eln
which is simply a statement of current conservation (see also Equation \ref{eq:divj} below) in a quasi-neutral plasma.

\subsection{Momentum}
The ion, electron and neutral momentum equations are shown below:
\bln
\bea
\frac{\partial}{\partial t} (m_i n\mb{V}_i) + 
\nabla \cdot (m_i n \mb{V}_i \mb{V}_i + \mbb{P}_i) & = &
 en(\mb{E} + \mb{V}_{i} \times \mb{B}) + m_{i}n\mb{g} + \mb{R}_i^{ie} + \mb{R}_{i}^{in} \nonumber \\
&  & + \Gamma^{ion} m_i \mb{V}_n -
\Gamma^{rec} m_i \mb{V}_i,  \label{eq:mom_i}\\
\frac{\partial}{\partial t} (m_e n \mb{V}_e) + 
\nabla \cdot (m_e n \mb{V}_e \mb{V}_e + \mbb{P}_e) & = &
 -en(\mb{E} + \mb{V}_{e} \times \mb{B}) + m_{e}n\mb{g} + \mb{R}_e^{ei} + \mb{R}_{e}^{en} \nonumber \\
&  & + \Gamma^{ion} m_e \mb{V}_n -
\Gamma^{rec} m_e \mb{V}_e, \label{eq:mom_e}\\
\frac{\partial}{\partial t} (m_n n_n \mb{V}_n) +  \nabla \cdot (m_n
n_n \mb{V}_n \mb{V}_n + \mathbb{P}_n) & = & m_{n}n_{n}\mb{g} + \mb{R}_n^{ne} + \mb{R}_n^{ni} \nonumber \\
&  & -\Gamma^{ion} m_i \mb{V}_n + \Gamma^{rec} \left( m_i \mb{V}_i + m_e \mb{V}_e \right). \label{eq:mom_n}
\eea \eln
The velocity and mass of species $\alpha$ are denoted $\mb{V}_{\alpha}$ and $m_{\alpha}$, respectively. 
The electric and magnetic field are denoted $\mb{E}$ and $\mb{B}$,
respectively.
The pressure tensor is 
 $ \mbb{P}_\alpha = P_\alpha \mbb{I} + \pi_\alpha$ where $P_{\alpha}$ is
the scalar pressure and $\pi_\alpha$ is the viscous stress tensor.
{\color{black} For the neutral fluid this is isotropic, but for electrons and ions this has elements that are 
functions of the particle magnetization (the magnetization is the ratio of the gyrofrequency to the collision frequency), and is anisotropic. The reader can find derivations of these 
viscous stress tensors in \citet{Braginskii}.}
The Coriolis force has been omitted 
for simplicity in further derivations, although its effects can be important in the I/T \cite[e.g.,][]{1984P&SS...32..469F}.
On the Sun, only 
long time-scale phenomena are affected by this force, for example, the rotation of sunspot groups and the 
large-scale solar dynamo. The transfer of momentum to species $\alpha$ 
due to a combination of identity-preserving collisions and charge-exchange collisions with species $\beta$ is given by  
\bln
\be
\mb{R}_\alpha^{\alpha\beta} \equiv m_{\alpha\beta}n_{\alpha}\nu_{\alpha\beta}(\mb{V}_{\beta}-\mb{V}_{\alpha}),
\label{eqn:R}
\ee
\eln
where $m_{\alpha\beta}=m_{\alpha}m_{\beta}/(m_{\alpha}+m_{\beta})$ and $\mb{R}_{\alpha}^{\alpha\beta} = - \mb{R}_{\beta}^{\beta\alpha}$.
The relative importance of charge-exchange collisions and identity-preserving collisions 
varies in the solar atmosphere and the I/T, but for the 
purposes of a comparison of the two plasma environments we 
combine the two types of collisions into one general ``interaction''. The collision frequency $\nu_{\alpha\beta}$ is then defined using a solid body approximation with a relevant choice of cross-section \citep[e.g.,][]{Leake2013}. The above equations are for a single ``average'' species of ions of mass $m_{i}$ and neutrals of mass $m_n$. 
 
\subsection{Energy}
The full derivation of the energy equations for all three components can be found in \citet{Meier12b}. The equation for the rate of change of the thermal plus kinetic energy $\epsilon_{\alpha} = \rho_{\alpha}{V}_{\alpha}^{2}/2 + P_{\alpha}/(\gamma_{\alpha}-1)$ of species $\alpha$, neglecting ionization and  recombination, is
\bln
\begin{equation}
\frac{\partial \epsilon_{\alpha}}{\partial t} +\nabla \cdot (\epsilon_{\alpha}\mb{V}_{\alpha} + \mb{V}_{\alpha}\cdot\mathbb{P}_{\alpha} + \mb{h}_{\alpha}) = \mb{V}_{\alpha}\cdot\left(q_{\alpha} n_{\alpha}\mathbf{E} +\sum_{\beta\ne\alpha}{ \mathbf{R}_{\alpha}^{\alpha\beta}} + m_{\alpha}n_{\alpha}\mb{g}\right) + \sum_{\beta\ne\alpha}{Q_{\alpha}^{\alpha\beta}} + S_{\alpha} + U_{\alpha} \label{energy_eqn}
\end{equation}
\eln
where $\rho_{\alpha}=m_{\alpha}n_{\alpha}$, $\gamma_{\alpha}$ is the ratio of specific heats, $q_{\alpha}$ is the charge,  
\bln
\be
Q_{\alpha}^{\alpha\beta} = \frac{1}{2}\mb{R}_{\alpha}^{\alpha\beta}\cdot(\mb{V}_{\beta}-\mb{V}_{\alpha}) + 3\frac{m_{\alpha\beta}}{m_{\alpha}}n_{\alpha}\nu_{\alpha\beta}k_{B}(T_{\beta}-T_{\alpha})
\ee 
\eln
is the heating of species $\alpha$ due to collisions with species $\beta$, $T_{\alpha}$ is the temperature of the fluid, $S_{\alpha}$ and $U_{\alpha}$ are radiative and chemical process contributions, respectively, and $\mb{h}_{\alpha}$ is the heat flux. {\color{black} This heat flux involves a thermal conductivity tensor that depends on the magnetization of the species , see \citet{Braginskii}.}

 
\subsection{Maxwell's Equations}
\label{maxwell_section}
 The equations relating changes in the electric and magnetic field are
 \bln
\begin{eqnarray}
\nabla \times \mb{E}  & = &  -\frac{\partial \mb{B}}{\partial t}, \label{eq:faraday}\\
 \nabla \times \mb{B}  & = &  \epsilon_{0}\mu_{0}\frac{\partial \mb{E}}{\partial t}+\mu_{0}\mb{J}, \label{eq:ampere} \\
 \nabla \cdot \mb{B} & = & 0, \label{eq:divB} \\
 \nabla \cdot \mb{E} & = &  \sigma / \epsilon_{0}.
\end{eqnarray}
\eln
Here $\sigma$ is the charge density, $\epsilon_{0}$ is the permittivity, and $\mu_{0}$ is the permeability. 
The second of these Maxwell equations can be written as an equation for $\mb{J}$:
\bln
\be
\mb{J} = \frac{\nabla \times \mb{B}}{\mu_{0}} - \epsilon_{0}\frac{\partial E}{\partial t},
\ee
\eln
the second term being the displacement current. Taking the ratio of the magnitude of the two terms on the right-hand side gives 
\bln
\be
\frac{|\frac{\nabla \times \mb{B}}{\mu_{0}} |}{|\epsilon_{0}\frac{\partial \mb{E}}{\partial t}|} \sim \frac{B_{0}t_{0}}{L_{0}\mu_{0}\epsilon_{0} E_{0}} \sim \frac{t_{0}^{2}c^{2}}{L_{0}^2}
\label{displacement}
\ee
\eln
where $c=(\mu_{0}\epsilon_{0})^{-1/2}$ is the speed of light, subscripts ``0'' indicate representative values for each variable, and $B_{0}/t_{0} \sim E_{0}/L_{0}$ is used from the first Maxwell equation (\ref{eq:faraday}).
From Equation (\ref{displacement}), the displacement current can be ignored in the equation for $\mb{J}$ if the system time scale $t_{0}$ is longer than the time it takes light to travel across the system $L_{0}/c$. This is a reasonable assumption in the Sun and the Earth's atmospheres, and leaves the equation for $\mb{J}$ in the static form of Amp\`ere's law:
\bln
\be
\mb{J} = \frac{\nabla \times \mb{B}}{\mu_{0}}.
\label{current}
\ee
\eln
For reasons discussed in detail later, in ionospheric literature, the current conservation equation  
\bln
\be
\nabla\cdot\mb{J}=0 \label{eq:divj}
\ee
\eln
is used, rather than Equation (\ref{current}), with the B field represented as a conservative field. We note that Equation (\ref{eq:divj}) follows immediately from Equation (\ref{current}) and the identity $\nabla\cdot(\nabla \times \mb{B}) = 0$, and is equivalent to Equation (\ref{eq:cont_diff}).
{\color{black} In cases where the characteristic time scale of variation is long compared to the Alfv\'{e}n wave travel time \citep{Vas_2012}, one can assume that $\partial \mb{B} / \partial t \approx 0$, $\nabla \times \mb{E} = 0$, and the electric field can be written as the gradient of a scalar, $\mb{E} = - \nabla \Phi$.

}

\subsection{Ohm's Law}
\label{genohmslaw}

{\color{black} 
The generalized Ohm's law can be interpreted as an equation for the temporal evolution of the current density 
$\partial \mb{J}/\partial t$. Thus, along with the momentum equations and Maxwell's equations the system equations contain time derivatives for all four variables $\mb{B}, ~ \mb{V}, ~\mb{E}, ~ \textrm{and} ~\mb{J}$. 
As will be shown in this section, under certain assumptions (e.g. low-frequency or long timescale), we can ignore $\partial \mb{J}/\partial t$ and rewrite the Ohm's law as a linear equation 
which relates the current density $\mb{J}$ to the electric field $\mb{E}^{*}$ in a specific rest frame using either the conductivity $\underline{\underline{\sigma}}$ or resistivity $\underline{\underline{\eta}}$ tensor. The time derivative $\partial  \mb{E}/\partial t$ can also be dropped for timescales longer than the light crossing time of the system (see above). This leaves just
the time derivatives for $\mb{B}$ and $\mb{V}$ (we will discuss this MHD approach in the following two sections).
Ionospheric physicists prefer the use of the conductivity formulation $\mb{J}= \underline{\underline{\sigma}}\cdot\mb{E}^{*}$, since electric fields can persist due to the generally low conductivity of the plasma, while solar physicists generally use the resistivity formulation $\mb{E}^{*}=\underline{\underline{\eta}}\cdot\mb{J}$. 

Below is a derivation of the generalized Ohm's law, valid for both the chromosphere and I/T. {\color{black} Firstly, we neglect the ionization and recombination terms in the three 
momentum equations (\ref{eq:mom_i}-\ref{eq:mom_n}). }
{\color{black} The generalized Ohm's law can be obtained by taking the momentum equation for each species $\alpha$ (\ref{eq:mom_i}-\ref{eq:mom_n}), multiplying by the charge to mass ratio $q_{\alpha}/m_{\alpha}$ ($q_{i}=e=-q_{e}$ and $q_{n}=0$), and summing over all species. Using the definition of $\mb{R}_{\alpha}^{\alpha\beta}$ in Equation (\ref{eqn:R}),
and the fact that for our singly ionized ion model $m_{n}=m_{i}+m_{e}$ so that $m_{in} \approx m_{i}/2$, $m_{ei}=m_{en} \approx m_{e}$, 
gives
\bln
\bea
\frac{\partial\mb{J}}{\partial t} = \frac{\partial (en\mb{V}_{i}-en\mb{V}_{e})}{\partial t} =&   & \frac{ne^{2}}{m_{e}}(\mb{E}+\mb{V}_{e}\times\mb{B} + \frac{m_{e}}{m_{i}}\frac{\mb{j}\times\mb{B}}{en})  
+ e\left(\frac{\nabla \cdot \mbb{\kappa}_{e}}{m_{e}} - \frac{\nabla \cdot \mbb{\kappa}_{i}}{m_{i}}\right) \nonumber \\
 & - &  (\nu_{ei}+\nu_{en})\mb{J} + en(\nu_{en}-\frac{\nu_{in}}{2})\mb{W}.
\label{eqn:ohmslaw_0} 
\eea
\eln
where we have neglected terms of order $1/m_{i}$ relative to terms of order $1/m_{e}$. The tensor $\mbb{\kappa}_{\alpha}\equiv \mbb{P}_{\alpha}+\rho_{\alpha}\mb{V}_{\alpha}\mb{V}_{\alpha}$.
We call Equation (\ref{eqn:ohmslaw_0}) the generalized Ohm's law, as it is derived from the momentum equations with minimal assumptions. In solar physics, the term "generalized Ohm's law" is often used to describe a linear relationship between the electric field and current density which includes more then just anisotropic resistivity. For example, fluid models which include Hall or Pedersen currents may all contain a ``generalized Ohm's law". In this paper, the term applies only to Equation (\ref{eqn:ohmslaw_0}) and any algebraic manipulations of it. When certain assumptions and simplifications are made to Equation (\ref{eqn:ohmslaw_0}), we just refer to the result as ``Ohm's law''.}

{\color{black} In ionospheric applications \citep[e.g.,][]{Vas_2012} Equation (\ref{eqn:ohmslaw_0}) can be recast so that the velocities on the right hand side are the plasma velocity $V_{p}$, defined by
\bln
\be
\mb{V}_{p}\equiv \frac{(m_{i}\mb{V}_{i}+m_{e}\mb{V}_{e})}{(m_{i}+m_{e})}\approx \mb{V}_{i} + \frac{m_{e}}{m_{i}}\mb{V}_{e} \approx \mb{V}_{i} - \frac{m_{e}}{m_{i}}\frac{\mb{J}}{en} \approx \mb{V}_{e} + \frac{\mb{J}}{en}.
\ee
\eln
where the approximations use $m_{e} \ll m_{i}$.
Some simple algebra shows that Equation (\ref{eqn:ohmslaw_0}) is equivalent to 
\bln
\bea
 \frac{\partial\mb{J}}{\partial t}& =  & \frac{ne^{2}}{m_{e}}(\mb{E}+\mb{V}_{p}\times\mb{B} + \frac{\mb{j}\times\mb{B}}{en})  
+ e\left(\frac{\nabla \cdot \mbb{\kappa}_{e}}{m_{e}} - \frac{\nabla \cdot \mbb{\kappa}_{i}}{m_{i}}\right) \nonumber \\
 & - &  (\nu_{ei}+\nu_{en}+\frac{m_{e}}{m_{i}}\frac{\nu_{in}}{2})\mb{J} + en(\nu_{en}-\frac{\nu_{in}}{2})(\mb{V}_{p}-\mb{V}_{n})
\label{eqn:ohmslaw_1} 
\eea
\eln
which is equivalent to the generalized Ohm's law presented in \citet{Vas_2012}, Equation (12), for a quasi-neutral plasma ($n_{i}=n_{e}$), though in that work the factor 1/2 is dropped from the $\nu_{in}$ terms.
}



It is worth deriving some initial quantities relevant to the electrodynamics at this point. 
The following magnetizations (the gyrofrequencies $\Omega_{\alpha}\equiv eB/m_{\alpha}$ divided by the collision frequencies $\nu_{\alpha\beta}$): 
\bln
\bea
k_{in} & = & \frac{eB}{\frac{m_{i}}{2}\nu_{in}}, \label{eqn:magn1} \\
k_{en} & = & \frac{eB}{m_{e}\nu_{en}},\\
k_{ei} & = & \frac{eB}{m_{e}\nu_{ei}}, \\
\frac{1}{k_{e}} & \equiv & \frac{1}{k_{en}}+\frac{1}{k_{ei}},
\label{eqn:magn_mult}
\eea
\eln
are measures of the ability of the ions and electrons to freely gyrate around the magnetic field. For example, if $k_{e} > 1$, then the gyration of the electrons around the magnetic field is largely unaffected by collisions with ions and neutrals. When $k_{e} < 1$, then the collisions decouple the electron from the field. The same situation applies to the ions. Note that ion-electron collisions hardly affect the gyration of the ions, so $k_{i}$, defined in the same way as $k_{e}$, is approximately $k_{in}$, and from hereon, we use $k_{in}$ instead of $k_{i}$.

{\color{black} Multiplying Equation (\ref{eqn:ohmslaw_0}) by $\frac{m_{e}}{e^{2}n}$, using the definitions of magnetizations above, noting that $k_{in}/k_{en} \sim \sqrt{m_{e}/m_{i}} \ll 1 $, and neglecting terms of order $\sqrt{m_{e}/m_{i}}$ 
relative to terms of order 1, we obtain 

\bln
\bea
 \mb{E} + (\mb{V}_{e}\times\mb{B}) & = & \left[\frac{1}{k_{ei}}+\frac{1}{k_{en}}\right]\frac{B}{en}\mb{J} - \frac{m_{e}}{m_{i}}\frac{B}{en}\mb{J}\times\hat{\mb{b}}
 \nonumber \\
& - &  \frac{B}{k_{en}}\mb{W} -  \frac{1}{en}\left(\nabla \cdot \mbb{\kappa}_{e} - \frac{m_{e}}{m_{i}}\nabla \cdot \mbb{\kappa}_{i}\right)
+ \frac{m_{e}}{e^{2}n}\frac{\partial \mb{J}}{\partial t}.
\label{eqn:ohmslaw_elec}
\eea
\eln
}
Later on we shall see how certain assumptions allow us to drop the term proportional to $\partial \mb{J}/\partial t$ to derive a low-frequency Ohm's law. This relates the electric field in a certain rest frame to the current density. Note that Equation (\ref{eqn:ohmslaw_elec}) contains the electric field in the rest frame of the electrons $\mb{E}+(\mb{v}_{e}\times\mb{B})$. As we shall see later on in this section, there are other choices of rest frame which affect the interpretation of different terms in the equation. 

{\color{black} To close the system of equations}, an equation is needed for the difference between the ion and neutral velocities, $\mb{W}=\mb{V}_{i}-\mb{V}_{n}$. This is obtained by taking the sum of the ion and electron momentum equations (\ref{eq:mom_i}+\ref{eq:mom_e}) divided by the plasma density $m_{i}n$, and subtracting the neutral momentum equation (\ref{eq:mom_n}) divided by the neutral density $m_{i}n_{n}$, again neglecting the ionization and recombination terms. This results in an equation for $d_{i}\mb{V}_{i}/dt - d_{n}\mb{V}_{n}/dt$, where $d_{\alpha}/dt = \partial / \partial t + \mb{V}_{\alpha}\cdot\nabla$ is the total Lagrangian derivative for a fluid of species $\alpha$. This equation  can be rearranged to express $\mb{W}$ as:
{\color{black} 
\bln
\be
\mb{W} = \frac{k_{in}}{k_{en}+k_{in}}\left\{
 \frac{\mb{J}}{en} + k_{en}\left[
 \frac{\xi_{n}\mb{J}\times\mb{\hat{b}}}{en} + \frac{\xi_{i}\nabla \cdot \mbb{P}_{n}}{eBn} - \frac{\xi_{n}\nabla \cdot \mbb{P}_{p}}{eBn} - \frac{\xi_{n}m_{i}}{eB} \frac{d^{*}\mb{W}}{dt}
 \right] 
 \right\}\label{eq:w}
\ee
\eln
}
where $\mbb{P}_{p}=\mbb{P}_{e}+\mbb{P}_{i}$ and $d^{*}\mb{W}/dt \equiv \left(\frac{d_{i}\mb{V}_{i}}{dt}-\frac{d_{n}\mb{V}_{n}}{dt}\right)$.
Using this equation for $\mb{W}$ in Equation (\ref{eqn:ohmslaw_elec}) yields
\{\color{black} \bln
\bea
 \mb{E} & + & (\mb{V}_{e}\times\mb{B})= \left[\frac{1}{k_{ei}}+\frac{1}{k_{en}+k_{in}}\right]\frac{B}{en}\mb{J} 
- \left[\frac{\xi_{n}k_{in}}{k_{en}+k_{in}}\right]\frac{B}{en}\mb{J}\times\hat{\mb{b}} +  \frac{m_{e}}{e^{2}n}\frac{\partial \mb{J}}{\partial t}
 \nonumber \\
& + &
\frac{k_{in}}{k_{en}+k_{in}}\left(\frac{\xi_{n}\nabla \cdot \mbb{P}_{p}-\xi_{i}\nabla \cdot \mbb{P}_{n}}{en}\right)  -  \frac{1}{en}\left(\nabla \cdot \mbb{\kappa}_{e} - \frac{m_{e}}{m_{i}}\nabla \cdot \mbb{\kappa}_{i}\right) 
 + \frac{\xi_{n}k_{in}}{k_{en}+k_{in}}\frac{m_{i}}{e}\frac{d^{*}\mb{W}}{dt}.
\label{gol_elec}
\eea
\eln
}

{\color{black} This is the generalized Ohm's law in the rest frame of the electrons. 
The first term that multiplies $\mb{J}$ on the RHS is conventionally called the Ohmic resistivity. It describes the collisions of electrons and ions which act to dissipate electric current. By substituting the equation for $\mb{W}$ (\ref{eq:w}) into the generalized Ohm's law, we introduced the effect of ion-neutral collisions into the Ohmic 
resistivity, this is because $\mb{W}$ is linearly related to $\mb{J}$ (for low frequency phenomena). Whereas in Equation (\ref{eqn:ohmslaw_elec}), the term in front of $\mb{J}$ only included electron collisions, now the Ohmic resistivity contains  two terms. The first $1/k_{ei} \sim \nu_{ei}$
describes direct collisions of electrons and ions. The second $1/(k_{en}+k_{in}) \sim \nu_{en}\nu_{in}/(\nu_{en}+\nu_{in})$ describes the combined effect of electron-neutral collisions ($k_{en}$) and ion-neutral ($k_{in}$) collisions. As the neutral density goes to zero, $k_{en}$ and $k_{in}$ tend to infinity and so the Ohmic resistivity becomes just direct electron-ion collisions.}

{\color{black} The generalized Ohm's law can be recast in the rest frame of the plasma using $\mb{V}_{e} \approx \mb{V}_{p} - \mb{J}/en$:
\bln
\bea
\mb{E}^{p} & \equiv & \mb{E}+(\mb{V}_{p}\times\mb{B}) \approx \left[\frac{1}{k_{ei}}+\frac{1}{k_{en}+k_{in}}\right]\frac{B}{en}\mb{J} 
+ \left[1-\frac{\xi_{n}k_{in}}{k_{en}+k_{in}}\right]\frac{B}{en}\mb{J}\times\hat{\mb{b}} + \frac{m_{e}}{e^{2}n}\frac{\partial \mb{J}}{\partial t}
\nonumber \\
& + & \frac{k_{in}}{k_{en}+k_{in}}\left(\frac{\xi_{n}\nabla \cdot \mbb{P}_{p}-\xi_{i}\nabla \cdot \mbb{P}_{n}}{en}\right)   
 -  \frac{1}{en}\left(\nabla \cdot \mbb{\kappa}_{e} - \frac{m_{e}}{m_{i}}\nabla \cdot \mbb{\kappa}_{i}\right) 
 + \frac{\xi_{n}k_{in}}{k_{en}+k_{in}}\frac{m_{i}}{e}\frac{d^{*}\mb{W}}{dt}.
\label{gol_ions1}
\eea
\eln
}

The rest frame of the plasma generally is not the best choice for a frame of reference in a weakly ionized mixture. 
In the I/T, where the neutral fraction is so much greater than the ion fraction, the neutral rest frame is more commonly used. A discussion on the choice of rest frame and its consequences for the discussion of frictional heating is given in \S \ref{joule}. {\color{black} Using 
\bln
\be
\mb{V}_{n} = \mb{V}_{i}-\mb{W} \approx \mb{V}_{p} + \frac{m_{e}}{m_{i}}\frac{\mb{J}}{en} - \mb{W}
\ee
\eln
and noting that when we substitute this into Equation (\ref{gol_ions1}), the resulting $\frac{m_{e}}{m_{i}}\frac{\mb{J}\times\hat{\mb{b}}}{en}$ can be neglected relative to $\frac{\mb{J}\times\hat{\mb{b}}}{en}$,
gives 
\bln
\begin{eqnarray}
\mb{E}^{*} & \equiv &  \mb{E}+(\mb{V}_{n}\times\mb{B}) \approx \mb{E}+(\mb{V}_{p}\times\mb{B}) +\frac{m_{e}}{m_{i}}\frac{(\mb{J}\times{\mb{B}})}{en} - (\mb{W}\times\mb{B})\nonumber \\
& =  &   \left[\frac{1}{k_{ei}} + \frac{1}{k_{en}+k_{in}}\right] \frac{B}{en}\mb{J} +
	        	          \left[\frac{k_{en}}{k_{en}+k_{in}}- \frac{\xi_{n}k_{in}}{k_{en}+k_{in}}\right]\frac{B}{en}\mb{J}\times\hat{\mb{b}}   + \frac{m_{e}}{e^{2}n}\frac{\partial \mb{J}}{\partial t}\nonumber \\
		          & - & \frac{B}{en}\left[\frac{\xi_{n}k_{en}k_{in}}{k_{en}+k_{in}}\right]\left(\mb{J}\times\hat{\mb{b}}\right)\times\hat{\mb{b}} - \frac{1}{en}\left(\nabla \cdot \mbb{\kappa}_{e} - \frac{m_{e}}{m_{i}}\nabla \cdot \mbb{\kappa}_{i}\right)  \nonumber \\
		      	&+ & \frac{k_{in}}{k_{en}+k_{in}}\left(\frac{\xi_{n}\nabla \cdot \mbb{P}_{p}-\xi_{i}\nabla \cdot \mbb{P}_{n}}{en}\right)
                                  + \frac{\xi_{n}k_{in}}{k_{en}+k_{in}}\frac{m_{i}}{e}\frac{d\mb{W}}{dt} \nonumber \\
			& + & \frac{k_{en}k_{in}}{k_{en}+k_{in}}\frac{\left(\xi_{n}\nabla \cdot \mbb{P}_{p}\times\hat{\mb{b}}-\xi_{i}\nabla \cdot \mbb{P}_{n}\times\hat{\mb{b}}\right)}{en}
		         +  \frac{\xi_{n}k_{en}k_{in}}{k_{en}+k_{in}}\frac{m_{i}}{e}\left(\frac{d^{*}\mb{W}}{dt}\right)\times\hat{\mb{b}}.		         \label{gol}
\end{eqnarray}
\eln
}
We note that when $\xi_{n}\rightarrow0$, the fully ionized generalized Ohm's law is recovered.

{\color{black}
In the chromosphere, where the ion fraction can become comparable to the neutral fraction as the plasma is heated up to 10000 K, the center of mass frame is often used \citep[see ][]{mitchner_1973}. 
The center of mass velocity is defined by 
\bln
\be
\mb{V}_{CM} \equiv \xi_{i}\mb{V}_{i}+\xi_{n}\mb{V}_{n} = \mb{V}_{i}-\xi_{n}\mb{W} \approx \mb{V}_{p} + \frac{m_{e}}{m_{i}}\frac{\mb{J}}{en} - \xi_{n}\mb{W}.
\ee
\eln
 Again, noting that when we substitute this into Equation (\ref{gol_ions1}), the resulting $\frac{m_{e}}{m_{i}}\frac{\mb{J}\times\hat{\mb{b}}}{en}$ can be neglected relative to $\frac{\mb{J}\times\hat{\mb{b}}}{en}$
gives

\bln
\begin{eqnarray}
\mb{E}^{CM} & \equiv &  \mb{E}+(\mb{V}_{CM}\times\mb{B}) \approx \mb{E}+(\mb{V}_{p}\times\mb{B}) + \frac{m_{e}}{m_{i}}\frac{(\mb{J}\times{\mb{B}})}{en} - \xi_{n}(\mb{W}\times\mb{B})\nonumber \\
& = &   \left[\frac{1}{k_{ei}} + \frac{1}{k_{en}+k_{in}}\right] \frac{B}{en}\mb{J} +
	        	          \left[1- \frac{2\xi_{n}k_{in}}{k_{en}+k_{in}}\right]\frac{B}{en}\mb{J}\times\hat{\mb{b}} + \frac{m_{e}}{e^{2}n}\frac{\partial \mb{J}}{\partial t}\nonumber \\
		          & - & \frac{B}{en}\left[\frac{\xi_{n}^{2}k_{en}k_{in}}{k_{en}+k_{in}}\right]\left(\mb{J}\times\hat{\mb{b}}\right)\times\hat{\mb{b}} - \frac{1}{en}\left(\nabla \cdot \mbb{\kappa}_{e} - \frac{m_{e}}{m_{i}}\nabla \cdot \mbb{\kappa}_{i}\right)
                                   \nonumber \\
		      	& + & \frac{k_{in}}{k_{en}+k_{in}}\left(\frac{\xi_{n}\nabla \cdot \mbb{P}_{p}-\xi_{i}\nabla \cdot \mbb{P}_{n}}{en}\right)
                                  + \frac{\xi_{n}k_{in}}{k_{en}+k_{in}}\frac{m_{i}}{e}\frac{d\mb{W}}{dt} \nonumber \\
			& + & \frac{\xi_{n}k_{en}k_{in}}{k_{en}+k_{in}}\frac{\left(\xi_{n}\nabla \cdot \mbb{P}_{p}\times\hat{\mb{b}}-\xi_{i}\nabla \cdot \mbb{P}_{n}\times\hat{\mb{b}}\right)}{en}
		         +  \frac{\xi_{n}^{2}k_{en}k_{in}}{k_{en}+k_{in}}\frac{m_{i}}{e}\left(\frac{d^{*}\mb{W}}{dt}\right)\times\hat{\mb{b}}.		         \label{gol_CM}
\end{eqnarray}
\eln

It is important to note the difference between the generalized Ohm's law for the center of mass frame (\ref{gol_CM}) and the generalized Ohm's law for the neutral frame (\ref{gol}).
The pre-factor in front of the $\left(\mb{J}\times\hat{\mb{b}}\right)\times\hat{\mb{b}}$ term, which will be discussed later in terms of the Pedersen conductivity or Pedersen resistivity, contains $\xi_{n}$ for the neutral frame equation and a $\xi_{n}^2$ for the center of mass equation, which is purely a result of the choice of frame. To allow for a collaborative discussion of the chromosphere and I/T in the context of the generalized Ohm's law, we will use the neutral frame equation (\ref{gol}) from hereon, but will refer to other frames where necessary. As will be seen in \S \ref{joule}, changing the reference frame leads to different terms appearing in the thermal energy equations for the species' temperatures, but cannot change the final values for these temperatures which determine the ionization faction.
}

\subsubsection{Low Frequency Approximation to the Ohm's Law}
\label{low_frequency_ohms_law}
{\color{black}
As stated above, the generalized Ohm's law (\ref{gol}) can be simplified to a linear equation relating $\mb{E}$ and $\mb{J}$, under certain assumptions. Let us look at the competing terms in Equation (\ref{gol}), but first define 
the following resistivities:
\bln
\bea
\eta_{\parallel} & \equiv & \frac{B}{en}\left[\frac{1}{k_{ei}} + \frac{1}{k_{en}+k_{in}} \right], \label{eqn:etas_start}\\
\eta_{C} & \equiv & \frac{B}{en}\left[\frac{\xi_{n}k_{en}k_{in}}{k_{en}+k_{in}}\right], \\
\eta_{P} & \equiv & \eta_{\parallel} + \eta_{C}, \\
\eta_{H} & \equiv  & \frac{B}{en}\left[\frac{k_{en}-\xi_{n}k_{in}}{k_{en}+k_{in}}\right],
\label{eqn:etas_end}
\eea
\eln
then Equation (\ref{gol}) becomes
\bln
\begin{eqnarray}
\mb{E}^{*} & = &  \mb{E}+\mb{V}_{n}\times\mb{B} =   \eta_{\|}\mb{J} + \eta_{H}\mb{J}\times\hat{\mb{b}}  -  \eta_{C}\left(\mb{J}\times\hat{\mb{b}}\right)\times\hat{\mb{b}} \nonumber \\
& + & \frac{m_{e}}{e^{2}n}\frac{\partial\mb{J}}{\partial t}  +\frac{\xi_{n}k_{in}}{k_{en}+k_{in}}\frac{m_{i}}{e}\frac{d^{*}\mb{W}}{dt} +  \frac{\xi_{n}k_{en}k_{in}}{k_{en}+k_{in}}\frac{m_{i}}{e}\left(\frac{d^{*}\mb{W}}{dt}\right)\times\hat{\mb{b}} - \frac{1}{en}\left(\nabla \cdot \mbb{\kappa}_{e} - \frac{m_{e}}{m_{i}}\nabla \cdot \mbb{\kappa}_{i}\right)
		          \nonumber \\
			& + &    \frac{k_{in}}{k_{en}+k_{in}}\left(\frac{\xi_{n}\nabla \cdot \mbb{P}_{p}-\xi_{i}\nabla \cdot \mbb{P}_{n}}{en}\right) + \frac{k_{en}k_{in}}{k_{en}+k_{in}}\frac{\left(\xi_{n}\nabla \cdot \mbb{P}_{p}\times\hat{\mb{b}}-\xi_{i}\nabla \cdot \mbb{P}_{n}\times\hat{\mb{b}}\right)}{en}
		         .		         \label{gol_2}
\end{eqnarray}
\eln
We can non-dimensionalize this equation by expressing each variable $A$ as $A=A_{0}\tilde{A}$ where $\tilde{A}$ has no dimensions. The dimensional constants $A_{0}$ are related to each other. For example, if $L_{0}$ is the system length, $B_{0}$ is the system magnetic field strength, and $n_{0}$ is the system number density of plasma, then $v_{0} = B_{0}/\sqrt{(\mu_{0}m_{i}n_{0})}$ is the system velocity, $J_{0} = B_{0}/\mu_{0}L_{0}$ is the system current density, and $f_{0} = v_{0}/L_{0}$ is the system frequency (or inverse time scale). We can then remove the dimensions from  Equation (\ref{gol_2}) by dividing it by $v_{0}B_{0}$. Using the following definitions of Lundquist number (S), ion inertial scale length ($d_{i}$), ion plasma frequency ($\omega_{p,i}$), electron inertial scale length ($d_{e}$), and electron plasma frequency ($\omega_{p,e}$):
\bln
\bea
S & = & \frac{\mu_{0}L_{0}v_{0}}{\eta_{\|}}, ~ d_{i} = \frac{c}{\omega_{p,i}}, ~ \omega_{p,i}^{2} = \frac{e^{2}n_{0}}{m_{i}\epsilon_{0}}, ~ d_{e} = \frac{c}{\omega_{p,e}}, ~ \omega_{p,e}^{2} = \frac{e^{2}n_{0}}{m_{e}\epsilon_{0}}
\eea
\eln
yields the dimensionless equation
\bln
\begin{eqnarray}
& & \tilde{ \mb{E}}  + (\tilde{\mb{V}}_{n}\times\tilde{\mb{B}})  =  \frac{1}{S}\tilde{\mb{J}} +\frac{d_i}{L_{0}}\tilde{\mb{J}}\times\hat{\mb{b}}  -  \frac{1}{S}\frac{\eta_{C}}{\eta_{\|}}\left(\tilde{\mb{J}}\times\hat{\mb{b}}\right)\times\hat{\mb{b}} 
 \nonumber \\
& + &  \left(\frac{d_e}{L_{0}}\right)^{2}\frac{\partial \tilde{\mb{J}}}{\partial \tilde{t}}+ \frac{\xi_{n}k_{in}}{k_{en}+k_{in}}\frac{f_{0}}{\Omega_{i}}\frac{f_{0}}{\nu_{in}}\left[\frac{d^{*}\tilde{\mb{W}}}{d\tilde{t}}+ k_{en}\frac{d^{*}\tilde{\mb{W}}}{d\tilde{t}}\times\hat{\mb{b}}\right]  -  \frac{d_{i}}{L_{0}}\left(\beta_{e}\tilde{\nabla} \cdot \tilde{\mbb{\kappa}_{e}} - \beta_{i}\frac{m_{e}}{m_{i}}\tilde{\nabla} \cdot \tilde{\mbb{\kappa}_{i}}\right)\nonumber \\
		          & + &  \frac{k_{in}}{k_{en}+k_{in}}\frac{d_i}{L_{0}}\left[\left(\xi_{n}\beta_{p}\tilde{\nabla}\cdot\tilde{\mbb{P}_{p}}-\xi_{i}\beta_{n}\tilde{\nabla}\cdot\tilde{\mbb{P}_{n}}\right) + k_{en}\left(\xi_{n}\beta_{p}\tilde{\nabla}\cdot\tilde{\mbb{P}_{p}}-\xi_{i}\beta_{n}\tilde{\nabla}\cdot\tilde{\mbb{P}_{n}}\right)\times\hat{\mb{b}}\right].  
		           \label{gol_3}
\end{eqnarray}
\eln

Assuming that $\tilde{\mb{E}} + (\tilde{\mb{V}}_{n}\times\tilde{\mb{B}})$ is of order 1, then we neglect terms on the RHS of Equation (\ref{gol_3}) which are much less than unity.
For length scales longer than the electron inertial scale $L_{0}\gg d_{e}$, the $\partial \tilde{\mb{J}}/\partial \tilde{t}$ term can be neglected. For the chromosphere, using the smallest density from Figure \ref{fig:densities} of $10^{17} ~ \textrm{m}^{-3}$ gives a maximum for $d_{e}$ of 17 cm. For the ionosphere, using the smallest density from Figure \ref{fig:densities} of $10^{9} ~ \textrm{m}^{-3}$ gives 170 m. Hence, it is safe to neglect the $\partial \tilde{\mb{J}}/\partial \tilde{t}$ term in the chromosphere, and appropriate in the I/T for lengths larger than a km.

The terms containing $d^{*}\tilde{\mb{W}}/d\tilde{t}$ can in general be neglected if the frequency of the system, $f_{0}$, is much smaller than the minimum of the ion-neutral collisional frequency $\nu_{in}$ and the ion gyrofrequency $\Omega_{i}$. 
\bln
\be
f_{0} \ll \min{(\Omega_{i}, \nu_{in})}.
\ee
\eln
This is the low frequency approximation mentioned above. This condition is nearly always true in the highly collisional chromosphere, and mostly true in local I/T models, except on the topside of the ionosphere and extending into the plasmasphere because the collision frequencies decrease rapidly with height. For frequencies comparable to the ion neutral frequency, the $d\tilde{\mb{W}}/d\tilde{t}$ terms cannot be neglected.
Note that there is an additional factor $\xi_{n}k_{in}/(k_{en}+k_{in})$ in front of one of these $d\tilde{\mb{W}}/d\tilde{t}$  terms, and $\xi_{n}k_{en}k_{in}/(k_{en}+k_{in})$ in front of the other. Also, $k_{en}/k_{in} \sim m_{i}\nu_{in}/m_{e}\nu_{en}$ and as we shall see later, $ m_{i}\nu_{in} \gg m_{e}\nu_{en}$ for both atmospheres, so that
$k_{en} \gg k_{in}$ and so $k_{in}/(k_{en}+k_{in}) \ll1 $ and $k_{en}k_{in}/(k_{en}+k_{in}) \approx k_{in}$. Hence one must be careful in the regime where $k_{in} \gg 1$. 

The last two terms in Equation (\ref{gol_3}) are pressure terms. The first, the electron and ion modified pressure term, 
$ \frac{d_{i}}{L_{0}}\left(\beta_{e}\tilde{\nabla} \cdot \tilde{\mbb{\kappa}_{e}} - \beta_{i}\frac{m_{e}}{m_{i}}\tilde{\nabla} \cdot \tilde{\mbb{\kappa}_{i}}\right)$, can be neglected if $\frac{di}{L_{0}}\beta_{e} \ll 1$.
The ion inertial scale $d_{i}$ has a maximum value of about 1 m in the chromosphere, and $\beta_{e}$ can be as large as 100 (see Figure \ref{fig:betas}), so for length scales $L_{0} \gg 100$ m, this pressure term can be neglected. For the I/T, the ion inertial scale can be as large as 2 km, but $\beta_{e}$ has a maximum of  $10^{-4}$ so that length scales larger than 1 m are sufficient to neglect the electron pressure.

The last pressure related term can be neglected in general when  $\frac{d_{i}}{L_{0}}\xi_{i}\beta_{n}$ and $\frac{d_{i}}{L_{0}}\xi_{n}\beta_{p}$ are much less than unity (again care must be taken when $k_{in} \gg 1$). The first of these two is satisfied easily as the ionization level $\xi_{i}$ is very small in most of the two atmospheres, and the second is of the order of the electron pressure term (see above).


Under these approximations, and noting that $\eta_{P} = \eta_{\|} + \eta_{C}$, the low frequency Ohm's law is simply

\bln
\bea
\mathbf{E}^{*} &= & \eta_{\parallel}\mb{J}_{\parallel} + \eta_{P}\mb{J}_{\perp}  + \eta_{H}\mb{J}\times\hat{\mb{b}} \nonumber \\
& +&  \mb{G}_{\|} + \mb{G}_{\perp} + \mb{G}_{\perp}\times\hat{\mb{b}}
\label{eq:gol3}
\eea
\eln
Here $\mb{J}_{\perp}= \hat{\mb{b}}\times\mb{J}\times\hat{\mb{b}}$ is the component of the current density perpendicular to the magnetic field. The quantity $\eta_{C}=\eta_{P}-\eta_{\parallel}$ is also referred to as the Cowling resistivity \citep{Cowling_1956}. The terms $\mb{G}_{\|}$, $\mb{G}_{\perp}$, $\mb{G}_{\perp}\times\hat{\mb{b}}$ are the grouped neglected terms, and are kept to show that they do not enter into the conductivities when we invert Equation (\ref{eq:gol3}) to 
 give $\mb{J} = \underline{\underline{\sigma}} \cdot \mb{E}^{*}$ in terms of the parallel, Hall and Pedersen conductivities: 
\bln
\bea
 \mathbf{J} & =&  \sigma_{\parallel}\mathbf{E}^{*}_{\parallel} + \sigma_{P}\mathbf{E}^{*}_{\perp} - \sigma_{H}\mathbf{E}^{*}_{\perp}\times\hat\mathbf{b} \nonumber\\
 & + &  \sigma_{\|}\mb{G}_{\|}  + (\sigma_{H}-\sigma_{P})(\mb{G}_{\perp}+\mb{G}_{\perp}\times\hat{\mb{b}}),
 \label{eq:gol4}
\eea
\eln
where}
\bln
\bea
\sigma_{\parallel} & = &  \frac{1}{\eta_{\parallel}}, \label{eqn:sigmas_start}\\
\sigma_{P} & = & \frac{\eta_{P}}{\eta_{P}^{2}+\eta_{H}^{2}}, \\
\sigma_{H} & = & \frac{\eta_{H}}{\eta_{P}^{2}+\eta_{H}^{2}}. \label{eqn:sigmas_end}
\eea
\eln
{\color{black} The neglected terms are terms which add to the electric field in the rest frame of the neutrals regardless of the presence of a current, and as such it is correct that they do not enter into the conductivities above.}

{\color{black} Ohm's law, whether expressed in the form $\mb{E}^{*} = \underline{\underline{\eta}} \cdot \mb{J}$ or $\mb{J} = \underline{\underline{\sigma}} \cdot \mb{E}^{*}$, can be derived in a number of different ways \cite[e.g.,][]{2001JGR...106.8149S,Vas_2012}. Here, we have derived it from the first-principles governing equations of motion for the electrons, ions, and neutrals. In the next section, we discuss applications of these general relations to the chromosphere and the I/T.  We note here that \citet{Song_2005} and \citet{Vas_2012} discuss the assumptions needed to reduce the generalized Ohm's law to a low-frequency Ohm's law equation which relates $\mb{E}$ and $\mb{J}$. In these papers, they state that the current given by the conventional ionospheric Ohm's law (Equation (\ref{eq:gol4}) above), is a stress-balance current, determined by a balance between Lorentz force and plasma-neutral collisional friction. Then $\mb{J}$ is physically related to $\mb{V}_{p}-\mb{V}_{n}$, and hence to $\mb{E}$ (as $\mb{E}$ is linearly related to $\mb{V}_{p}-\mb{V}_{n}$). Looking at the sum of the momentum equations for ions and electrons, then a balance of Lorentz force and plasma-neutral collisional friction is equivalent to the neglect of all inertial terms, as well as pressure and gravity terms. This is not exactly the same as our assumptions made here, as we neglect terms proportional to $\partial\mb{J}/\partial t$ and  $d^{*}\mb{W}/dt$, based on length and frequency arguments, rather than neglecting $\partial(m_{i}n\mb{V}_{i})/dt$ and $\partial(m{i}n_{n}\mb{V}_{n})/dt$ individually, a fact which allows us to retain individual inertial terms later during discussion of electromagnetic energy transfer into thermal and kinetic energy. } 

{\color{black} Note that in the limit of vanishing collisions, the resistive terms in the generalized Ohm's law (\ref{eqn:ohmslaw_elec}) tend to zero, but that in the linear Ohm's law (\ref{eq:gol3}) the Pedersen resistivity tends to infinity.
This behavior originates in our solution for W in Equation (\ref{eq:w}), which assumes that collisions dominate over inertia.  This cannot occur, of course, in the limit of vanishing collision frequency.  }

\section{Magnetization, Mobility, and Resistive Properties}
\label{electrodynamics}
\subsection{Magnetization Domains}
\begin{figure}[h]
\includegraphics[width=\textwidth]{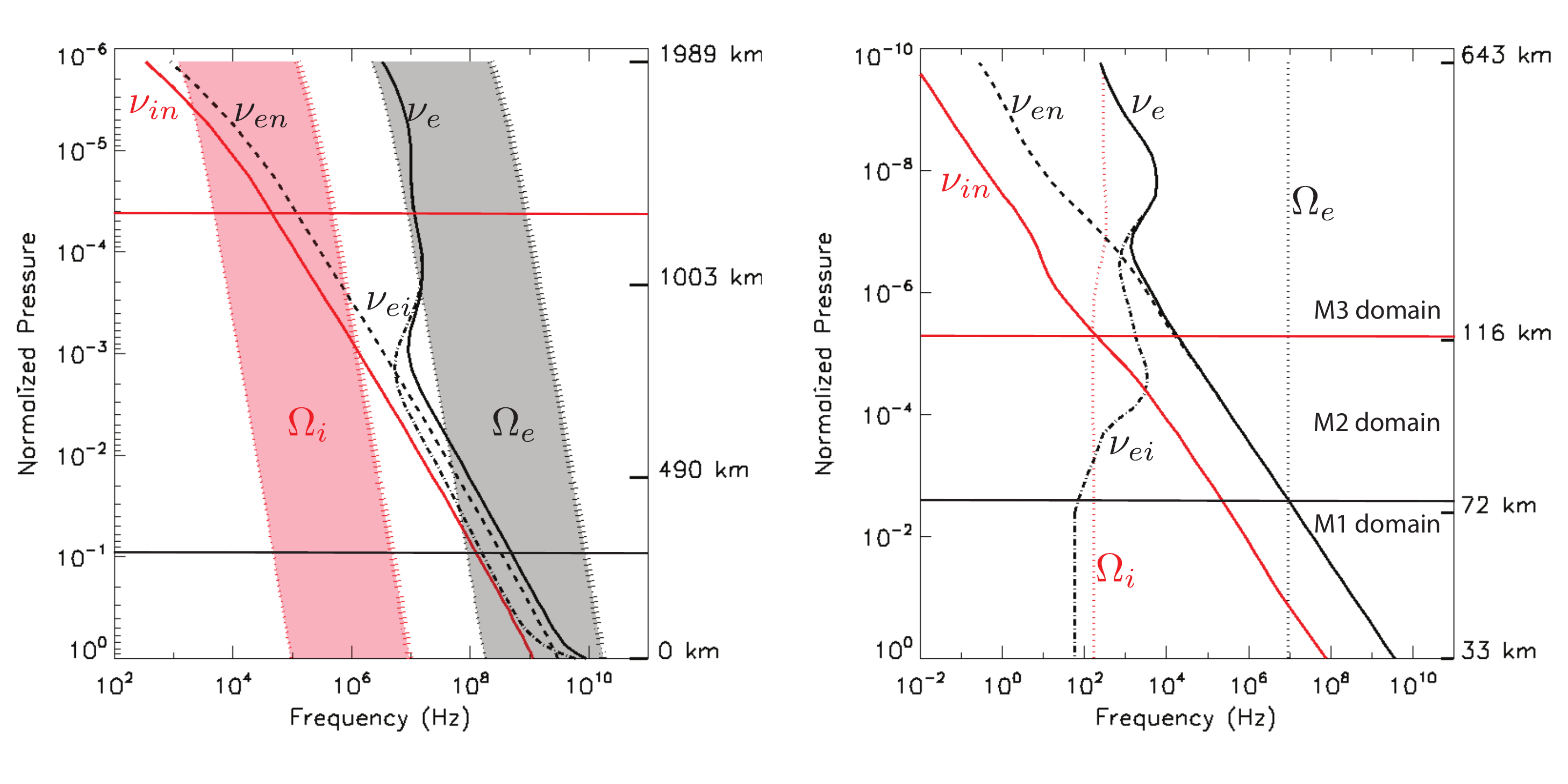}
\caption{\color{black} Collision frequencies and gyrofrequencies (Hz) in the ALC7 chromosphere (left) and TIMEGCM I/T (right). The collision frequencies are shown as solid red lines ($\nu_{in})$, dashed lines ($\nu_{en}$), dot-dashed lines ($\nu_{ei}$), and solid black lines ($\nu_{e}$). The gyrofrequencies are shown as dotted lines for ions ($\Omega_{i}$: red) and electrons ($\Omega_{e}$: black). For the chromosphere, where gyrofrequencies depend on height and there is a choice of three surface field values, the lines for the two extreme values of $B_{0}=10$ G (thin dotted line) and $B_{0}=1000 $ G (thick dotted line) in Equation (\ref{eqn:chromB}) are shown, with a shaded area between them. In the I/T (right panel), the horizontal lines mark altitudes at which $\nu_{e}=\Omega_{e}$ (black line) and  $\nu_{in}=\Omega_{i}$ (black line). In the chromosphere model (left panel), these heights do not have one single value, but a representative line is shown for these transitions.}
{\label{fig:frequencies}     }  
\end{figure}

Having derived Ohm's law in terms of the plasma magnetizations, it is useful to now compare and contrast these
fundamental parameters in the two environments, {\color{black} as was originally done by \citet{Goodman2000, Goodman2004a}.} Figure \ref{fig:frequencies} shows the contributing collision frequencies 
($\nu_{ei}$, $\nu_{en}$, $\nu_{e} \equiv \nu_{en} + \nu_{ei}$, $\nu_{in}$) and the electron and ion gyrofrequencies 
($\Omega_e, \Omega_i$) as functions of height in the chromosphere and I/T.

The frequencies of collisions of charged particles with neutrals are similar at the base of both environments, and are very high, in the range $10^{8}$--$10^{10}$ Hz. They decline to about 1 kHz at the top of the chromosphere, and much farther, to below 1 Hz, at the top of the I/T. The charged and neutral fluids may be considered to be strongly coupled in response to waves and transient phenomena whose frequencies fall well below these collision frequencies at any particular height.  Independent motions of the charged particles and neutrals can occur at frequencies well above those thresholds. Some consequences of this frequency-varying coupling on the propagation of Alfv\'{e}n waves, and the relative motions of plasma and magnetic field generally, are considered in more detail below in \S\ref{frozen_slip}.

  
\begin{figure}[h]
\includegraphics[width=\textwidth]{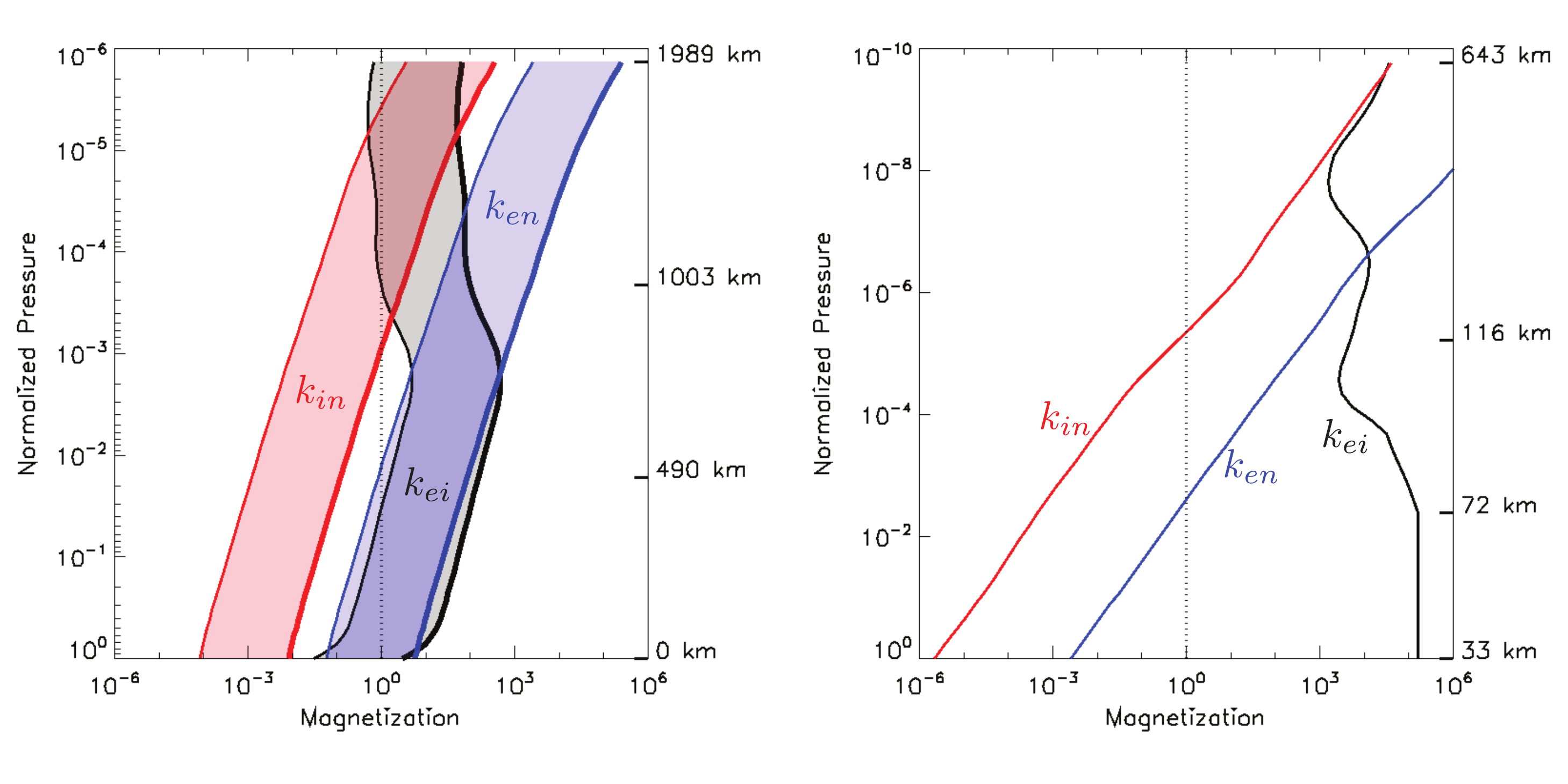}
\caption{ {\color{black} Individual magnetizations in the chromosphere (left) and I/T (right). The magnetizations are shown as blue lines ($k_{en}$), red lines ($k_{in}$) and black lines ($k_{ei}$).
For the chromosphere (left panel), where magnetizations depend on magnetic field strength, the lines for the two extreme values of $B_{0}=10$ G (thin lines) and $B_{0}=1000 $ G (thick lines) in Equation (\ref{eqn:chromB}) are shown, with a shaded area between them.}
\label{fig:magnetization}  }     
\end{figure}

\begin{figure}[h]
\includegraphics[width=\textwidth]{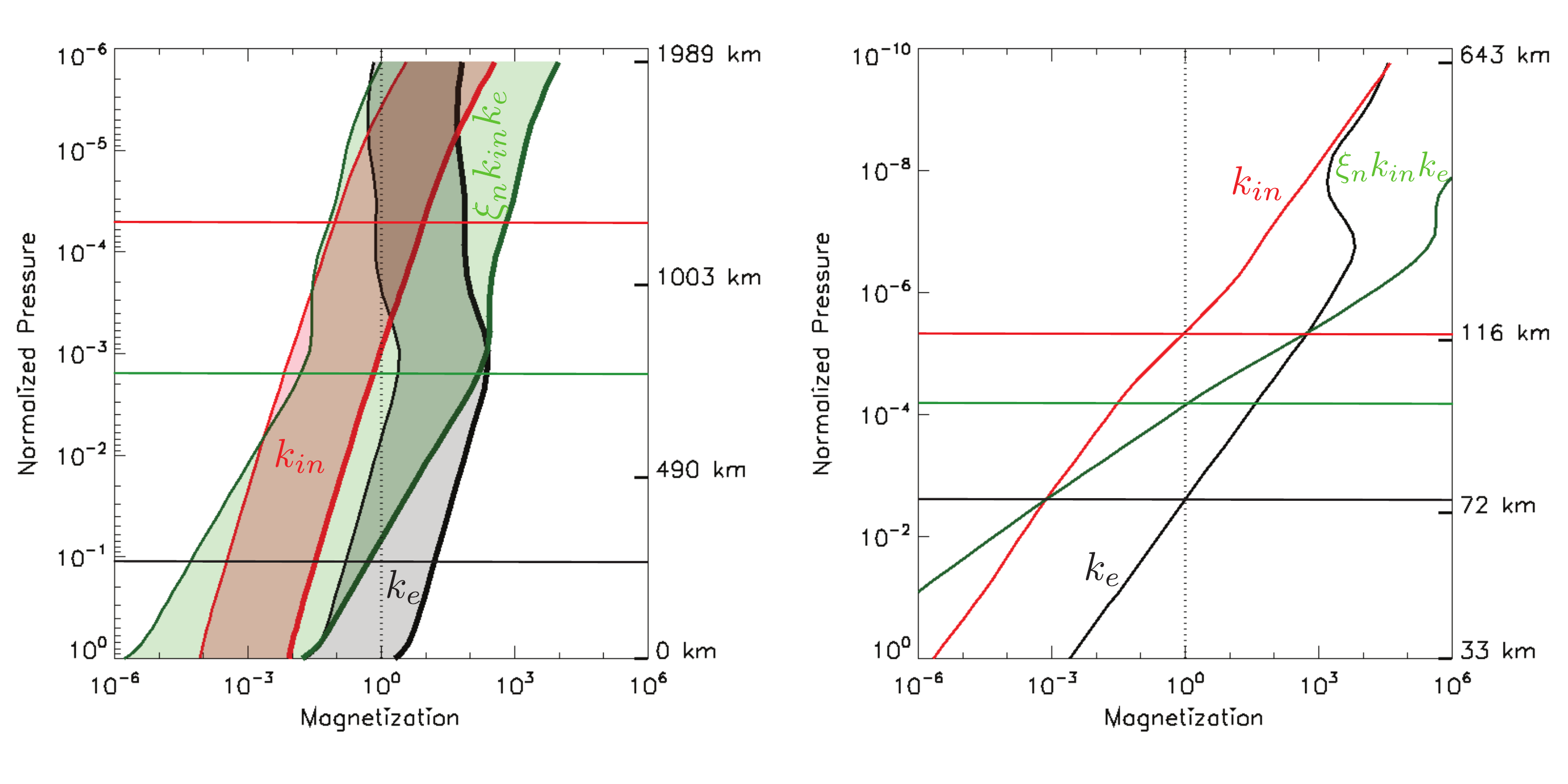}
\caption{{\color{black} Combined magnetizations in the chromosphere (left) and I/T (right). The magnetizations are shown as black lines ($k_{e}$), red lines ($k_{in}$) and green lines ($\xi_{n}k_{in}k_{e}$).
For the chromosphere (left panel), where magnetizations depend on magnetic field strength, the lines for the two extreme values of $B_{0}=10$ G (thin lines) and $B_{0}=1000 $ G (thick lines) in Equation (\ref{eqn:chromB}) are shown, with a shaded area between them. In the I/T (right panel), the horizontal lines mark altitudes at which $k_{e}=1$ (black line), $k_{in}=1$ (red line), and $\xi_{n}k_{in}k_{e}=1$ (green line). In the chromosphere model (left panel), these heights do not have one single value, but a representative line is shown for these transitions.}
\label{fig:magnetization2} }   
\end{figure}

{\color{black} In the I/T (right panel), the horizontal lines mark altitudes at which $\nu_{e}=\Omega_{e}$ (black line) and  $\nu_{in}=\Omega_{i}$ (red line). These locations are where $k_{e}=1$ and $k_{in}=1$, respectively, and represent transitions from unmagnetized to magnetized. In the chromosphere model (left panel), these respective heights do not have one single value, but a representative line is shown for these transitions.
The magnetizations, defined in Equations (\ref{eqn:magn1}-\ref{eqn:magn_mult}), associated with these collision frequencies and gyrofrequencies are shown in Figures \ref{fig:magnetization} and \ref{fig:magnetization2} for both the chromosphere and I/T. 
This analysis was originally conducted by \citet{Goodman2000,Goodman2004a} using a different semi-implicit model for the chromosphere, and the profiles obtained in this review are quantitatively comparable to those of \citet{Goodman2000,Goodman2004a} and those obtained in a more recent analysis by \citet{2013A&A...554A..22V}. From Figure \ref{fig:magnetization2}, one can define three distinct regions in the I/T:
\bln
\begin{eqnarray}
\textrm{M1 domain} ~~~ & \textrm{both ions and electrons are unmagnetized}  &  ~~~ k_{in},k_{e} < 1 \nonumber \\
\textrm{M2 domain} ~~~ & \textrm{ions are unmagnetized; electrons are magnetized} & ~~~  k_{in} < 1,k_{e} > 1 \nonumber \\
\textrm{M3 domain} ~~~ & \textrm{both ions and electrons are magnetized}  &  ~~~ k_{in},k_{e} > 1\nonumber
\end{eqnarray}
\eln
We point out that the magnetization domains are not directly related to the D, E, and F layers of the I/T, which are 
characterized by the dominant ion and neutral chemistry that occurs in the corresponding altitude bands. Our M1, M2, and M3 domains instead 
highlight commonalities in the charged-particle dynamics within the chromosphere and I/T. For the chromosphere, where there is a range of locations where these transitions occur, these three magnetization regions do not have specific heights. For example, for strong field strengths, e.g., above sunspots, the electrons can be magnetized $k_{e}>1$ at all heights (solid black line in Figure \ref{fig:magnetization}), but for low field strengths (thin black line) they can be magnetized above about 500 km and unmagnetized beneath. For ions in the chromosphere, there is a range between about 600 and 1600 km where the ions can transition from being unmagnetized ($k_{in} < 1$) to magnetized ($k_{in} > 1$), depending on the field strength.
In general, however, both the chromosphere and I/T are stratified, 
partially ionized mixtures in which the charged particles undergo a transition from completely unmagnetized (lower M1) 
to completely magnetized (upper M3), with a central region (M2) in which ions are unmagnetized and electrons are magnetized. 
Figure \ref{fig:magnetization2} also shows the function $\Gamma \equiv \xi_{n}k_{e}k_{in}$ (green line) which is equal to unity near the 
center of the M2 domain. As will be discussed later, the transition from  $\Gamma < 1 $ to $\Gamma > 1$ is a transition  from isotropic transport processes to anisotropic. This transition can be important for both the 
electrodynamics and magnetohydrodynamics, and the heating of the chromosphere (if not the I/T).
In the chromosphere, the locations of the M1, M2, and M3 domains change as the magnetic field strength changes, and so can be quite different over active regions compared to quiet regions of the Sun.  In the I/T, in contrast, the magnetic field is nearly static and the collision frequencies are dictated by the vertical distribution of the neutral 
gas. Thus, the locations of the M1, M2 and M3 domains in the I/T change only slowly with respect to altitude and latitude. 
}

Figure \ref{fig:frequencies} shows that electron-ion collisions are much less frequent than electron-neutral
collisions in the M1 and M2 domains of the I/T, but not so in the chromosphere. This is due to the much larger ionization 
fraction in the chromosphere relative to the I/T. In the M3 domains of both the chromosphere and I/T, on the other hand, 
electron-ion collisions dominate as the neutral density falls off. 
Nevertheless, a common assumption in ionospheric 
physics is that electron-ion collisions do not play a part in the electrodynamics \citep{2001JGR...106.8149S}. We will discuss this assumption in relation to the mobility of charged particles below.
When electron-ion 
collisions are negligible, then $k_{ei}$ becomes very large (as is evident in Figure \ref{fig:magnetization} in the lower I/T) 
and $k_{e} \approx k_{en}$ in Equations (\ref{eqn:magn1}--\ref{eqn:magn_mult}). Also, the neutral fraction $\xi_{n}$ is very 
close to 1, which is valid for all of the I/T and all but the highest altitudes in the chromosphere. In this limit, the 
resistivities (\ref{eqn:etas_start}--\ref{eqn:etas_end}), simplify to 
\bln
\bea
\eta_{\parallel} & \approx & \frac{B}{en} \frac{1}{k_{en}+k_{in}}, \label{eqn:itmus_start}\\
\eta_{C} & \approx & \frac{B}{en} \frac{k_{en}k_{in}}{k_{en}+k_{in}},\\
\eta_{P} & \approx & \frac{B}{en} \frac{1+k_{en}k_{in}}{k_{en}+k_{in}},\\ 
\eta_{H} & \approx & \frac{B}{en} \frac{k_{en}-k_{in}}{k_{en}+k_{in}}.
\label{eqn:itmus_end}
\eea
\eln
Substituting these expressions into (\ref{eqn:sigmas_start}--\ref{eqn:sigmas_end}) and combining terms yields the corresponding conductivities 
\bln
\bea
\sigma_{\parallel} & \approx & \frac{en}{B} \left( k_{en}+k_{in} \right), \label{eqn:itsigmas_start}\\
\sigma_{P} & \approx & \frac{en}{B} \left( \frac{k_{in}}{1+k_{in}^{2}} + \frac{k_{en}}{1+k_{en}^{2}} \right), \\
\sigma_{H} & \approx & \frac{en}{B} \left( \frac{1}{1+k_{in}^{2}} - \frac{1}{1+k_{en}^{2}} \right). \label{eqn:itsigmas_end}
\eea
\eln
These expressions (\ref{eqn:itsigmas_start}--\ref{eqn:itsigmas_end}) are identical to those in Equations (27--29) of \citet{2001JGR...106.8149S}, 
who derived them under the same assumptions : $\nu_{ei} \rightarrow 0$ and $\xi_{n} \rightarrow 1$.

\subsection{Charged Particle Mobilities and Electrical Currents }
\label{mobilities}

We can extract the common pre-factor $\left( en/B \right)$ from the conductivities (\ref{eqn:itsigmas_start}--\ref{eqn:itsigmas_end}):
\bln
\bea
\mu_{\parallel} \equiv \sigma_{\parallel} \frac{B}{en} & \approx & k_{en}+k_{in}, \label{eqn:mobs1}\\
\mu_{P} \equiv \sigma_{P} \frac{B}{en} & \approx & \frac{k_{en}}{1+k_{en}^{2}} + \frac{k_{in}}{1+k_{in}^{2}}, \\
\mu_{H} \equiv \sigma_{H} \frac{B}{en} & \approx & \frac{1}{1+k_{in}^{2}} - \frac{1}{1+k_{en}^{2}}.
\label{eqn:mobs2}
\eea
\eln
These definitions also were introduced by \citet{2001JGR...106.8149S}. The mobilities $\mu$ (\ref{eqn:mobs1}--\ref{eqn:mobs2}) are explicit functions of the magnetizations $k$ (\ref{eqn:magn1}-\ref{eqn:magn_mult}) that show the relative contributions of electrons and ions to the electric current.

\begin{figure}[h]
\includegraphics[width=\textwidth]{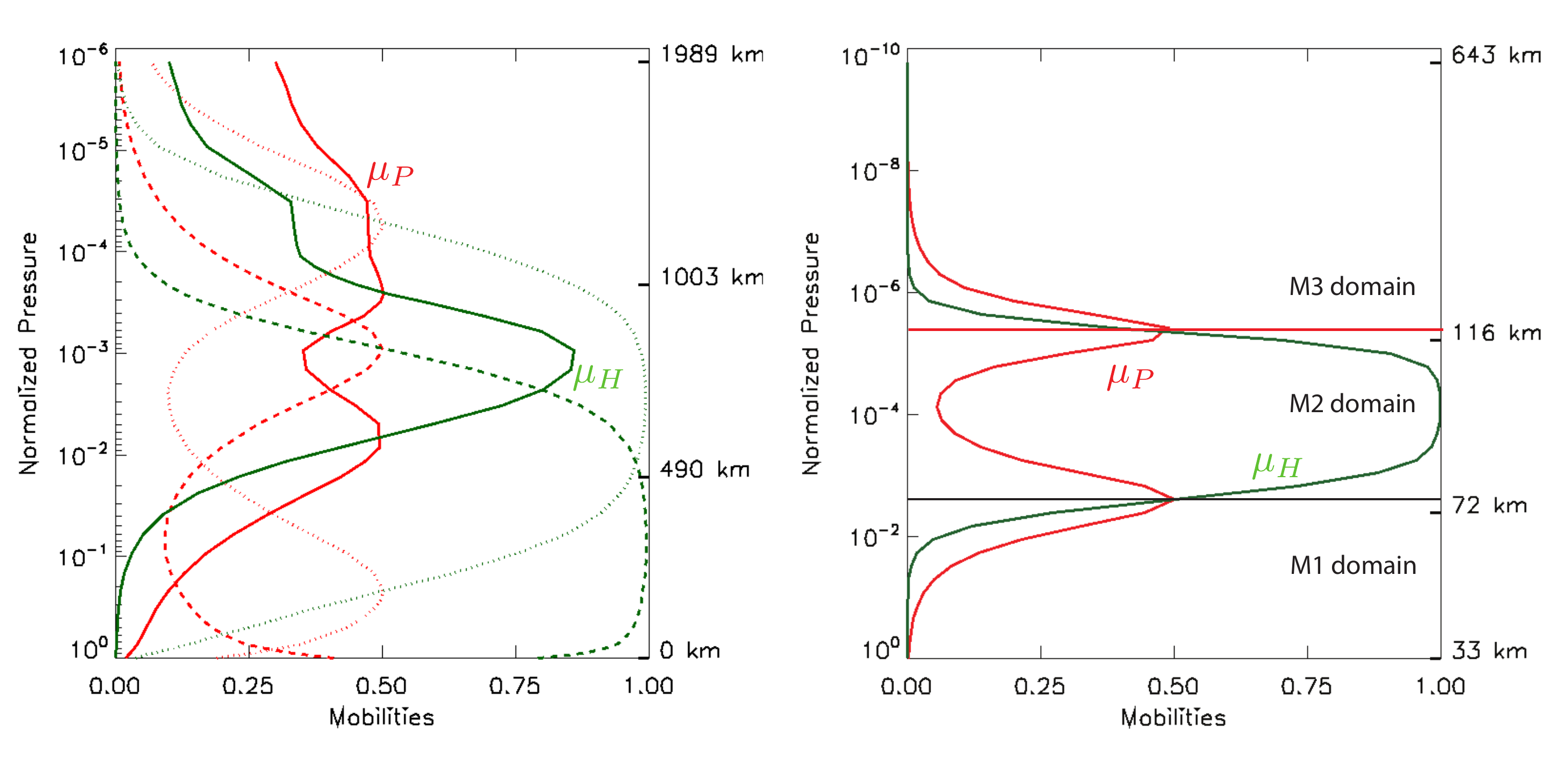}
\caption{ {\color{black} Pedersen ($\mu_{P}$; red) and Hall ($\mu_{H}$; green) mobilities in the chromosphere (left) and I/T (right). For the chromosphere, the solid, dotted and dashed lines represent the profiles using the three different values of $B_{0}=10,100,1000 ~ \textrm{G}$, respectively, in Equation (\ref{eqn:chromB}) for the magnetic field.}
\label{fig:mobs} }     
\end{figure}

Figure \ref{fig:mobs} shows the mobilities $\mu_{P}$ and $\mu_{H}$ for 
the chromosphere and I/T. The plots for the full mobilities and those for the approximations assuming $\nu_{ei}=0$ and $\xi_{n}=1$ (\ref{eqn:mobs1}-\ref{eqn:mobs2}) overlay each other, validating the assumption that electron-ion collisions do not contribute to the electrodynamics. 
\begin{figure}[h]
\includegraphics[width=\textwidth]{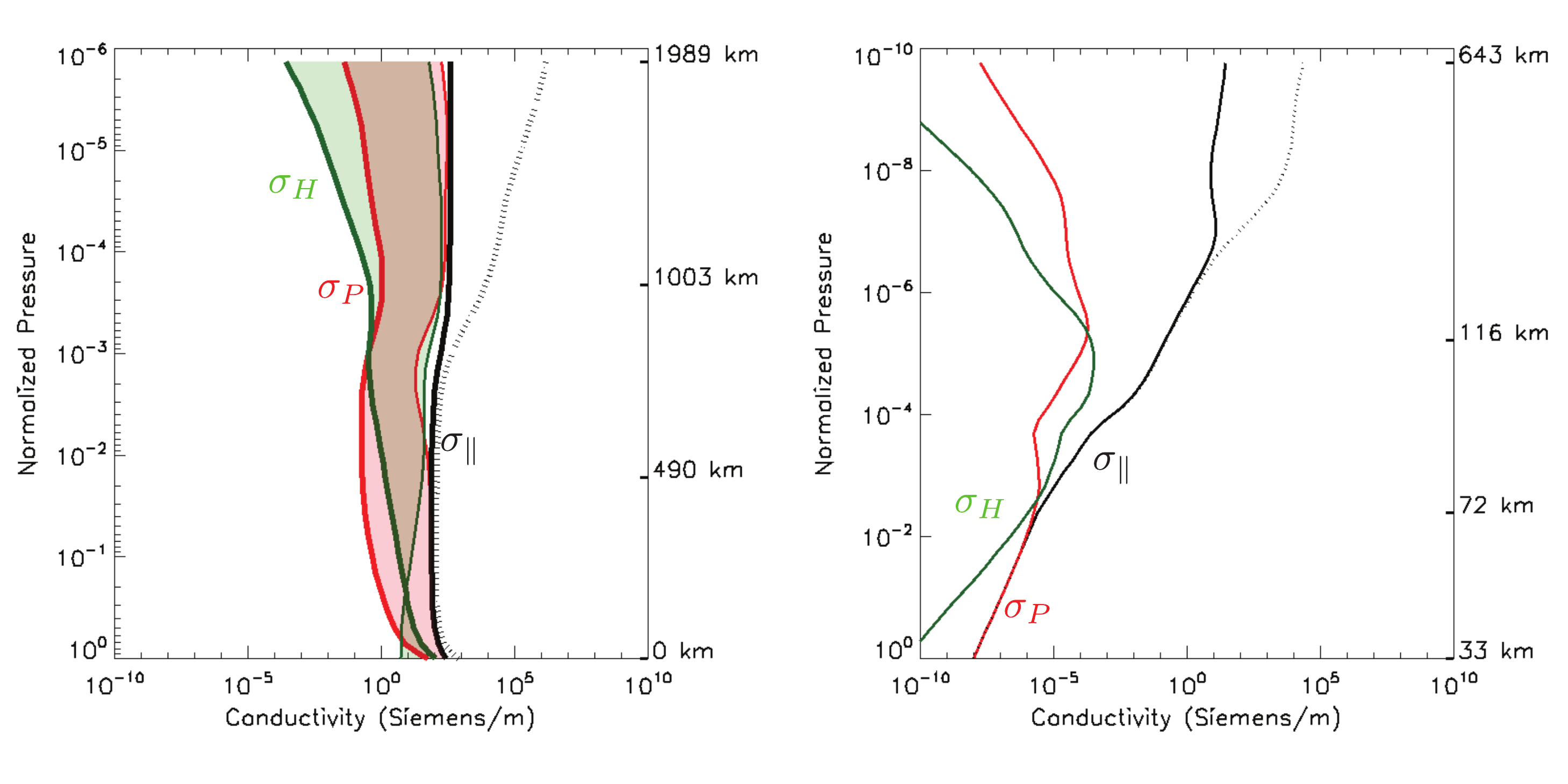}
\caption{{\color{black}Parallel ($\sigma_{\parallel}$; black), Pedersen ($\sigma_{P}$; red), and Hall ($\sigma_{H}$; green) conductivities (in Siemens/m) in the chromosphere (left) and I/T (right). Dashed lines show the corresponding values when electron-ion collisions are ignored. For the chromosphere (left panel), where the Pedersen and Hall conductivity depend on magnetic field strength, the lines for the two extreme values of $B_{0}=10$ G (thin lines) and $B_{0}=1000 $ G (thick lines) in Equation (\ref{eqn:chromB}) are shown, with a shaded area between them. The parallel conductivity is the only conductivity that appears to be affected by the assumption that electron-ion collisions can be ignored.}
\label{fig:cond}}      
\end{figure}
{\color{black}
The overall shape of the mobility curves in the chromosphere and in the I/T is very similar, though the vertical extent of these curves in the chromosphere depends on the magnetic field model used. Three curves for each mobility are shown in the left panel, one for each of the three magnetic field surface strengths in Equation (\ref{eqn:chromB}). 
In the chromosphere and I/T, the Pedersen mobility $\mu_{P}$ (red line in Figure \ref{fig:mobs}) is double peaked, with maxima very near the transitions from domains M1 to M2 (where $k_{e} = 1$) and from M2 to M3 (where $k_{in} = 1$). 
The Hall mobility $\mu_{H}$ (green line) is single peaked, with its maximum between the two peaks in the Pedersen mobility. The lower peak for the weakest chromospheric magnetic field model is missing from the plot as it appears below the lower boundary of the atmosphere. We shall discuss the mobility of the ions and electrons and their contribution to currents below.}

The conductivities (including the electron density and representative magnetic field strength for the chromosphere and I/T) are shown in Figure \ref{fig:cond}. The calculations using the full expressions are shown as solid lines, while those neglecting electron-ion collisions are shown as dashed lines.  Convolving the electron density with the relative mobilities causes some of the structure with altitude in Figure \ref{fig:mobs} to be less obvious in Figure \ref{fig:cond}. In particular, for the I/T, the lower Pedersen conductivity peak is significantly reduced due to the strongly diminished electron density at those altitudes. The figure also reveals that the entire I/T above 70 km is subject to anisotropic electrodynamics ($\sigma_{P} \ne \sigma_{\parallel}$).
In the chromosphere, for weak fields, $\sigma_{P}$ can be equal to $\sigma_{\|}$ up to about 500km before anisotropy develops, but for strong fields, the anisotropy can be evident for the entire chromosphere.
An important role of electron-ion collisions in limiting the parallel conductivity ($\sigma_{\parallel}$) is obvious at high altitudes (compare the solid and dashed black lines), throughout the upper half of the chromosphere and also above 120 km in the ionosphere, as can be seen in Figure \ref{fig:cond}. All three conductivities, especially the Pedersen and Hall conductivities, are far larger in the chromosphere than in the I/T. 

{\color{black} 
It is worth further study of the contribution of electrons and ions to the Pedersen and Hall currents responding to a generic electric field in the rest frame of the neutrals $\mb{E}^{*}$, via the Pedersen and Hall mobilities. Examination of the curves for the I/T allows a simple discussion, but one must note that the chromosphere has a variable magnetic field, and so the transitions discussed here, such as unmagnetized to magnetized plasma, will vary with height depending on the field strength. 

First recall the low frequency Ohm's law
\bln
\be
\mb{J} = \sigma_{\|}\mb{E}_{\|} + \sigma_{P}\mb{E}^{*}_{\perp} - \sigma_{H}\mb{E}^{*}_{\perp}\times\hat{\mb{b}}
\ee
\eln
and so the current perpendicular to the magnetic field vector ($\mb{J}_{\perp}$) is a combination of Hall 
($\sigma_{H}\mb{E}^{*}_{\perp}\times\hat{\mb{b}}$) and Pedersen ($\sigma_{P}\mb{E}^{*}_{\perp}$) currents. We shall see later how the relative contribution of these
two currents affects the efficiency of heating due to plasma-neutral collisions. Let us first consider the variation with altitude of these mobilities, conductivities, and currents.
A complete description would involve examining the momentum equations for ions and electrons, and solving for velocity, and then using $\mb{J} = en(\mb{V}_{i}-\mb{V}_{e})$ to 
define the current. For brevity, we discuss the motions of the ions and electrons and do not present such a derivation, though the results are consistent with such a method.

\begin{itemize}
\item{At the lowest altitudes in the I/T, and near the solar surface in weak field regions, collisions with neutrals are high, and the mobility of electrons and ions are low and they are unmagnetized, $k_{en},~ k_{in} \ll 1$ (see Figure \ref{fig:magnetization}). Thus $\mu_{p} \approx \mu_{H} \approx 0$, and $\mb{J}_{\perp}\approx0$.}

\item{With increasing altitude, just below the lower peak in the Pedersen mobility (Figure \ref{fig:mobs}), 
electrons become more mobile $k_{en} > k_{in}$ and drive a Pedersen current in the $\mb{E}^{*}_{\perp}$ direction. Here $\mu_{P} > \sigma_{H}$ and $\sigma_{P} \sim \sigma_{\|} > \sigma_{H}$ (see Figure \ref{fig:cond}).}

\item{There is a height at which $k_{en} \approx 1$, while $k_{in} \ll 1$. This is the lower peak in the Pedersen mobility in Figure \ref{fig:mobs}. Here $\mu_{P}\approx\mu_{H}\approx 1/2$ and the  perpendicular current 
is equally contributed to by the Hall and Pedersen currents. 
At this point, there is still isotropy in the conductivity: $\sigma_{P}\approx \sigma_{H} \approx \sigma_{\|}$, as seen in Figure \ref{fig:cond}.
}

\item{At higher altitudes, the electrons become completely magnetized, but the ions remain unmagnetized . Here $k_{en} \gg 1$ while $k_{in} \ll 1$, and $\mu_{H}\approx 1$, while $\mu_{P} < \mu_{H}$ and $\sigma_{p} < \sigma_{H} < \sigma_{\|}$. This is the peak in the Hall mobility in Figure \ref{fig:mobs}). The electron motion drives a current predominantly in the Hall direction $ - \sigma_{H}\mb{E}^{*}_{\perp}\times\hat{\mb{b}}$.}

\item{The next altitude of interest is where ions start to become mobile and $k_{in}\approx 1$ and $k_{en} \ll1$. This is the higher peak in the Pedersen mobility, where $\mu_{P}=\mu_{H}=1/2$, and the perpendicular current is equally Pedersen and Hall current.
However, unlike at the lower peak in the Pedersen mobility, at this upper peak $\sigma_{P}\approx \sigma_{H} \ll \sigma_{\|}$ and there is significant anisotropy in the conductivities (this will be important for plasma-neutral heating, as discussed in later sections).}

\item{Just above the upper Pedersen mobility peak, $\mu_{P}>\mu_{H}$, $\sigma_{P}>\sigma_{H}$,  and the perpendicular current is mainly the Pedersen current in the $\mb{E}^{*}_{\perp}$ direction and $\sigma_{H}<\sigma_{P}\ll \sigma_{\|}$. }

\item{Finally, at the highest altitudes where collisions with neutrals have fallen sufficiently, the electrons and ions are completely magnetized, and $k_{en}, ~ k_{in} \gg 1$. In this region $\mu_{p} \sim \mu_{H} \sim 0$, 
  $\mb{J}_{\perp}\approx0$
and $\sigma_{H}< \sigma_{P}<\sigma_{\|}$. }
\end{itemize}
This discussion of currents not only highlights the similarities between the two atmospheres, but allows for a discussion of the conversion of electromagnetic energy into thermal and kinetic energy, via the $\mb{E}\cdot\mb{J}$ term, as will be discussed in a later section on energy transfer.
}

\subsection{Perpendicular vs.\ Parallel Current Resistivity}
\label{current_diss}
{\color{black} As we have discussed in some detail, collisional drag between the charged-particle and neutral fluids introduces a substantial anisotropy in the resistivities and conductivities of a partially ionized mixture.}
At high altitudes of the chromosphere and I/T, the parallel and Pedersen conductivities increasingly diverge from one another, with $\sigma_{P}<\sigma_{\|}$, or 
equivalently $\eta_{P}>\eta_{\|}$.
This reveals a preference in both atmospheres toward relatively fast dissipation of currents that are directed perpendicular to the magnetic field and so contribute to magnetic forces, but relatively slow dissipation of currents that are directed along the magnetic field and so are force-free. On the Sun, during the emergence of magnetic flux from below the photosphere into the overlying atmosphere, the anisotropic resistivities in the chromosphere reshape the current profile and transition the magnetic field to the essentially force-free configuration (${\bf J}_\perp \approx 0$) that it must assume in the very low-beta corona \citep{2006A&A...450..805L,2007ApJ...666..541A,Arber_2009,Leake_2013}. \citet{Goodman2000,Goodman2004a} discusses the relationship between anisotropy and the force-free nature of the field for current dissipation driven by wave motions in the chromosphere.
  At Earth, the anisotropic conductivities in the I/T enable magnetic-field-aligned currents to neutralize the parallel electric field (${\bf E}_\parallel = \mb{b}\cdot\mb{E} \approx 0$) while perpendicular currents and electric fields can be sustained when driven by, for example, cross-field neutral winds. As mentioned earlier, a slowly evolving electric field can be written as the gradient of a scalar potential, $\mb{E} = -\nabla \Phi$. Then 
${\bf E}_{\|} \approx 0 \Rightarrow \mb{b}\cdot\nabla\Phi \approx 0$, and magnetic field lines are equipotential paths through the I/T {\color{black} if it is not coupled to the magnetosphere.}

\begin{figure}[h]
\includegraphics[width=\textwidth]{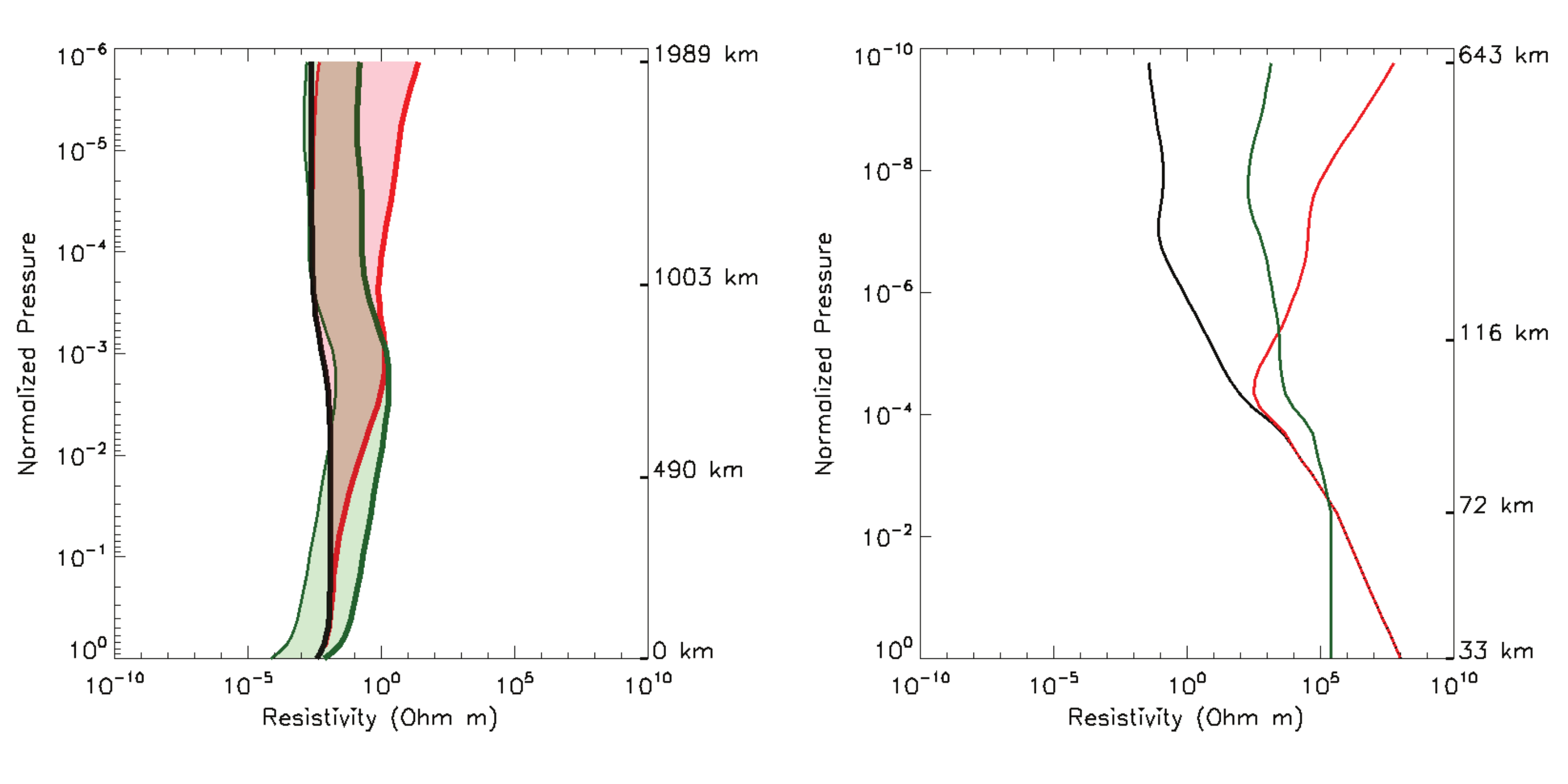}
\caption{ {\color{black} Parallel (black), Pedersen (red), and Hall (green) resistivities in the chromosphere (left) and I/T (right). 
 For the chromosphere (left panel), where the resistivities depend on magnetic field strength, the lines for the two extreme values of $B_{0}=10$ G (thin lines) and $B_{0}=1000 $ G (thick lines) in Equation (\ref{eqn:chromB}) are shown, with a shaded area between them. }
\label{fig:res}  }     
\end{figure}

Another way of expressing this anisotropy in the resistivities is to use the magnetizations. In general, one would expect that $m_{e}\nu_{en} \ll m_{i}\nu_{in}$, since $m_{\alpha}\nu_{\alpha n} \propto \sqrt{m_{\alpha} T_{\alpha}}$. However, this approximation may not hold throughout the I/T at all times \citep{2001JGR...106.8149S}. Assuming that $m_{e}\nu_{en} \ll m_{i}\nu_{in}$, and noting that this implies $k_{in} \ll k_{en}$, the resistivities (\ref{eqn:etas_start}--\ref{eqn:etas_end}) in Ohm's law (\ref{eq:gol3}) become 
\bln
\begin{eqnarray}
\eta_{\parallel} & \approx & \frac{B}{en} \frac{1}{k_{e}}, \\ 
\eta_{C} & \approx & \frac{B}{en} \xi_{n}k_{in}, \\
\eta_{P} & \approx & \frac{B}{en} \frac{1+\xi_{n}k_{e}k_{in}}{k_{e}}, \\
\eta_{H} & \approx & \frac{B}{en}.
\label{eq:gol_chrom}
\end{eqnarray}
\eln
The corresponding expressions for $\sigma_{\parallel}$, $\sigma_{P}$ and $\sigma_{H}$, after using Equations (\ref{eqn:sigmas_start}--\ref{eqn:sigmas_end}), are exactly the same as  Equations (23--25) in \citet{2001JGR...106.8149S},  derived under the same assumptions. The parallel, Pedersen, and Hall resistivities for the chromosphere and I/T are shown in Figure \ref{fig:res}. The plots calculated under the assumption $m_{e}\nu_{en} \ll m_{i}\nu_{in}$ overlay the full calculations, verifying the assumption. 

An important quantity that sets the anisotropy of the resistivities is the ratio of the Cowling to parallel resistivity:
\bln
\be
\frac{\eta_{C}}{\eta_{\parallel}} \approx \xi_{n}k_{e}k_{in}.
\ee
\eln
{\color{black} Note that this was denoted $\Gamma$ in \citet{Goodman2004a}, where the Ohm's law was cast in the 
center of mass frame and thus had the form $\xi_{n}^{2}k_{e}k_{in}$.}
Figure \ref{fig:magnetization2}  shows the product $\xi_{n}k_{e}k_{in}$ in the chromosphere and I/T. Increasing with height from the bottom, the electrons become magnetized first ($k_{e} > 1$), then the ions do so ($k_{in}>1$). Between these two points, $\xi_{n}k_{e}k_{in}=1$ within the M2 domain. At lesser heights, the current density is isotropic, $\eta_{c}/\eta_{\|} \ll 1$, or $\eta_{\parallel}\approx\eta_{P}$ and $\mb{E^{*}} \approx \eta_{\parallel}\mb{J}$, while at greater heights it is anisotropic, $\eta_{P}\gg \eta_{\|}$ and $\mb{E^{*}} = \eta_{\parallel}\mb{J}_{\parallel}+\eta_{P}\mb{J}_{\perp} + \eta_{H}\mb{J}\wedge\mb{\hat{b}}$. As will be discussed later in this paper, this transition is believed to be important with respect to the heating of the chromosphere.

\subsection{Summary}

{\color{black} Despite a large disparity in electron number density between the chromosphere and I/T, the governing physics of the two regions display some remarkable similarities. 
These similarities are mainly related to the transition from unmagnetized to magnetized plasma, and the variation of the electron and ion mobilities with altitude. The decrease in collisional frequency with height in both environments causes a change in magnetization with altitude, such that a central region is created where the ions are unmagnetized and electrons are magnetized. This is characterized by the region $\Gamma\equiv\xi_{n}k_{e}k_{in}\approx 1$. 
As $\Gamma$ increases with altitude, the ions become more magnetized and can drive Pedersen currents. These dissipative currents are perpendicular to the field, and are a possible source of heating via the $\mb{E}\cdot\mb{J}$ term (frictional heating, of which some is Joule heating, see the discussion in the following section).

However,  a large electron number density difference between the two regions leads to a large difference in the magnitudes of the conductivities and resistivities. As we shall see in the next section this leads to the I/T being resistively dominated, while the chromosphere is only resistive for scales less than a km. Also, the nature of the drivers in the two environments is also very different.  The drivers of the low frequency (relative to the collision frequency) electric fields in the I/T are mainly neutral winds colliding with ions and electrons, and externally imposed electric fields from the magnetosphere. In the chromosphere, the drivers exist on a large range of time scales, as long as weeks (sunspot and prominence formation and persistence) and as short as seconds (magnetic reconnection events such as jets and flares, and high-frequency waves).}

\section{Magnetohydrodynamic and Electrodynamic Processes}
\label{mhd_ed_processes}
\subsection{Frozen-In vs.\ Resistive-Slip Evolution}
\label{frozen_slip}
A key parameter governing the coupling of a fully ionized plasma to the magnetic field is the Lundquist number \citep{1952ArFy...5..297L}, 
\bln
\begin{equation}
  S \equiv \frac{\mu_0^{1/2} B_0 \ell}{\rho_0^{1/2} \eta_{\|}} = \frac{\mu_0 V_{A0} \ell}{\eta_{\|}},\label{EqNum12}
\end{equation}
\eln
where $\rho_0=m_{i}n$ is the mass density, $\eta_{\|}$ is the resistivity, 
$\ell$ is a characteristic length scale of variation, and $V_{A0} = B_0 / \left( \mu_0 \rho_0 \right)^{1/2}$ is the Alfv\'en speed. This dimensionless ratio measures the relative importance of convection and resistive diffusion in the evolution of the magnetic field. Where it is large, convection dominates, and the plasma moves with the magnetic field so that the field lines are ``frozen'' into the fluid \citep{1963coel.book.....A}; where it is small, resistivity dominates, and the plasma slips through the magnetic field lines. A derivation of the Lundquist number demonstrating this fundamental property is instructive and guides its generalization to multi-fluid situations with plasma-neutral coupling.

{\color{black} The motion of the fully ionized plasma induced by fluctuations in the magnetic field} is estimated by balancing inertia against Lorentz forces in the equation of motion,
\bln
\begin{equation}
  \rho_0 \frac{\partial {\bf{V}_{p}}}{\partial t} \approx \frac{1}{\mu_0} \left( \nabla \times \delta {\bf B} \right) \times {\bf B}_0,\label{EqNum13}
\end{equation}
\eln
whence 
\bln
\begin{equation}
  \omega \rho_0 V_{p} \approx \frac{B_0 \delta B}{\mu_0 \ell},\label{EqNum14}
\end{equation}
\eln
where $\omega$ is the frequency associated with the length $\ell$. The evolution of the magnetic field ${\bf B}$ is governed by Faraday's law, which requires Ohm's law. 
{\color{black} 
In the limit of a fully ionized plasma, and for low frequency phenomena ($f_{0}\ll\nu_{in}, ~ \Omega_{i}$), the Ohm's law is
\bln
\begin{equation}
  {\bf E} + ({\bf{V}_{p}} \times {\bf B}) = \eta_{\|} {\bf J} = \frac{\eta}{\mu_0} \nabla \times {\bf B},\label{EqNum15}
\end{equation}
\eln
consistent with \citet{1963coel.book.....A}, where we have also assumed that the length scale of interest is larger than the ion and electron inertial scales 
$l \gg d_{e},d_{i}$, as discussed in \S \ref{genohmslaw}.}

{\color{black} FaradayÕs law then yields the standard MHD induction equation:}
\bln
\begin{equation}
  \frac{\partial {\bf B}}{\partial t} = \nabla \times \left( {\bf{V}_{p}} \times {\bf B} 
\right) + \frac{\eta_{\|}}{\mu_0} \nabla^2 {\bf B}.\label{EqNum16}
\end{equation}
\eln
Balancing the time derivative of the fluctuating field against the convection of the ambient field, which is assumed to dominate the effects of resistivity acting on the fluctuating field, we find
\bln
\begin{equation}
  \omega \delta B \approx \frac{1}{\ell} V_{p} B_0 \gg \frac{\eta_{\|}}{\mu_0 \ell^2} \delta B.\label{EqNum17}
\end{equation}
\eln
Using Equation (\ref{EqNum17}) to eliminate $\omega$ from Equation (\ref{EqNum14}) and solving for $V_{p}$ gives 
\bln
\begin{equation}
  V_{p}^{2} \approx \frac{{\delta B}^2}{\mu_0 \rho_0},\label{EqNum18}
\end{equation}
\eln
so that the velocity $V_{p}$ is just the Alfv\'en speed evaluated at the perturbed field strength $\delta B$. The ratio of the retained convection term to the neglected resistivity term in Equation (\ref{EqNum17}) then becomes 
\bln
\begin{equation}
  \frac{\mu_0 \ell B_0}{\eta_{\|}} \frac{V_{p}}{\delta B} = \frac{\mu_0^{1/2} B_0 \ell}{\rho_0^{1/2} \eta_{\|}} = S,\label{EqNum19a}
\end{equation}
\eln
the Lundquist number. Thus, as claimed, convection dominates resistivity if $S \gg 1$. Reversing the inequality in Equation (\ref{EqNum17}) and carrying through the rest of the analysis, the unimportance of the convection term relative to the resistivity term in that case is measured by the ratio 
\bln
\begin{equation}
  \frac{\mu_0 \ell B_0}{\eta_{\|}} \frac{V_{p}}{\delta B} = \frac{\mu_0 B_0^2 \ell^2}{\rho_0 \eta_{\|}^2} = S^2.\label{EqNum19b}
\end{equation}
\eln
Therefore, again as claimed, resistivity dominates convection if $S \ll 1$.

Now we generalize these considerations to a partially ionized mixture. First we note that the Lundquist number is directly proportional to the characteristic length $\ell$ of the variations in the magnetic and velocity fields. By combining Equations (\ref{EqNum14}) and (\ref{EqNum17}), we find in the convection-dominated case that $\ell$ is just the reciprocal wavenumber of an Alfv\'{e}n wave at frequency $\omega$,
\bln
\begin{equation}
  \ell = \frac{B_0}{\mu_0^{1/2} \rho_0^{1/2} \omega} = \frac{V_{A0}}{\omega}.\label{EqNum20}
\end{equation}
\eln
In a fully ionized plasma, the waves span a continuum of frequencies and wavenumbers whose character changes from frozen-in, oscillatory motions at large $S$ (large $\ell$, small $\omega$) to resistive-slip, damped motions at small $S$ (small $\ell$, large $\omega$). In a partially ionized mixture, the collisions between plasma particles and neutrals modify the response of the gas to the magnetic field and also raise the resistivity acting on perpendicular currents $\mb{J}_\perp$ from the parallel value $\eta_{\|}$ to the Pedersen value $\eta_{P}$, as in Equation (\ref{eq:gol3}). Two limiting frequency ranges are particularly illustrative. {\color{black} At high frequencies $\omega >\nu_{in}$, the plasma-neutral coupling is weak, and the waves propagate at the Alfv\'en speed $V_{Ap0}$ determined by the plasma mass density $\rho_{p0}$ alone. We note here that the low frequency Ohm's law may not exactly apply in this high frequency situation, and there will be a term related to $d^{*}\mb{W}/dt$ that contributes to the electric field in the rest frame of the plasma, as in Equation (\ref{gol_ions1}). As discussed in \S \ref{low_frequency_ohms_law} the importance of this $d^{*}\mb{W}/dt$ term depends on the product of the three ratios $\omega/\nu_{in}$, $\omega/\Omega_{i}$ and $k_{in}/k_{en}$, the first of which is large in this high frequency regime, and the last of which is always small. Choosing a magnetized regime where $\nu_{in} < \Omega_{i}$, we can still have $\omega > \nu_{in}$ but be able to apply the low frequency limit of the Ohm's law (\ref{eq:gol3}).} Using the transition frequency $\omega = \nu_{in}$ in Equation (\ref{EqNum20}) to set $\ell$, the Lundquist number in Equation (\ref{EqNum12}) then takes the value 
\bln
\begin{equation}
  S_{in}' \equiv \frac{\mu_0 V_{Ap0}^2}{\eta_P \nu_{in}}.\label{EqNum21a}
\end{equation}
\eln
At low frequencies $\omega < \nu_{ni}$, on the other hand, the plasma-neutral coupling is strong, and the waves propagate at the Alfv\'en speed $V_{At0}$ determined by the total (plasma+neutral) mass density $\rho_{t0}=\rho_{p0}+\rho_{n0}$ \citep[e.g.,][]{Song_2005}. Using the transition frequency $\omega = \nu_{ni}$ for these waves, we obtain the Lundquist number 
\bln
\begin{equation}
  S_{ni}' \equiv \frac{\mu_0 V_{At0}^2}{\eta_P \nu_{ni}}.\label{EqNum21b}
\end{equation}
\eln
The ratio of the convection to the resistivity terms in the plasma-neutral-modified MHD induction equation at the transition frequency $\nu_{in}$ ($\nu_{ni}$) is $S_{in}'$ ($S_{ni}'$). Thus, the motions of the plasma in response to magnetic field fluctuations at that frequency are frozen-in or resistive-slip, respectively, according to whether $S' \gg 1$ or $S' \ll 1$. We point out that the values of $S_{in}'$ and $S_{ni}'$ are nearly equal, since they are inversely proportional to the products $n_{p0} n_{n0}$ and $(n_{n0}+n_{p0}) n_{p0}$, respectively, whereas $n_{p0} \ll n_{n0}$ through much of both atmospheres (Figure \ref{fig:densities}).

\begin{figure}[h]
\includegraphics[width=\textwidth]{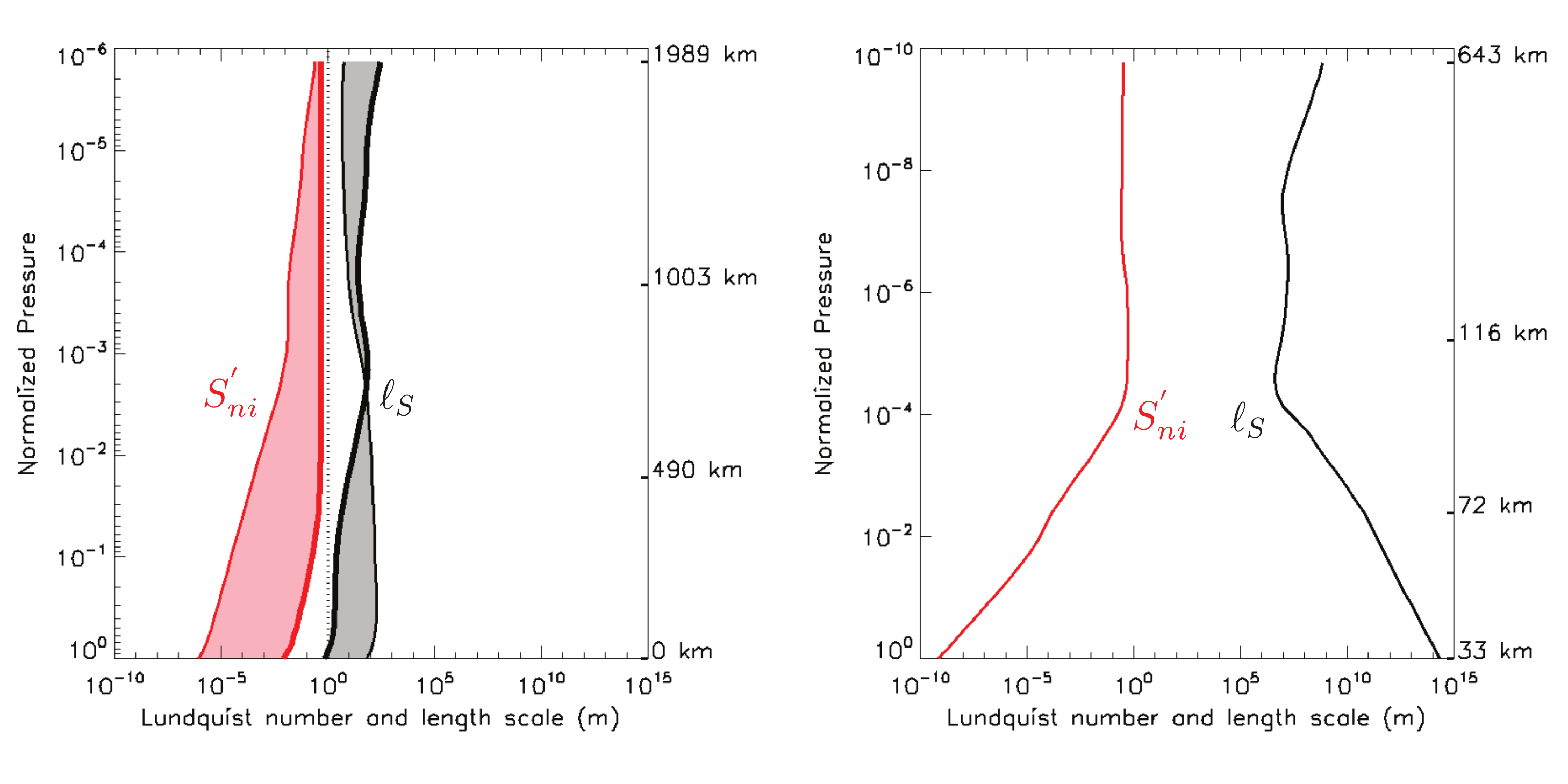}
\caption{{\color{black} Values of the Lundquist number $S_{ni}'$ (red line), Equation (\ref{EqNum21b}), for Alfv\'en waves at the neutral/ion collision frequency $\nu_{ni}$, and of the Lundquist scale (black line) $\ell_S$, in m, Equation (\ref{EqNum22}), using the Pedersen resistivities in the ALC7 chromosphere (left) and the TIMEGCM I/T (right).  For the chromosphere (left panel), where the values depend on magnetic field strength, the lines for the two extreme values of $B_{0}=10$ G (thin lines) and $B_{0}=1000 $ G (thick lines) in Equation (\ref{eqn:chromB}) are shown, with a shaded area between them. }
\label{fig:lehnerts}}      
\end{figure}

The values of the Lundquist number $S_{ni}'$ at the lower transition frequency $\nu_{ni}$ are shown (red dashed curves) for the ALC7 chromosphere (left) and TIMEGCM I/T (right) in Figure \ref{fig:lehnerts}.  For the chromosphere (left panel), where the values depend on magnetic field strength, the lines for the two extreme values of $B_{0}=10$ G (thin lines) and $B_{0}=1000 $ G (thick lines) in Equation (\ref{eqn:chromB}) are shown, with a shaded area between them. The numbers are very small at low altitudes and approach, but do not quite attain, unity at high altitudes, in both cases. Consequently, the field lines resistively slip through the gas very readily at low altitudes, and they are not well frozen-in to the gas motions anywhere in the two atmospheres, for frequencies at or above the neutral/ion collision frequency $\nu_{ni}$. Flux-freezing will occur only at still lower frequencies $\omega \le \nu_{ni} S_{ni}'$, where the condition $S \ge 1$ can, in principle, be met. This is mostly likely to happen at high altitudes. The condition $S = 1$ is met at the Lundquist scale $\ell_S$, defined using Equations (\ref{EqNum12}) and then (\ref{EqNum21b}) as 
\bln
\begin{equation}
  \ell_S \equiv \frac{\eta_P}{\mu_0 V_{At0}} = \frac{V_{At0}}{\nu_{ni}} \frac{1}{S_{ni}'}.\label{EqNum22}
\end{equation}
\eln
The values of the Lundquist scale $\ell_S$ (in m) also are shown (black solid curves) for the ALC7 chromosphere (left) and TIMEGCM I/T (right) in Figure \ref{fig:lehnerts}. We emphasize the very sharp contrast in the Lundquist scales for the two environments.  In the chromosphere, in quiet average conditions, $\ell_S$ is everywhere below 1 km; only shorter-wavelength disturbances are resistivity-dominated, whereas an extended range of longer-wavelength disturbances on the Sun is convection-dominated, with increasingly frozen-in motions of gas and magnetic field.  In the I/T, on the other hand, $\ell_S$ everywhere exceeds 9$\times$10$^4$ km, or 15 $R_E$; therefore, {\it all} Alfv\'enic disturbances at Earth are resistivity-dominated, with easy slippage of the partially ionized mixture relative to the magnetic field. This implies that any fluctuations in the geomagnetic field of significant amplitude must be driven by some external agent, rather than from within the I/T itself. Thus, excepting disturbances incident from the overlying magnetosphere, Earth's magnetic field is approximately static. Meanwhile, both the photosphere and chromosphere serve as very active sources of strong magnetic-field fluctuations on the Sun. This confluence of circumstances -- the disparate small/large Lundquist scales and the dynamic/static character of the chromospheric/ionospheric magnetic field -- is responsible for the prevalence of magnetohydrodynamics in conceptualizing and quantifying processes in the chromosphere, on the one hand, and of electrodynamics in the I/T, on the other.

\subsection{Neutral-Wind Driven Dynamos: An example of the E-J and v-B paradigms}
\label{neutral_wind_dynamos}

{\color{black} In this section we discuss the phenomena of neutral-wind driven dynamos, and use it as an example of how the I/T and chromospheric communities use two different paradigms to explain the same phenomena. Another example will be given in \S \ref{rayleightaylor}.}

At Earth, hydrodynamic forcing of the thermosphere neutral gas through pressure-gradient forces generated by differential solar radiative heating, Coriolis forces, and ion-drag forces create a global circulation of neutral winds. Additional forces on the neutral gas from Joule, collisional, and particle heating and from vertically propagating waves complicate the circulation. This neutral circulation is impressed upon the ionospheric plasma through collisions, causing the ions and electrons to undergo differential motion and leading to the production of currents and electric fields \citep[e.g.,][]{2000GMS...118..131R}.

Neutral winds on the Sun that, in principle, might drive I/T-like dynamo action through collisional coupling to the plasma originate in (1) the randomly shifting near-surface convection cells in the neutral-dominated photosphere and (2) the global atmospheric oscillations produced by acoustic and gravity waves. Hence, both the chromosphere and I/T may be subject to dynamo action in which inhomogeneous flows of neutrals couple to the plasma and drive persistent electric currents. 

For clarity, we point out that the usual usage of the term ``dynamo'' in solar physics refers to amplification of seed magnetic fields to greater strengths by convective turnover, twisting, and differential shearing of the plasma alone. Such dynamo action is widely accepted to occur deep in the Sun's fully ionized convective zone, producing the sunspot cycle of strong magnetic fields \citep[e.g.,][]{2010LRSP....7....3C}, and also may occur within the near-surface convective layers to produce flux concentrations at much smaller scales \citep[e.g.,][]{2012LRSP....9....4S}. Neither of these dynamos relies upon plasma-neutral coupling to generate magnetic fields; indeed, neutrals generally are not considered in these models, especially in the deep Sun.

 Below is a discussion of the neutral wind driven dynamo action from two points of view, dynamic and static electrodynamics, which highlights the differing approaches of the chromospheric and I/T communities to  common phenomena.

{\color{black}
Proceeding from the full time-dependent equations of electromagnetics and plasma physics, \citet{Vas_2012} describes the sequence of events that establishes the distributed neutral-wind driven dynamo as follows:
\begin{enumerate}
  \item Plasma motions $\mb{V}_p$ are induced locally by collisions with the neutral wind $\mb{V}_n$;
  \item The resultant bulk plasma flow $\mb{V}_{p}$ produces a persistent electric field, $\mb{E}_{\perp} = - \mb{V}_p \times \mb{B}_0$;
  \item Gradients in $\mb{E}_{\perp}$ along and across $\mb{B}_0$ generate magnetic perturbations $\delta \mb{B}$;
  \item Electric currents $\mb{J}$ arise due to gradients in the magnetic perturbations $\delta \mb{B}$;
  \item Lorentz forces $\mb{J} \times \mb{B}_0$ drive MHD waves propagating away from the dynamo region, propagating flows and currents;
\end{enumerate}
}
From the perspective of solar chromospheric physics, this chain of processes is intuitively appealing and non-controversial, although the details undoubtedly would be debated and additional study would be warranted to confirm its essential correctness. The chromosphere exhibits unceasing vigorous flows and strong magnetic variability. Therefore, explaining any chromospheric phenomenon based on a neutral-wind driven dynamo demands consistency with the full time-dependent equations and their implications from the outset. Exploratory work in this direction has been done on the quasi-static structuring of so-called network magnetic fields in the lanes of chromospheric convection cells by \citet{1991A&A...241..613H}, who call this process a ``photospheric dynamo'' \citep[see also][]{1997A&A...318..947H}. In addition, \citet{2011ARep...55.1132K} has proposed that the highly intermittent, collimated chromospheric upflows known as spicules, which are ubiquitous in the inter-cellular lanes, are powered by Alfv\'en waves driven by neutral-wind dynamo action. One of the necessary conditions for this ``photospheric dynamo" is that the electrons are magnetized but the ions are not. This condition is realized in the 1D static model used here at about 500 km above the solar surface; for stronger background fields, this critical height is reduced due to the increase in the electron gyrofrequency. \citet{Krasnoselskikh_2010} recently suggested that intense currents can be generated when magnetized electrons drift under the action of electric and magnetic fields induced in the reference frame of ions moving with the neutral gas, and that the resistive dissipation of these currents may be important for chromospheric heating.

{\color{black} From the perspective of} ionospheric electrodynamics, Ohm's law works very well on time scales longer than the ion-neutral collision time, which for the ionospheric M2 and M3 domains where primary dynamo activity occurs is only a few hundredths of a second to a few seconds.  For periods exceeding this time scale, steady-state electrodynamic behavior results and electrostatic electric fields and divergence-free current densities can be assumed. Consequently, a static approach is often applied to describing the I/T's neutral wind dynamo process:

{\color{black}
\begin{enumerate}[font=\itshape]
  \item Collisions with neutrals create plasma motions $\mb{V}_p$ along $\mb{V}_{n}$. At the same time, oppositely directed 
  $\mb{R}_{i}^{in}\times\mb{B}$ and $\mb{R}_{e}^{en}\times\mb{B}$  ion and electron drifts are created;
    \item Electric currents $\mb{J}$ resulting from the drifts drive charge separation between the ions and electrons;
  \item An electric field $\mb{E}$ is established in the dynamo region due to the charge separation;
  \item Potential mapping along magnetic field lines due to rapid electron motions along $\mb{B}_{0}$ extends the electric field to increasingly remote regions;
  \item $\mb{E} \times \mb{B}$ drifts create bulk plasma flows $\mb{V}_p$ outside the dynamo region;
\end{enumerate}

}
{\color{black} The first sequence above asserts that bulk plasma flows drive the electric field and distribute the dynamo along magnetic field lines through the intermediary of magnetic fluctuations having associated currents; thus, $\mb{B}$ and $\mb{V}$ have primary roles, while $\mb{E}$ and $\mb{J}$ are secondary. The second sequence, on the other hand, states that current-generated electric fields distribute the dynamo along magnetic field lines and drive the bulk plasma flows; thus, $\mb{E}$ and $\mb{J}$ have primary roles, while $\mb{B}$ and $\mb{V}$ are secondary.

\citet{Vas_2011,Vas_2012} criticizes the electrodynamic interpretation as originating from extending intuition developed from the simple steady-state relations among the four variables -- $\mb{E}, ~ \mb{J}, ~\mb{B}, ~\mb{V}$ -- into time-varying situations, without due regard for the causal relationships implied by the dynamical equations.  At scales long compared to the electron plasma oscillation period and large compared to the electron Debye length, the displacement current can be neglected in Amp\'{e}reÕs law and the time derivative of the particle current can be neglected in the generalized OhmÕs law, so there are no dynamical equations for $d\mb{E}/dt$ and $d\mb{J}/dt$ to be solved \citep{Vas_2005a}.  Instead, 
$\mb{J}$ is related to $\mb{B}$ through 
Amp\'{e}reÕs law, and $\mb{E}$ is related to $\mb{B}$, $\mb{V}$, and $\mb{J}$ through OhmÕs law.  The dynamical equations 
that remain then contain only $d\mb{B}/dt$ and $d\mb{V}/dt$ among the four variables.  

These dynamical issues also have been addressed more broadly and in several specific applications by Parker (1996a,b, 2007).  He emphasizes that the converse procedure, eliminating $\mb{B}$ and $\mb{V}$ from the dynamical relations in favor of $\mb{E}$ and 
$\mb{J}$, culminates in nonlinear integrodifferential equations that are nearly intractable to solve and obscure the underlying physical processes.  The two alternatives are characterized as the ÔB-V vs.\ E-J paradigms.Õ  He concludes that applying insights from laboratory experimental configurations -- in which electric fields drive currents and generate magnetic fields -- to anything beyond symmetric, static situations in the solar and terrestrial atmospheres is fraught with difficulty.  The result has been much misunderstanding, and even misdirection of effort, in the two communities, according to Parker.  

For the application to wind driven dynamos discussed here, we find that the difference in perspectives between the solar and ionospheric approach is driven by a difference in the plasma parameters. The chromosphere is ideal (non-resistive) on length scales much larger than a km, and so the electric field in the rest frame of the neutrals 
$\mb{E}+(\mb{V}_{n}\times\mb{B})$ is small. This leads to a magnetic perturbation and wave-propagation interpretation of the dynamo. In the resistive I/T however, there is an electric field in the rest frame of the neutrals created by charge separation due to ion electron drifts. This locally generated electric field propagates electrostatically along the field leading to a different interpretation of the phenomena. 

The disparate perspectives imparted by these frameworks raise substantial barriers to communication and to making collaborative progress on otherwise similar phenomena.  An explicit example of the challenges posed is provided in \S \ref{rayleightaylor} on the Rayleigh-Taylor instability, which we analyze from both magnetohydrodynamic (chromospheric) and electrodynamic (ionospheric) perspectives.
}

\subsection{Farley-Buneman Instability}
As mentioned above, in the M2 domains of the chromosphere and I/T neutral flows perpendicular to the magnetic field can drive ions along with them but not electrons, and this produces currents. 
In the I/T, these currents comprise the electrojet current systems that occur in equatorial and auroral regions \citep{Kelley_1989,Schunk_Nagy_2000}. The equatorial current system is driven by tidal E-layer neutral winds generated by daytime solar heating 
{\color{black} (the electric field is radial and the current is azimuthal)}. It peaks near the magnetic equator as a consequence of the nearly horizontal field lines. The auroral current system is associated with field-aligned, high-latitude currents driven by the solar-wind/magnetosphere interaction, as well as by precipitating energetic particles \citep{Kamide_1982}. 

The M2 domains in both the I/T and chromosphere environments are an ideal place for the occurrence of the Farley-Buneman instability \citep{1995JGR...10014605D,2005A&A...442.1099F,2006JGRA..111.3302O,2008A&A...480..839F,2013JGRA..118.1306O,2013arXiv1308.0305M}, which is a two-stream kinetic instability in which the ions are unmagnetized while the electrons are tied to the magnetic field \citep{Farley_1963, Buneman_1963}. Its thermodynamic consequences may not be very important compared to frictional heating, especially in the chromosphere where the spatial scales associated with the currents are so small that Joule heating outweighs the instability-related heating \citep{2009ApJ...706L..12G}.  However, there may be transient situations, such as the exhaust regions of magnetic reconnection in the chromosphere, where electron and ion fluids separate on small scales and the Farley-Buneman instability could be thermodynamically important. In the Earth's electrojet current systems, the plasma waves generated by the Farley-Buneman instability can provide an anomalous resistivity, which in turn modifies the ambient current systems and electric fields \cite[e.g.][]{1995JGR...100.9653H}. {\color{black} When plasma waves are excited, the low frequency Ohm's law is incorrect, and as in the work of \citet{1995JGR...100.9653H}, terms such as the electron and ion inertia should be included in the model.}

\subsection{Summary}

{\color{black} Factors such as the resistive nature and the static magnetic field of the I/T have lead to a common approach where the electric field ($\mb{E}$) and electric current ($\mb{J}$) are the primary variables, and these are often used to describe how the I/T is driven (the E-J paradigm). An example was performed in our previous discussion of mobility of ions and electrons in an electric field. However, only very slow changes in the I/T can be accommodated by an electrostatic field. Furthermore, while for long time scales we can remove the dynamical equations for $d\mb{E}/dt$ and $d\mb{J}/dt$ and have $\mb{B}$ and $\mb{V}$ be the primary variables (V-B paradigm), the converse is not true.

At the same time, using the E-J paradigm may allow one to arrive at an equivalent result as the V-B paradigm. This was achieved in a general sense for wind-driven dynamos and will be shown to be possible for the Rayleigh-Taylor instability in a following section. }




 
\section{Energy Transfer}
\label{EM}

{\color{black} In this section we first discuss the major contributions to the energy balance in both the chromosphere and I/T, then go on to discuss the general process of conversion of electromagnetic energy to thermal and kinetic energy. Then we present ideas on the flux and dissipation of electromagnetic energy in the two environments, and the role of plasma-neutral collisions in this process.}
\subsection{Major Contributions to Energy Balance}

\subsubsection{Chromosphere}

{\color{black} The bulk of the chromosphere} is a few thousand degrees hotter than the underlying photosphere.  It also is far cooler than the corona above, but is so much denser that it requires roughly ten times more heat input than the corona (when measured as a height-integrated rate of change of energy density) to maintain its elevated temperature.  Hydrogen ionization in the chromosphere acts to
balance heating even as the density drops with height. This is because the large abundance and ionization energy of hydrogen
allows the ionization of hydrogen to absorb energy, while this ionization creates free electrons which excite
heavier species such as iron, magnesium and calcium, resulting in steady radiative cooling. The heating of the chromosphere, 
which must account for these radiative losses, comes from a combination of collisional effects (Joule, frictional, and viscous heating) and compressional heating. 
The Joule and viscous heating mechanisms become relatively more important on progressively smaller scales, such as 
those associated with current sheets, shocks, or wave motions. The first of these phenomena is a DC mechanism, while the last two are AC in nature. 

A large class of potential heating mechanisms for the chromosphere is derived from the fact that the turbulent convection zone below is capable of supplying a flux of wave energy into the chromosphere. 
Previously suggested heating mechanisms for the chromosphere have included acoustic wave dissipation, though it
is unclear if acoustic waves can propagate high enough to deposit their energy in the chromosphere  \citep{1946NW.....33..118B,1948ApJ...107....1S,1990ASPC....9....3U,2005ESASP.600E..17F, 2005Natur.435..919F,2006ApJ...646..579F, Kalkofen_2007}. MHD waves have more recently been investigated, particularly Alfv\'{e}n waves as they can propagate upwards along magnetic fieldlines into the upper chromosphere \cite[e.g.][]{2007AAS...210.9415D, 2007AGUFMSH21A0289T}. However,
although plasma-neutral collisions and viscous effects can potentially dissipate  wave energy in the chromosphere, it is not fully understood whether the dissipation and heating of the waves provided by the convection zone is sufficient to counter the radiative losses  \citep{DePontieu_1999, Leake_2005, De_Pontieu_2007,2008ApJ...680.1542H, Goodman_2011, Song_2011}, {\color{black} although recent simulations of Joule-dissipated reflected Alfv\'{e}n waves in the lower chromosphere (below 400 km) by \citet{Tu_2013} suggest that the dissipated wave energy can account for the losses there}. 

Recently, the dissipation of Alfv\'{e}n wave energy by other mechanisms has been considered, such as non-linear interactions, mode conversion, and resonant heating. \citet{1990SSRv...54..377N} present a more  comprehensive literature review of these types of investigations. Low frequency mechanisms, such as the dissipation of currents perpendicular to the magnetic field by Pedersen resistivity  \citep{2006A&A...450..805L, 2007ApJ...666..541A, Arber_2009, khomenko, 2012ApJ...753..161M, Leake_2013}, magnetic reconnection \citep{1983ApJ...264..635P, 1988ApJ...326..407P, Dahlburg03, Dahlburg05, 2006SoPh..234...41K, Goodman_judge_2012}, and neutral-wind dynamos \citep{Krasnoselskikh_2010}, have also been investigated.
Clearly, there is a ``zoo'' of possible chromospheric heating mechanisms, and a complete review of the chromospheric heating 
problem is one which is outside the scope of this review. What is also clear is that this heating is occurring in a region of highly coupled ionized plasma and neutral gas, and that plasma-neutral interactions are vital to heating of the chromosphere by bulk motions. 


\subsubsection{Ionosphere}

The contributions to the energy transfer in the ionosphere/thermosphere are better understood than those in the chromosphere.   However, the large variability in time and space of these mechanisms and the relative contributions to the overall heating and cooling remain challenges to understanding the thermal properties of the system. The lower amount of plasma-neutral and electron-ion coupling in the I/T (based on the collision frequencies), compared to the chromosphere, means that the individual heating of the constituents and their thermal coupling must be considered. The dominant heating term for the neutral gas is absorption of UV/EUV radiation. 
 The UV/EUV energy flux from the Sun is $\sim$ 4 orders of magnitude less than at visible wavelengths \cite[e.g,,][]{2000JASTP..62.1233T}, but is of sufficient energy to form the ionosphere through photoionization (by contrast, ionization is predominantly collisional excitation in the chromosphere), produce secondary ionization, energize gas emissions in airglow, and raise neutral and plasma temperatures to more than one thousand degrees. The UV/EUV flux is 
is nearly completely absorbed between 80 and 200 km altitude. However, the most dynamic term of the energy equation given in Equation (\ref{energy_eqn}) is the collisional exchange term between ions and neutrals. This frictional heating of the neutral gas by collisions with ions can be as significant as solar heating during geomagnetic storms and represents the most variable source of energy to quantify in the I/T energy equation. Other internal processes within the weakly ionized mixture, such as thermal exchange between the plasma and neutral gas, lead to differing thermal structure with height for the plasma and neutral gas, as indicated in Figure \ref{fig:temperatures}. 
 
 It is worth examining high latitude heating further as it is central to the aims of this paper. It has been recognized for some time that the observed thermal structure of the  I/T is only adequately represented when energy resulting from solar wind interactions with the Earth's magnetosphere is included in the  I/T energy budget. This energy to the  I/T is manifested in the form of Poynting flux (see below) and particle kinetic energy flux. The partitioning of auroral kinetic energy (KE) flux in the 80-200 km altitude range is roughly 50 \% heating, 45 \% ionization, and 5 \% optical production \cite[e.g.,][]{2004JASTP..66..807T}. Typically the energy flux by particles is less than that due to electromagnetic processes. 
 {\color{black} Electromagnetic fields transfer energy from the plasma to the neutral gas on the neutral-ion collisional timescale.  Other processes, at times, prove important in the energy balance of the plasma or neutral gas in the I/T. Ion frictional heating at high latitudes is an effective process that can lead to high ion temperatures in the M2 region. This heat is transferred to the neutral gas by thermal differences between the neutrals and ions, contributing to the heating of the neutral gas. Farley-Buneman (F-B) instability in the M2 region is a common occurrence in the I/T. At high latitudes, where strong electric fields are present, the F-B instability can lead to very efficient plasma wave heating of electrons in the M2 region with temperatures exceeding 2000 K. However, this does not lead to any significant change in the ion and neutral temperature as they have much greater thermal heat capacity.}
 

\subsection{{\color{black} Frictional Heating}}
\label{joule}
{\color{black} One important contribution to the energy balance in both environments is the conversion of electromagnetic energy to thermal and kinetic energy, $\mb{E}\cdot\mb{J}>0$. This is commonly expressed as 
the sum of two terms: {frictional (including Joule) heating} and work done \cite[e.g.][]{Lu_1995, Thayer_1995, Fujii_1999, Thayer_2000, 2000GMS...118..131R, Goodman2000, Goodman2004a,Vas_Song_2005}. 
In a non-relativistic, quasi-neutral plasma, $\mb{E}$ is frame-dependent while  $\mb{J}$ is not, hence the relationship between $\mb{E}$ and $\mb{J}$ is frame-dependent. The electric field in a frame moving with velocity $\mb{V}$ is $\mb{E}^{V} \equiv \mb{E}+(\mb{V}\times\mb{B})$ and 
\bln
\be
\mb{E}\cdot\mb{J} = \mb{E}^{V}\cdot\mb{J}-(\mb{V}\times\mb{B})\cdot\mb{J} = \mb{E}^{V}\cdot\mb{J}+(\mb{J}\times\mb{B})\cdot\mb{V}
\ee
\eln
thus, the electromagnetic energy conversion rate $\mb{E}\cdot\mb{J}$ is the sum of a { heating} term $\mb{E}^{V}\cdot\mb{J}$ and a term which is the rate work is done by the Lorentz force on the flow $\mb{V}$.

However, the choice of rest frame ($\mb{V}$) for the term $\mb{E}^{V}\cdot\mb{J}$ can alter the interpretation of frictional heating in the energy equation. As shown in \S \ref{genohmslaw}, the equation for $\mb{E}^{V}$ contains different terms depending on the choice of rest frame (plasma, neutral, or center of mass).  One should also note that in the upper chromosphere where the ionization level increases, the center of mass frame differs significantly from the neutral frame, which changes the equation for $\mb{E}^{V}$. Hence the term {frictional heating} can have physically different meanings depending on the rest frame chosen. For example, choosing the rest frame of the plasma, and using the low-frequency approximation to the generalized Ohm's law in \S \ref{genohmslaw}, Equation (\ref{gol_ions1}), the frictional heating is 
\bln
\be
\mb{E}^{p}\cdot\mb{J} = (\mb{E}+\mb{V}_{p}\times\mb{B})\cdot\mb{J} =  \eta_{\|}\mb{J}^{2},
\label{eq:jh1}
\ee
\eln
and is just the Ohmic \textit{Joule heating} term, related to the parallel resistivity (electron-ion collisions plus coupled electron-neutral and ion-neutral collisions). It should be noted that care must be taken to ensure that the
assumptions made to obtain the low frequency limit are not violated when discussing frictional heating for a 
particular phenomena, e.g., wave propagation and dissipation (where the frequency must be smaller than the ion-neutral collision frequency).
Choosing the rest frame of the neutrals the frictional heating is (using Equation (\ref{gol}) this time)
\bln
\be
\mb{E}^{*}\cdot\mb{J} = (\mb{E}+\mb{V}_{n}\times\mb{B})\cdot\mb{J} =  \eta_{\|}\mb{J}^{2} + \eta_{C}\mb{J}_{\perp}^{2}.
\label{eq:jh2}
\ee
\eln

This $\mb{E}^{*}\cdot\mb{J}$, often called Joule heating by I/T researchers, is related to the Ohmic Joule heating (in the rest frame of the plasma), $\mb{E}^{p}$, by 
\bln
\bea
\mb{E}^{*}\cdot\mb{J} = \mb{E}^{p}\cdot\mb{J} - \left[(\mb{V}_{p}-\mb{V}_{n})\times\mb{B}\right]\cdot\mb{J} & = & \mb{E}^{p}\cdot\mb{J} + \left(\mb{J}\times\mb{B}\right)\cdot(\mb{V}_{p}-\mb{V}_{n}) \nonumber \\
& \approx & \mb{E}^{p}\cdot\mb{J} + \rho_{i}\nu_{in}|\mb{V}_{p}-\mb{V}_{n}|^{2}
\label{Joule_Heating}
\eea
\eln
which assumes that  Lorentz and plasma-neutral drag forces dominate in the plasma (ion + electron) equation of motion ($\mb{J}\times\mb{B} \approx \mb{R}_{in}$), and inertial, pressure and gravity terms can be neglected, as discussed in \citet{Vas_Song_2005}. Hence the conventional ionospheric Joule heating, in the rest frame of the neutrals, is the heating in the rest frame of the plasma plus heating due to plasma-neutral collisions. The first term, the plasma Joule heating, is strictly \textit{Ohmic} Joule heating in the sense that the electric field in the rest frame of the plasma accelerates electrons and ions, creating a current proportional to the electric field, and this current is limited mainly be electron-neutral collisions. \citet{Vas_Song_2005} explain that this Ohmic Joule heating  goes into the thermal energy of the plasma, while the term due to to plasma-neutral collisions distributes energy between the plasma and neutrals equally. The last term in Equation (\ref{Joule_Heating}) can also be thought of as a term proportional to the difference in work done by the Lorentz force on the plasma relative to work done by the Lorentz force on the neutrals.}

In addition to the frame-dependence issue stated above, care must be taken when discussing energy exchange (thermal and kinetic) between the electron, ion and neutral components, and the energy balance of the mass-averaged plasma-neutral mixture. This is discussed in detail in \citet{Vas_Song_2005}. 

Figure \ref{fig:res} shows that $\eta_{C} \gg \eta_{\|}$ for most of the ionosphere, and for a reasonable proportion of the chromosphere. Using this fact, and comparing Equations (\ref{eq:jh1}) and (\ref{eq:jh2}), tells us that for most of the ionosphere and much of the chromosphere, the frictional heating has only a tiny contribution from the  Ohmic Joule heating, and is dominated by the contribution from collisions of plasma with neutrals. The relationship of ionospheric Joule heating to neutral gas frictional heating was demonstrated in descriptions put forward by \citet{St_Maurice_1981}.

A common approach in chromospheric physics is to assume a single-fluid model.
Recall the individual species ($\alpha$) total energy equation:
\bln
\begin{equation}
\frac{\partial \epsilon_{\alpha}}{\partial t} +\nabla \cdot (\epsilon_{\alpha}\mb{V}_{\alpha} + \mb{V}_{\alpha}\cdot\mathbb{P}_{\alpha} + \mb{h}_{\alpha}) = \mb{V}_{\alpha}\cdot\left(q_{\alpha} n_{\alpha}\mathbf{E} + \mathbf{R}_{\alpha}^{\alpha\beta} + m_{\alpha}n_{\alpha}\mb{g}\right) + \sum_{\beta\ne\alpha}{Q_{\alpha}^{\alpha\beta}} + S_{\alpha} + U_{\alpha}
\end{equation}
\eln
where $Q_{\alpha}^{\alpha\beta}$ is the rate of change of total energy of the species $\alpha$ due to collisions with species $\beta$, and
$\epsilon_{\alpha} = \rho_{\alpha}v_{\alpha}^{2}/2 + P_{\alpha}/(\gamma-1)$.
{\color{black} Summing over all species $\alpha$ gives the total energy equation for the mixture:
\bln
\begin{equation}
\frac{\partial\epsilon_{CM}}{\partial t} + \nabla\cdot\left( \epsilon_{CM}\mb{V}_{CM} + \mb{V}_{CM}\cdot\mbb{P}_{CM} + \mb{h}_{CM}  + \right) 
= \mb{E}\cdot\mb{J} + \rho_{CM}\mb{V}_{CM}\cdot\mb{g}  + S_{CM} + U_{CM} 
\label{eq:total_int_energy0}
\end{equation}
\eln
where the following mass-averaged quantities are defined in a similar vein as \citet{Vas_Song_2005}:
\bln
\bea
\rho_{CM} & = & \sum_{\alpha}{\rho_{\alpha}}, ~ \mb{V}_{CM} =  \frac{\sum_{\alpha}{\rho_{\alpha}\mb{V}_{\alpha}}}{\sum_{\alpha}{\rho_{\alpha}}} \nonumber \\
\epsilon_{CM} & = & \frac{\rho_{CM}V_{CM}^{2}}{2} + \frac{P_{CM}}{\gamma-1} + \frac{\rho_{CM}\xi_{n}(1-\xi_{n})W^{2}}{2}, ~ P_{CM} =   \sum_{\alpha}{P_{\alpha}} \nonumber\\
\mbb{P}_{CM} & = & \sum_{\alpha}{\mbb{P}_{\alpha}}+ \rho_{CM}\xi_{n}(1-\xi_{n})\mb{W}\mb{W} \nonumber \\
h_{CM} & = &  \sum_{\alpha}{h_{\alpha}} + \xi_{n}\mb{W}\left[ \left(\frac{1}{\xi_{n}}-1\right)P_{n}-(P_{i}+P_{e})
-(1-\xi_{n})\frac{\rho_{CM}^{2}\mb{W}^{2}}{2}\right] \nonumber \\
& - & \xi_{n}\mb{W}\cdot\left[\left(\frac{1}{\xi_{n}}-1\right)\mbb{P}_{n}-(\mbb{P}_{i}+\mbb{P}_{e})\right] \nonumber
\eea
\eln
}

All the terms in the single-fluid equations reflect the total description of the plasma-neutral mixture. However, the individual speciesÕ equations contain terms that reflect the species description within the mixture, which may not affect the behavior of the mixture as a whole. Thus, terms such as the collisional terms $\mb{V}_{\alpha}\cdot \mb{R}_{\alpha}^{\alpha\beta}$ and $Q_{\alpha}^{\alpha\beta}$ cancel when summed over the whole mixture. The individual species approach is a path-dependent system that accounts for the processes between the different species that lead to macroscopic changes in the system. The single-fluid approach is independent of the path by which energy is transferred amongst the species in its description of the macroscopic behavior of the mixture. 

The term $\mb{E}\cdot\mb{J}$  comes from $\sum_{\alpha} \mb{V}_{\alpha}\cdot(q_{\alpha} n_{\alpha}\mathbf{E})$. As discussed above, the term $\mb{E}\cdot\mb{J}$ can be written as
\bln
\be
\mb{E}\cdot\mb{J}  = \mb{E}^{CM}\cdot\mb{J}+(\mb{J}\times\mb{B})\cdot\mb{V}_{CM}
\label{eq:E.J}
\ee
\eln
where $ \mb{E}^{CM} = \mb{E} + (\mb{V}_{CM}\times\mb{B})$ is the electric field in the rest frame of the mass-averaged plasma-neutral mixture. 
The term $\mb{E}\cdot\mb{J}$, valid for any frame of reference, goes into the mass-averaged total (kinetic + thermal) energy of the plasma-neutral mixture. {\color{black} In the typical analysis of the total energy of a fluid, one can generally remove the kinetic part by subtracting the momentum equation dotted into the velocity to obtain the thermal energy equation. Looking at Equation (\ref{eq:E.J}) above, 
one may think that 
$\mb{E}\cdot\mb{J}$ can be split into a heating term $\mb{E}^{CM}\cdot\mb{J}$ which goes into the thermal energy of the center of mass fluid, and a work term $(\mb{J}\times\mb{B})\cdot\mb{V}_{CM}$, which goes into the kinetic energy.
 However, as was discussed by \citet{Vas_Song_2005}, and looking at Equation (\ref{eq:total_int_energy0}) and the mass averaged energy, pressure, and heat flux defined above, one can see that the single-fluid has a ÒthermalÓ energy density that includes a contribution from kinetic energy of relative motion between plasma and neutrals. It also has a non-isotropic pressure and a non-zero heat flux even if these are absent in the individual plasma and  neutral fluids.


}

As we did for the Joule heating in the rest frame of the neutrals,
 we can write the heating term $\mb{E}^{CM}\cdot\mb{J}$ as the sum of the  Ohmic Joule heating term and frictional heating due to the relative flow velocity of the mass-averaged mixture and the ionized plasma
\bln
\bea
\mb{E}^{CM}\cdot\mb{J} = \mb{E}^{p}\cdot\mb{J} - \left[(\mb{V}_{p}-\mb{V}_{CM})\times\mb{B}\right] \cdot\mb{J}  = \mb{E}^{p}\cdot\mb{J} + \left(\mb{J}\times\mb{B}\right)\cdot(\mb{V}_{p}-\mb{V}_{CM}).
\eea
\eln
Due to the low ionization level in the I/T and the lower and middle chromosphere, this is approximately the same as Equation (\ref{Joule_Heating}) as $\mb{V}_{CM}\approx \mb{V}_{n}$ in these regions. Hence the frictional heating of the plasma-neutral mixture is dominated by collisions between plasma and neutrals, with a very small contribution from electron collisions (Ohmic) Joule heating.

Having seen how the electromagnetic energy is converted into thermal and kinetic energy, we can now examine the evolution of this energy.
The use of PoyntingÕs theorem, derived directly from MaxwellÕs equations, can relate the total energy equation with the equation for electromagnetic energy when written in the form
\bln
\be
\frac{\partial}{\partial t}\left(\frac{B^2 + E^2/c^2}{2\mu_{0}}\right) + \nabla\cdot\left(\frac{\mb{E}\times\mb{B}}{\mu_{0}}\right) + \mb{E}\cdot\mb{J} = 0 \label{eq:poynting}
\ee
\eln
The first term is the time-rate-of-change of electromagnetic energy density. The second term is the divergence in the Poynting vector $\mb{S}_{p}\equiv\mb{E}\times\mb{B}/\mu_{0}$, or the electromagnetic energy flux, and the last term is the rate of electromagnetic energy transferred to the medium. Clearly the last term couples this equation with the single fluid total energy equation described above, and connects electromagnetic energy changes with changes in the total energy of the gas mixture. PoyntingÕs theorem applies to all types of electromagnetic interactions, ranging from electromagnetic waves to steady-state fields. 
It is this electromagnetic exchange between the fields and the gas mixture that is of interest in contrasting the chromosphere and the I/T.

\subsection{The Role of Plasma-neutral Coupling in Energy Transfer}

{\color{black} As mentioned above, heating $Q \equiv \mb{E}^{V}\cdot\mb{J}$ where $\mb{E}^{V}$ is in the neutral or center of mass frame of the chromosphere and I/T contains a component due to plasma-neutral collisions. This component can dominate over Ohmic Joule heating when $\eta_{P}>\eta_{\|}$, or $\sigma_{P}<\sigma_{\|}$. Hence plasma-neutral collisions could play a major role in the conversion of electromagnetic energy into thermal energy.  }

{\color{black} We can discuss the efficiency of this heating without referring to a particular mechanism for the generation of currents. In \S \ref{mobilities} we presented a description of the altitude variation of electrical currents of a given electric field. In particular, we looked at how the contributions to the current perpendicular to the magnetic field $\mb{J}_{\perp}$ by Pedersen ($\mb{J}_{P} \equiv \sigma_{P}\mb{E}^{*}_{\perp}$) and Hall ($\mb{J}_{H} \equiv -\sigma_{H}\mb{E}^{*}_{\perp}\times\hat{\mb{b}}$) currents varied as the mobilities of the electrons and ions varied. Let us now look at the contributions to the heating $(\mb{E}^{*}\cdot\mb{J}$). \citet{Goodman2004a} showed that when the heating is written as
\bln
\bea
Q  \equiv  \mb{E}^{*}\cdot\mb{J} 
& = & \frac{J_{\|}^{2}}{\sigma_{\|}} + \frac{\sigma_{P}J_{\perp}^{2}}{\sigma_{P}^{2}+\sigma_{H}^{2}} \equiv Q_{\|} + Q_{P}
\label{eq:Q}
\eea
\eln
then the efficiency of Pedersen heating $Q_{P}$, the ratio of Pedersen heating to its maximal value, obtained when the perpendicular current $\mb{J}_{\perp}$ is all Pedersen current and no Hall current, can be expressed as
\bln
\be
R_{Q}\equiv Q_{P}/Q_{P,max} = \frac{\sigma_{P}^{2}}{\sigma_{P}^{2}+\sigma_{H}^{2}}.
\ee
\eln 
This tells us how efficiently the mechanism that generates $\mb{E}^{*}_{\perp}$ heats the atmosphere. We discuss the general nature of $R_{Q}$ here and return to such mechanisms later in this section. 

Figure \ref{fig:heating} shows this efficiency for both chromosphere and I/T, including the results for the three different magnetic field models for the chromosphere. Also shown is the temperature (we show the neutral temperature only for the I/T).

Due to the above definition of $R_{Q}$, when $\mb{J}_{\perp}\approx 0$, $R_{Q}\approx1$. Otherwise, $R_{Q}$ is close to 1 when the perpendicular current is dominated by Pedersen current. 
Recall from \S \ref{mobilities} that this occurs just below the lower peak, and just above the upper peak, in the Pedersen mobility. Near the lower peak, the conductivity (and resistivity) is isotropic and Joule heating and Pedersen heating are the same. Near the upper peak, $\eta_{P} > \eta_{\|}$ and Pedersen heating dominates over Joule heating. Between these two regions of $R_{Q}\approx 1$ is a region where the Hall current dominates the perpendicular current and so $R_{Q}$ is minimal. Thus the transition of interest is from the minimal efficiency near the Hall mobility peak up to the region above the upper peak in the Pedersen resistivity. 
This transition altitude occurs somewhere around the temperature minimum. In fact, the efficiency increases from its minimum to 1 with altitude as the term $\Gamma \equiv \xi_{n}k_{e}k_{in}$ increases from below to above 1. This can be observed by comparing Figure \ref{fig:heating} and Figure \ref{fig:magnetization2}. 

\citet{Goodman2000,Goodman2004a} proposed that the increase in Pedersen heating as $\Gamma $ undergoes the transition from much less than 1 to much more than 1 may explain the source of chromospheric heating that has so puzzled solar physicists. Moreover, theoretical work by \citet{Song_2011} and numerical simulations 
by \citet{Tu_2013} have found that given an atmospheric profile similar to the one used here, the heating due to damping of propagating Alfv\'{e}n waves is predominantly Ohmic Joule heating below the temperature minimum, and plasma-neutral frictional (or Pedersen) heating above. These recent simulations suggest a possible explanation for the presence of the temperature minimum.  However, such 
simulations, which contain no energy equation and thus no self-consistent heating, do not produce the observed temperature structure of the chromosphere, but use an initial condition which is designed to look like observationally inferred 1D profiles. The physics of thermal conduction from the hot corona, as well as radiative processes, must be included for the simulations to obtain self-consistent heating rates. Future simulations that investigate heating mechanisms must be able to create the correct temperature profile self-consistently (for example of attempts to do so see the papers of \citet{Carlsson_2012} and  \citet{Abbett_2007}) as well as including well-resolved physical 
mechanisms (such as Alfv\'{e}n wave wave heating). This is the largest obstacle to identifying chromospheric heating mechanisms through numerical and theoretical investigations.

Pedersen current dissipation is a general mechanism for dissipating the energy in electric currents which are orthogonal to the magnetic field. Therefore any process which drives such currents is damped by this mechanism. To actually estimate $Q_{P}$ and not just $R_{Q}$, a process must first be identified and a physical model of the process must then be created. The degree of the damping of the process which drives perpendicular currents depends on two things: the reservoir of energy 
to drive the currents, and the amount of energy which is used to generate and maintain electric fields to support the currents. 
In the I/T, the actual value of $Q_{P}$ is small compared to the dominant processes of heating. 
Thus, we just consider the chromosphere in this discussion. In the chromosphere transient electric fields are created by general time-dependent flows, such as convection zone flows, wave motions, and magnetic reconnection sites. 

\begin{figure}[h]
\includegraphics[width=\textwidth]{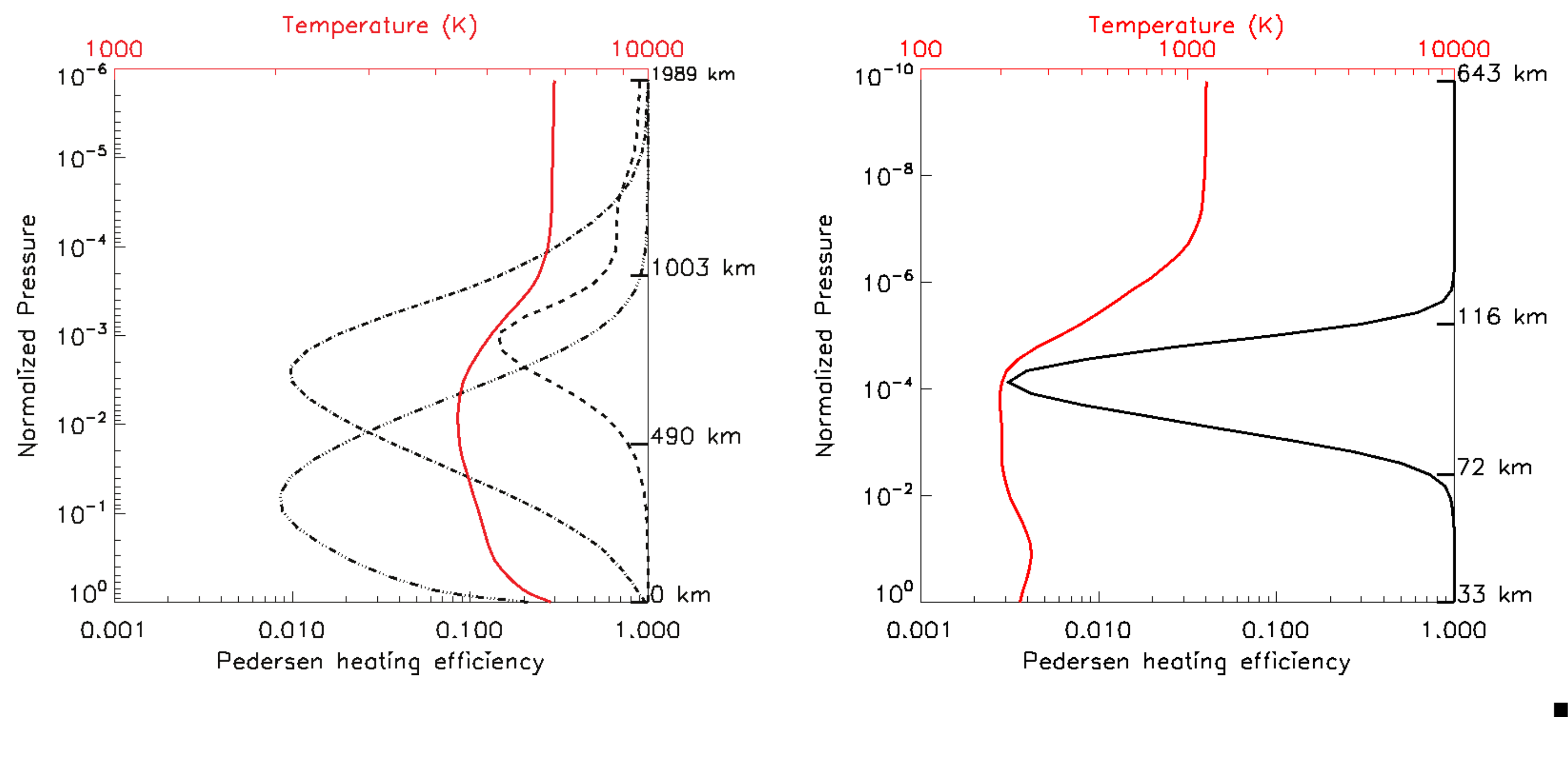}
\caption{{\color{black} Pedersen heating efficiencies (black lines) in the ALC7 chromosphere (left) and in the TIMEGCM I/T (right). Also shown is the electron temperature (red line). For the chromosphere, the three black lines show the heating efficiencies for the three magnetic field strengths of 10 G (dashed),100 G (dot-dashed), and 1000 G (dot-dot-dashed). }
\label{fig:heating}}      
\end{figure}


}

The chromosphere and I/T are atmosphere regions commonly characterized as weakly ionized mixtures permeated by magnetic field lines. {\color{black} These regions are subject to electromagnetic energy flux from neighboring regions of electrical energy generation. On the Sun, the turbulent convection zone beneath the chromosphere creates a spectrum of oscillations, in the range 0.5 to 1000 minutes, or $10^{-5}$ to 0.03 Hz \citep[e.g.,][]{Cranmer_2005}.} In the high plasma $\beta_{p}$, highly ionized  convection zone there are three types of fluid (non-kinetic) waves:  Alfv\'{e}n waves, isotropic sound waves, and magneto-acoustic waves that are guided by the magnetic field \citep{2011A&A...529A..82Z}. The last two of these types are compressional and most likely do not propagate into the upper chromosphere. However, the changing conditions with height change the nature of these waves as they propagate into the chromosphere. Alfv\'{e}n waves from the convection zone cause the oscillation of ionized plasma in the presence of a magnetic field which provides a mechanism for the conversion of the kinetic energy of convection into electrical energy in a manner equivalent to a magnetohydrodynamic (MHD) electrical generator. In this case, $\mb{E}\cdot\mb{J}$ is negative and mechanical energy is converted to electromagnetic energy which propagates with the waves into the chromosphere. Thus there is an influx of electromagnetic energy from the convection zone into the chromosphere. Much work has been done on the propagation and dissipation of Alfv\'{e}n waves into the chromosphere from the convection zone \cite[e.g.][]{DePontieu_1999, Leake_2005, De_Pontieu_2007,2008ApJ...680.1542H, Goodman_2011, Song_2011, Tu_2013}. 

A similar situation occurs above the I/T. The Earth's magnetosphere undergoes convection due to the interaction of the solar wind with the magnetosphere. There is a net down-flow of energetic particles and Alfv\'{e}n waves. Thus the 
 Earth's magnetosphere can also be considered an MHD electrical generator converting solar wind mechanical energy to electrical energy that is transferred to the I/T along highly conducting magnetic field lines
\citep{2004JASTP..66..807T, Song_2011, Tu_2011}.

Both atmospheres exhibit an inflow of electromagnetic energy in the form of waves from outside regions. 
The chromosphere experiences a large drop in density, and this implies that for a given field strength, the Alfv\'{e}n speed rapidly increases. If the length scale of the increase is comparable to the wavelength of an upwardly propagating wave then reflection can occur, similar to the situation in the Ionospheric Alfven Resonator \citep[see][and references therein]{1981Ge&Ae..21..816P}. Thus the chromosphere has downwardly propagating reflected waves. High frequency (1-100 Hz) Alfv\'{e}n waves generated from coronal reconnection sites \cite[e.g.][]{2002SoPh..206..285V,2010PASJ...62..993K,2011AGUFMSH31A1990E} may also propagate downward into the chromosphere, as well as leaking coronal loop oscillations caused by disturbances in the corona \cite[e.g.][]{1999Sci...285..862N,2002ApJ...568L.135O}. In the I/T, Alfv\'{e}n waves and quasi-static fields imposed by the magnetosphere are manifested in the form of auroral arcs, field-aligned currents and electric fields that constitute the electromagnetic energy flux into the region \cite[e.g.][]{1994GeoRL..21.1855E,2001JGR...106.3603T,2003Sci...299..383K,2013JGRD..118.3933D}

{\color{black} The chromosphere and Earth's I/T are conductors (the M2-domain in particular) and can convert the electrical energy of the waves to thermal and mechanical energy of the neutral gas ($\mb{E}\cdot\mb{J}>0$).  As shown previously, the plasma-neutral collisions dominate over electron-ion Ohmic Joule heating, and must play a vital role in the dissipation of EM wave energy in both environments. 
For a finite superposition of waves with commensurate frequencies, one can use Poynting's theorem, Equation (\ref{eq:poynting}), averaged over a wave period, to give 
\bln
\be
<\nabla \cdot \mb{S}_{p}> = <\mb{E}\cdot\mb{J}>.
\ee
\eln
Integrating over a volume of atmosphere (be it chromosphere or I/T), one can use the integral divergence theorem to express the rate at which electromagnetic 
energy is converted into both thermal energy and bulk flow kinetic energy in the volume as  
\bln
\be
Q \equiv \int_{V}{<\mb{E}\cdot\mb{J}>dV} = \int_{S}{<\mb{S}_{p}\cdot\hat{\mb{n}}>}
\ee
\eln
i.e., the surface integral of the Poynting vector normal to the surface of the volume. Thus the convergence of Poynting flux into the volume can tell us something about the frictional heating rate inside the volume. Examples of this have been performed in polar regions of the Earth's atmosphere \cite[e.g.][]{Kelley_1991,2004JASTP..66..807T} and for simple models of the Sun's chromosphere \citep{Leake_2005, Goodman_2011, Song_2011, Tu_2013}.

Looking at the volume integral of the thermal energy equation (\ref{eq:total_int_energy0}), one can see other terms other than frictional heating which add to the thermal energy in the volume V. The terms represent processes such as the thermal flux through the boundary S of V due to work done by compressive and viscous forces acting at S on the fluid in V. This is in addition to the dissipative  mechanisms which are resistive, viscous and compressive, such as the shock absorption of magneto-acoustic waves that are created by conversion of Alfv\'{e}n waves, and viscous dissipation of non-linearly interacting Alfv\'{e}n waves, are also important in the chromosphere, if not for the I/T \citep{1990SSRv...54..377N}. The various forcing and dissipation mechanisms and the scales on which they operate present a significant challenge for the use of convergence of the Poynting flux to understand the heating of the two environments. In addition there are exchange processes for other quantities, such as particle kinetic energy (from the magnetosphere into the I/T and from flares in the corona into the chromosphere), which complicate matters further. However, even if plasma-neutral collisions are not a main contributor to the heating in the chromosphere and I/T, they significantly affect the propagation of waves {\color{black} \cite[e.g.][]{Song_2005,2011A&A...529A..82Z}, and thus must be included in theoretical and numerical investigations into wave heating.}}

\section{Rayleigh-Taylor Instability}
\label{rayleightaylor}

\subsection{Occurrence in the Sun's and Earth's atmospheres}
\label{rayleightaylor_occurrence}

The unsteady transfer of energy and mass in the dynamic atmospheres of both the
Earth and the Sun often creates configurations in
which dense matter overlies tenuous matter.  In the presence of
gravity, such configurations can be unstable to disturbances
that exchange the overlying heavy fluid with the light fluid below.
This evolution produces falling spikes of the former and rising
bubbles of the latter, and the atmosphere evolves toward a state of
lower gravitational potential energy.  The linear stability of such
unstable hydrostatic equilibria has been examined both for broadly
distributed layers of continuously upward-increasing mass density, in
which the characteristic wavelength of the disturbances is comparable
to or smaller than the layer thickness \citep{Rayleigh_1882}, 
and for very narrow layers of effectively discontinuous upward jumps 
in mass density, in which the characteristic wavelength of the 
disturbances is much larger than the thickness of the layer 
\citep{Taylor_1950}.

At Earth, a continuously distributed negative gradient in density of charged plasma 
is created by the nonuniform production and loss of ions.  At high
altitudes, relatively few atoms are available to be ionized by the
absorption of EUV energy from the Sun, so low-density plasma is
created; at low altitudes, more ions are created because of the higher
neutral density, but rapid recombination with electrons again yields
low-density plasma.  The result is a plasma density that peaks at an
intermediate height of about 300 km (see Figure \ref{fig:densities}), and 
at altitudes below that it 
 can be unstable to the Rayleigh-Taylor instability.  Typically this occurs after 
sunset.  Disturbances generate large-scale density depletions in the lower 
I/T that can rise to high altitudes (over 1000 km). Severe disruptions 
in the radio transmission characteristics of Earth's atmosphere are one 
consequence of onset of this  phenomenon, commonly known as ``spread-F'' \citep[see 
review by][]{2009AnGeo..27.1915W}.  Additional fine structure is induced 
at the sides of the rising bubbles by non-uniformities in the winds of the 
coexisting neutral component of the atmosphere.

On the Sun, negative density gradients can be produced by 
 the plasma pressure deficit present in regions of strong magnetic field,
 which compensates for the magnetic pressure enhancement
there and maintains an overall balance of the total pressure force
across the magnetic region. The magnetic Rayleigh-Taylor instability creates
filamentary structures in newly emerged flux regions that are long
parallel to the field lines (to avoid bending them) and short
perpendicular to the field (to grow as fast as possible); see 
\citet{2005Natur.434..478I} and \citet{2006PASJ...58..423I} for these results 
and applications to arch filament systems on the Sun, and \citet{2007ApJ...666..541A} for 
an application to the dynamic emergence of a new active region.  The Rayleigh-Taylor 
instability also has been invoked in observations and modeling of 
hedgerow solar prominences \citep{2010ApJ...716.1288B,2011ApJ...736L...1H,2012ApJ...746..120H,2012ApJ...756..110H}.
 Prominences \citep[e.g.,][]{1995ASSL..199.....T} are large-scale clouds of 
relatively cold and dense material, consisting mostly of neutral hydrogen atoms, 
suspended on magnetic field lines within the surrounding hot and tenuous coronal 
plasma, comprised primarily of fully ionized hydrogen.
In these observations and models of hedgerow solar prominences, bubbles of low-density, 
high-temperature coronal plasma 
well up from below and intrude into the high-density, low-temperature 
body of the prominence above.

\begin{figure}[h]
  \begin{center}
    \includegraphics[width=1.8in]{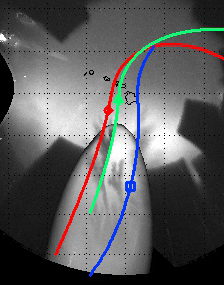}
    \includegraphics[width=2.3in]{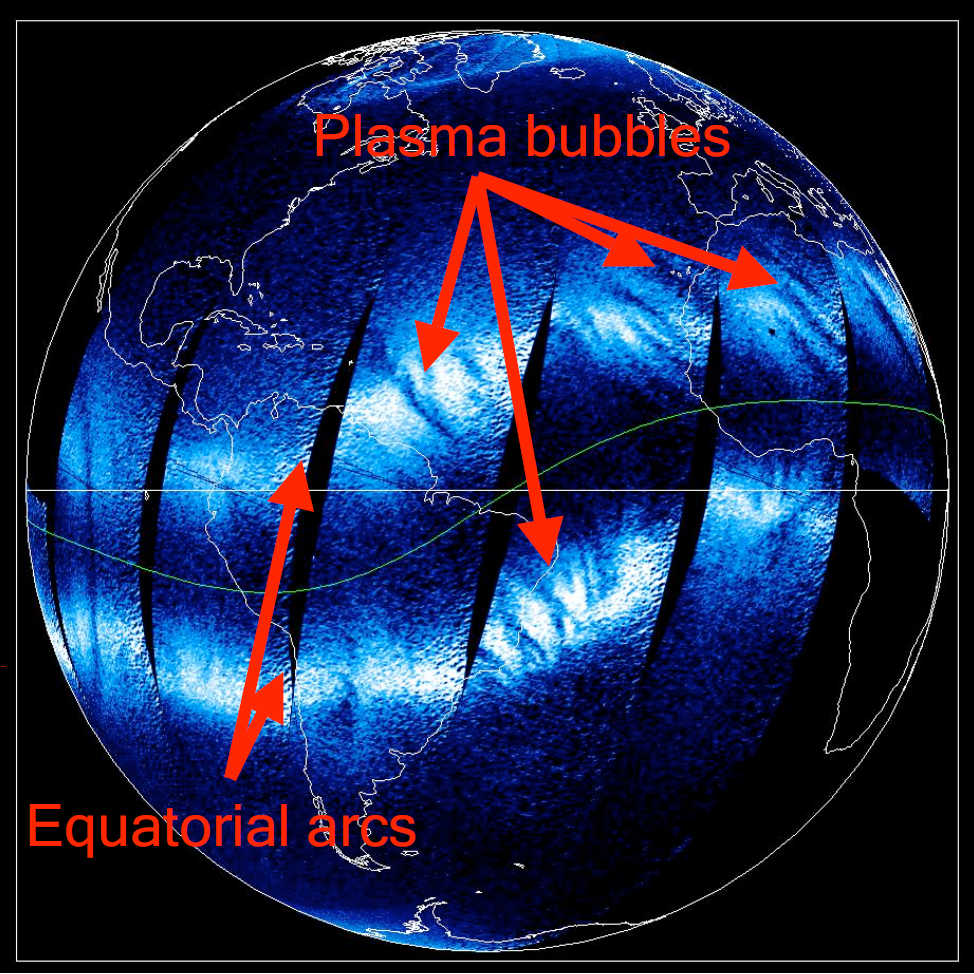}
    \caption{Left: Rayleigh-Taylor instabilities in the heliosphere.
             {\color{black} Airglow measurements showing evidence of the Rayleigh-Taylor instability in the terrestrial I/T. The dark shapes in the gray color-scale are expanding bubbles of plasma due to the instability. Colored tracks (red/green/blue) are paths of Global Positioning System (GPS) satellites. Figure
             is reprinted courtesy of J.\ Makela and \textit{Annales 
             Geophysicae}.} Right: Global 135.6 nm brightness map indicating plasma depletions. Figure is reprinted courtesy of F.\ Kamalabadi and \textit{Annales Geophysicae}
    \label{fig:rayleightaylor1}     }  
  \end{center}
\end{figure}

\begin{figure}[h]
  \begin{center}
  	\includegraphics[width=2.4in]{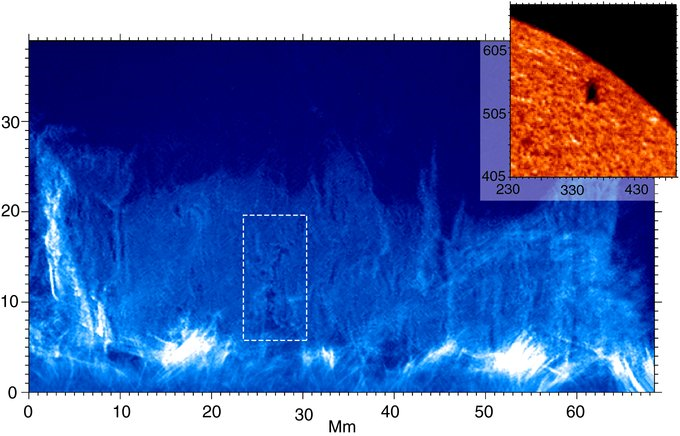}
    \includegraphics[width=2.2in]{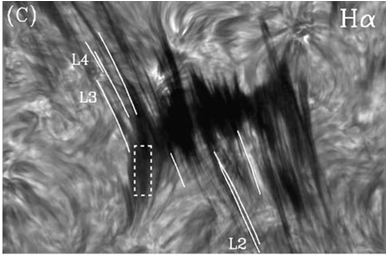}
    \caption{Left: Magneto-convection in a solar 
             hedgerow prominence (blue/white color shading, right panel). Figure
             and movies are reprinted courtesy of T.\ E.\ Berger and the \textit{Astrophysical 
             Journal}. 
             Right: Filamentary structure in a quiescent solar 
             prominence, a possible consequence of Rayleigh-Taylor 
             instability in the corona. Figure is reprinted courtesy 
             of Y. Lin and \textit{Solar Physics}.
    \label{fig:rayleightaylor2}   }    
  \end{center}
\end{figure}

The qualitative similarities in these phenomena at the Earth and Sun
are illustrated by the images in Figures \ref{fig:rayleightaylor1} and \ref{fig:rayleightaylor2}.
The left panel in Figure \ref{fig:rayleightaylor1} shows data from airglow measurements made at Haleakala, 
Hawaii during the interval 29-30 September 2002 of spread-F bubbles 
in the I/T \citep{2004AnGeo..22.3109M}.  Colored tracks (red/green/blue) 
are paths of Global Positioning System (GPS) satellites.  Notice the intricate 
fingering of the dark bubbles, which are regions of very low electron density. 
This image is taken from a movie (Sep29\_30\_GPS\_7774.mov) that illustrates more 
completely the dynamics of these features.  The left panel in Figure \ref{fig:rayleightaylor2} shows 
data from the {\it Hinode} Solar Optical Telescope on 30 November 2006 of a solar 
hedgerow prominence at the solar limb \citep[][Figure 1]{2010ApJ...716.1288B}.  
The dashed white box outlines a region of a dark bubble forming, rising, fingering, 
and fragmenting among the shimmering strands of the prominence gas. Note that this 
measured emission originates in the Balmer series of neutral hydrogen atoms, marking 
cool material; the dark bubble is hot material in which the hydrogen is fully ionized 
and so does not radiate in this line \citep{2011Natur.472..197B}. The inset image 
at the upper right shows the prominence as a dark feature on the otherwise bright solar disk three days earlier. The 
{\it Hinode} image is part of a movie \citep[][ApJ330604F1.mov]{2010ApJ...716.1288B} 
that shows the full evolution of this bubble and many others.  Both the terrestrial 
and solar movies are included in this paper as supplementary material.

Another apparent consequence of Rayleigh-Taylor instability in solar prominences is the 
organization of the cool material into long, thin threads that are extended along magnetic 
field lines.  A striking example of this structure \citep[Figure 1c]{2007SoPh..246...65L} 
is shown in Figure \ref{fig:rayleightaylor2}, as observed at high resolution with the 
Swedish Solar Telescope on 22 August 2004. The measured thread diameter is comparable 
to the resolution limit ($\sim$ 200 km) of the telescope's adaptive optics, while the 
thread lengths range up to at least 10 Mm. The excellent thermal insulation across the 
prominence magnetic field implies that the cross-thread scale length separating the cool 
(10$^4$ K) prominence and hot (10$^6$ K) corona is very small ($\sim$ 10 km or less) 
compared to the thread diameter. 
During the onset of the instability, the horizontal 
magnetic field exerts a stabilizing force that favors long wavelengths parallel to the 
field, while the motion feeding the growth favors short wavelengths perpendicular to 
the field \citep{1961hhs..book.....C}. This implies a predisposition toward long, thin, 
field-aligned threads, like those observed. Similar features have been found in simulations 
of magnetized Rayleigh-Taylor instability in astrophysical contexts 
\citep[e.g.,][]{2007PhFl...19i4104S,2007ApJ...671.1726S}. The implications of the 
instability for understanding the observed thread structure in solar prominences 
are now being investigated \citep[e.g.,][]{2013SPD....44...41D}.

Images of the Rayleigh-Taylor instability at Earth similarly show elongation along the
magnetic field. In this case it is the plasma density depletions that appear as dark features in
EUV images \cite[see Figure \ref{fig:rayleightaylor1} and][]{Kamalabadi_2009} such as produced by the Global
Ultraviolet Imager (GUVI) on the NASA TIMED satellite \citep{Christensen_2003}. The elongation
occurs because, in the I/T, the instability is governed by an electrostatic potential that varies
across but not along the geomagnetic field \citep{Haerendel_1992}. This potential produces ExB
drifts that act everywhere along a given geomagnetic flux tube, moving low-density plasma
upward and outward, to higher altitudes and latitudes. Ion dynamics within these field-aligned
plasma density bubbles has only recently been simulated \citep{Huba_2009a, Huba_2009b, 
Krall_2010}. 

\subsection{Stability Analysis}
\label{rayleightaylor_growth}

The occurrence of Rayleigh-Taylor instabilities is important in both the 
chromospheric and ionospheric contexts and has a common physical origin 
in the negative density stratification of a fluid in the presence of gravity, 
as discussed above. However, the mathematical manipulations performed and 
the language used to describe the underlying physics are quite different 
in the two communities of investigators. In the following subsections, 
we present simple derivations of the dispersion relation for unstable 
Rayleigh-Taylor disturbances within the contexts of chromospheric 
magnetohydrodynamics (\S \ref{rayleightaylor_growth_bv}) and ionospheric 
electrodynamics (\S \ref{rayleightaylor_growth_ej}). 
The first uses $\mb{B}$ and $\mb{V}$ as primary
variables, while the second uses $\mb{E}$ and $\mb{J}$.
In the end, however, the approaches are complementary, 
and the same dispersion relation is obtained irrespective of how the analysis 
is framed. In the concluding subsection (\S \ref{rayleightaylor_growth_limit}), we
simplify the general dispersion equation for regimes in which the plasma
is strongly coupled to the neutrals by collisions. This is the case in
the solar chromosphere and prominences, and in the I/T. This example illustrates an important physical process that is 
common to the two environments, but is conceptualized and analyzed in very 
different ways, therefore, in our experience, impeding mutual understanding 
across disciplines.


\subsubsection {Magnetohydrodynamic (Chromospheric) Context}
\label{rayleightaylor_growth_bv}
The basic equations used in the analysis are those for continuity 
and momentum of the neutral gas and the one-fluid plasma. The plasma 
equations are the mass-weighted sum of the ion and electron continuity 
equations (\ref{eq:cont_i} and \ref{eq:cont_e}) and the sum of the 
momentum equations for the ions and electrons (\ref{eq:mom_i} and 
\ref{eq:mom_e}). Ionization and recombination terms are ignored, as 
are electron inertia effects; in particular, we assume that $m_e 
\nu_{en} \ll m_i \nu_{in}$ (see \S \ref{current_diss}) . Consider a simple neutral-plasma configuration 
in a slab geometry, ${\bf B} = B_0 \bf{\hat{e}}_z$, ${\bf g} = g \bf{\hat{e}}_x$,
${\bf V}_{n0} = {\bf V}_{p0} = 0$, $\rho_n (x) = m_n n_n(x)$, and $\rho_p(x) = 
m_p n_p(x)$, where the subscript 0 indicates an equilibrium value. Perturbed 
variables $\tilde q$ are assumed to vary as $\tilde q \exp (iky-i\omega t)$, 
where $\omega = \omega_r + i\gamma$ and the disturbances propagate in the 
$y$ direction. The neutral and plasma flows are assumed to be incompressible, 
$\nabla \cdot {\bf V} = 0.$ For the uniform equilibrium magnetic field assumed, 
the ideal MHD induction equation, Equation (\ref{EqNum16}) with $\eta = 0$, then 
yields ${\bf \tilde B} = 0$, so the magnetic field remains undisturbed. Since 
we neglect the $x$ dependence of the perturbed quantities, incompressibility 
requires $\tilde V_{ny} = \tilde V_{py} = 0$. Physically, the time scale 
associated with the instability is required to be much longer than those 
associated with compressive acoustic or magnetosonic waves.

The linearized continuity equation (\ref{eq:cont_n}) and the $x$ component 
of the linearized momentum equation (\ref{eq:mom_n}) for the neutrals then 
take the forms 
\bln
\bea
  -i \omega \tilde n_n 
  & = & - \tilde V_{nx} \frac{d n_{n0}}{d x},
  \label{eq5c} \\
  - i \omega n_{n0} m_n \tilde V_{nx}
  & = & + \tilde n_n m_n g - n_{n0} m_n \nu_{ni} \left( \tilde V_{nx}-\tilde V_{px} \right).
   \label{eq6c}
\eea
\eln
The corresponding equations for the plasma, from (\ref{eq:cont_i}--\ref{eq:cont_e}) 
and (\ref{eq:mom_i}--\ref{eq:mom_e}), are 
\bln
\bea
  - i \omega \tilde n_p  
  & = & - \tilde V_{px} \frac{d n_{p0}}{d x},
  \label{eq7c} \\
  - i \omega n_{p0} m_p \tilde V_{px}
  & = & + \tilde n_p m_p g - n_{p0} m_p \nu_{in} \left( \tilde V_{px}-\tilde V_{nx} \right).
   \label{eq8c}
\eea
\eln
Solving for $\tilde n_n$ and $\tilde n_p$ in Equations (\ref{eq5c}) and
(\ref{eq7c}) and substituting into Equations (\ref{eq6c}) and 
(\ref{eq8c}), respectively, yields, after some rearrangement,
\bln
\bea
  \left( \omega^2 + i \omega \nu_{ni} - \frac{g}{L_{n}} \right) \tilde V_{nx} & = & 
  i \omega \nu_{ni} \tilde V_{px},
  \label{eq9c} \\
  \left( \omega^2 + i \omega \nu_{in} - \frac{g}{L_{p}} \right) \tilde V_{px} & = & 
  i \omega \nu_{in} \tilde V_{nx},
  \label{eq10c}
\eea
\eln
where $L_{n}^{-1} = d\ln n_{n0}/dx$ and $L_{p}^{-1} = d\ln n_{p0}/dx$ are the local 
neutral and plasma density scale heights.
Combining Equations (\ref{eq9c}) and (\ref{eq10c}), we arrive at the
dispersion equation 
\bln
\begin{equation}
  \left( \omega^2 + i \omega \nu_{ni} - \frac{g}{L_{n}} \right)
  \left( \omega^2 + i \omega \nu_{in} - \frac{g}{L_{p}} \right)
  + \omega^2 \nu_{ni} \nu_{in} = 0,
  \label{eq11c}
\end{equation}
\eln
which can be expanded to read 
\bln
\begin{equation}
  \omega^4 +i \omega^3 \left( \nu_{in} + \nu_{ni} \right) - 
  \omega^2 \left( \frac{g}{L_{p}} + \frac{g}{L_{n}} \right) -
  i \omega \left( \frac{\nu_{ni}g}{L_{p}} + \frac{\nu_{in}g}{L_{n}} \right) +
  \frac{g^2}{L_{n}L_{p}} = 0.
  \label{eq12c}
\end{equation}
\eln
Rewritten in terms of the Brunt-V\"ais\"al\"a frequencies $N_n$ and $N_p$, Equation 
(\ref{EqNum2}), the dispersion relation (\ref{eq12c}) becomes 
\bln
\begin{equation}
  \omega^4 +i \omega^3 \left( \nu_{in} + \nu_{ni} \right) - 
  \omega^2 \left( N_p^2 + N_n^2 \right) -
  i \omega \left( \nu_{ni} N_p^2 + \nu_{in} N_n^2 \right) +
  N_n^2 N_p^2 = 0.
  \label{eq13c}
\end{equation}
\eln
{\color{black} Approximate solutions to this equation in the strong-coupling limit reveal the instability growth rates, as discussed in \S
 \ref{rayleightaylor_growth_limit} below.}

\subsubsection {Electrostatic (Ionospheric) Context}
\label{rayleightaylor_growth_ej}

{\color{black} The set of three-fluid equations used to analyze the Rayleigh-Taylor instability in the F layer of 
the EarthÕs I/T consists of those for electron continuity} and current conservation, electron and ion momentum, and neutral 
continuity and momentum \citep{ossakow81}. The equilibrium state has $n_{e0}(x) 
= n_{i0}(x)$, ${\bf E}_0=0$, ${\bf B} = B_0 \bf{\hat{e}}_z$, and ${\bf V}_{e0} 
= {\bf V}_{i0}= {\bf V}_{n0} = 0.$ The equilibrium is perturbed such that 
all disturbances are proportional to $\exp(iky-i\omega t).$ We invoke the 
local approximation that $\partial / \partial x \ll k$ with regard to the 
perturbed variables. In conjunction with Faraday's law, this implies that 
$\tilde E_x = 0$ in order for ${\bf \tilde B} = 0$ to be maintained.

Neglecting inertia, gravity, and collisional coupling due to the small 
electron mass, the perturbed electron velocity from the linearized Equation 
(\ref{eq:mom_e}) is simply the ${\bf E} \times {\bf B}$ drift due to the 
instability, 
\bln
\begin{equation}
  \tilde V_{ex} = \frac{\tilde E_y}{B_0},\quad \tilde V_{ey} = 0.
  \label{eq5a}
\end{equation}
\eln
Applying the local approximation to the current conservation constraint, Equation 
(\ref{eq:divj}), we find that it must be the case that $\tilde J_y = 0$. Hence, 
\bln
\begin{equation}
  \tilde V_{iy} = \tilde V_{ey} = 0,
  \label{eq6a}
\end{equation}
\eln
where the second equality follows from Equation (\ref{eq5a}). In the $y$ component of 
the ion momentum equation (\ref{eq:mom_i}), only the drag term due to collisions 
with neutrals now remains.  Thus, we also must have that 
\bln
\begin{equation}
  \tilde V_{ny} = \tilde V_{iy} = 0.
  \label{eq7a}
\end{equation}
\eln
The three conditions (\ref{eq5a}--\ref{eq7a}) imply that $\nabla \cdot {\bf V}_e 
= \nabla \cdot {\bf V}_i = \nabla \cdot {\bf V}_n = 0$, i.e., the flows of all 
three fluids are incompressible in this approximation. {\color{black} This demonstration justifies 
{\it a posteriori} the incompressibility assumption that we made at the outset of \S 
\ref{rayleightaylor_growth_bv}.}

A second deduction that follows from the current conservation constraint and 
$\tilde J_y = 0$ is that $\tilde J_x$ must be uniform along $x$,
\bln
\begin{equation}
  \frac{\partial \tilde J_x}{\partial x} = \left( \tilde V_{ix} - \tilde V_{ex} \right) e \frac{d n_e}{d x} = 0.
  \label{eq8a}
\end{equation}
\eln
Consequently, we also must have that 
\bln
\begin{equation}
  \tilde V_{ix} = \tilde V_{ex} = \frac{\tilde E_y}{B_0}.
  \label{eq9a}
\end{equation}
\eln
The bulk flows of the electrons and ions due to the instability are, therefore, 
identical, ${\bf \tilde V}_e = {\bf \tilde V}_i = {\bf \tilde E} \times {\bf B}$. {\color{black} This is consistent with a 
vanishing electric field in the frame of the electrons, ${\bf \tilde E} + {\bf \tilde V_e} \times {\bf B_0} = 0,$ 
which was tacitly assumed in the magnetohydrodynamic analysis in \S \ref{rayleightaylor_growth_bv}, where ${\bf \tilde E} $
plays no explicit role.  In addition, the total perturbed current vanishes, ${\bf \tilde J} = 0$, consistent with 
${\bf \tilde B}$ = 0 and with the earlier analysis.}

From the linearized electron continuity equation (\ref{eq:cont_e}), the perturbed 
electron density satisfies 
{\color{black} 
\bln
\be
  -i \omega \tilde n_e =  \frac{d n_{e0}}{d x} \tilde V_{ex} =  - \frac{d n_{i0}}{d x} \tilde V_{ix} = i \omega \tilde n_i,
  \label{eq10a}
\ee
\eln}
after using first $n_{i0} = n_{e0}$ and $\tilde V_{ix} = \tilde V_{ex}$, and then 
the ion continuity equation (\ref{eq:cont_i}). Substituting the definition $L_p 
= d \ln n_{i0} / d x = d \ln n_{e0} / d x$, and the ${\bf E} \times {\bf B}$ drift 
velocity from Equation (\ref{eq9a}), we obtain the (equal) electron and ion density 
perturbations 
{\color{black}
\bln
\begin{equation}
  \frac{\tilde n_e}{n_{e0}} = \frac{\tilde n_i}{n_{i0}} = \frac{i}{\omega} \frac{1}{L_p} {\tilde V}_{ex} = 
  \frac{i}{\omega} \frac{1}{L_p} {\tilde V}_{ix} = - \frac{i}{\omega} 
  \frac{1}{L_p} \frac{\tilde E_y}{B_0}.
  \label{eq11a}
\end{equation}
\eln
}
{\color{black} Equations (\ref{eq10a}) and (\ref{eq11a}) are equivalent to Equation (\ref{eq7c}) 
in \S \ref{rayleightaylor_growth_ej}, since $n_e = n_i = n_p/2$ and $\bf{V}_e = \bf{V}_i = \bf{V}_p$.}

Finally, after recalling that $\tilde E_x = 0$ and $\tilde V_{iy} = 0$, the $x$ component 
of the linearized ion momentum equation (\ref{eq:mom_i}) yields
\bln
\begin{eqnarray}
  \nu_{in} \tilde V_{nx} &=& \left( - i \omega + \nu_{in} \right) \tilde V_{ix} 
  - g \frac{\tilde n_i}{n_{i0}}\\
                         &=& \left( - i \omega + \nu_{in} + \frac{i}{\omega} 
  \frac{g}{L_p} \right) \tilde V_{ix}\\
                         &=& \left( - i \omega + \nu_{in} + \frac{i}{\omega} 
  \frac{g}{L_p} \right) \frac{\tilde E_y}{B_0}.
  \label{eq12a}
\end{eqnarray}
\eln
The linearized continuity and momentum equations for the neutrals are combined as in 
\S \ref{rayleightaylor_growth_bv} to obtain (cf.\ Equation \ref{eq9c}) 
\bln
\begin{equation}
  \left( -i \omega + \nu_{ni} + \frac{i}{\omega} \frac{g}{L_{n}} \right) \tilde V_{nx} 
  = \nu_{ni} \tilde V_{ix}
  = \nu_{ni} \frac{\tilde E_y}{B_0}.
  \label{eq13a}
\end{equation}
\eln
Eliminating $\tilde V_{nx}$ from Equations (\ref{eq12a}) and (\ref{eq13a}) leads to the desired 
dispersion equation,
\bln
\begin{equation}
  \left( \omega^2 + i \omega \nu_{ni} - \frac{g}{L_{n}} \right)
  \left( \omega^2 + i \omega \nu_{in} - \frac{g}{L_{p}} \right)
  + \omega^2 \nu_{ni} \nu_{in} = 0.
  \label{eq14a}
\end{equation}
\eln
This is identical to the magnetohydrodynamic result, Equation (\ref{eq11c}), {\color{black} and will be analyzed further in the next section.}

{\color{black} 
Our derivations of the dispersion equation in the electrodynamic and magnetohydrodynamic contexts culminate 
in the same result, albeit by following different paths.  The principal difference is the role assigned to the electric 
field, which is primary in the ionospheric context and in the literature of the I/T community, but is all but invisible
 in the chromospheric context and in much of the literature of the solar community (excepting situations of rapidly 
 changing magnetic fields associated with flares and other transient behavior).  As the derivations highlight, it is
  commonly said about the I/T that the electric field gives rise to $\bf{E} \times \bf{B}$ drifts, implying that $\bf{E} $
  drives $\bf{V}$.  In contrast, the same is rarely, if ever, said about the chromosphere or corona, where $\bf{E}$ is 
  principally a consequence of plasma flow $\bf{V}$ across the magnetic field $\bf{B}$, so $\bf{V}$ drives $\bf{E}$.  
  The normal mode analysis performed here does not distinguish between these two perspectives; only an 
  initial value analysis can do that.  
 }

For completeness, we note that this analysis of the Rayleigh-Taylor instability in the ionospheric 
context is predicated on a local evaluation of plasma and neutral variables, i.e., it applies in a 
restricted region in space. In this limit, as we have seen, the dispersion equation is the same 
as in the solar context. However, in the I/T, as discussed in \S \ref{maxwell_section}, the 
magnetic field lines are assumed to be equipotentials and the electric field generated by the 
instability extends along the entire magnetic flux tube. Thus, the instability is affected by 
the plasma on the flux tube that encompasses both the E and F layers of the I/T and a ``flux-tube 
integrated'' analysis of the instability is required. A discussion of this type of analysis as it 
applies to Rayleigh-Taylor instability in the I/T is given by \citet{sultan96}. {\color{black}  
 The result is a growth rate averaged all along an equipotential field line, which in order of magnitude 
has a growth time of about 15 min.  In contrast, the Alfv\'{e}n travel time is only about 10 s.  
Thus, as argued by \citet{Vas_2012} (see our \S \ref{neutral_wind_dynamos}), the non-potential component of the electric field is 
small, and the electrostatic approximation is quite good.  A similar conclusion follows from a more complete 
electromagnetic analysis of the Rayleigh-Taylor instability, including Alfv\'{e}n waves and the Pedersen resistivity of the 
I/T, performed by \citet{Basu_2005}.  The result found is that the magnetic field perturbations are very small 
due to resistive slip between the plasma and magnetic field (see our \S \ref{frozen_slip}), so that, again, the electrostatic  approximation is well justified.}

\subsubsection {Strong-coupling Limit}
\label{rayleightaylor_growth_limit}

The general dispersion equation (\ref{eq11c}) or (\ref{eq14a}) trivially factors 
into distinct neutral and plasma modes in the limit of weak collisions, $\nu_{\alpha\beta}
\rightarrow 0$:
\bln
\begin{equation}
  \omega^2 = N_n^2 = \frac{g}{L_n};\quad \omega^2 = N_p^2 = \frac{g}{L_p}.
  \label{EqNumW}
\end{equation}
\eln
{\color{black} In \S \ref{numbers}, Figure \ref{fig:incollisions}, on the other hand, we noted that the characteristic 
Brunt-V\"ais\"al\"a frequencies $N \equiv (g/L)^{1/2}$ in the solar and 
terrestrial atmospheres are much smaller than the ion-neutral collision 
frequencies $\nu_{in}$. We can scale these quantities by setting $\nu_{in} = \mathcal{O}(1)$, and $N_{p},N_{n} = \mathcal{O}(\epsilon)$
where $\epsilon \ll 1$. We also have $\nu_{ni} = \mathcal{O}(\epsilon)$ in a weakly ionized plasma.
We can now look for strongly-coupled solutions to the  
general quartic dispersion relation given in Equation (\ref{eq13c}), i.e.
 $\omega_{r} \ll \nu_{in}$ where $\omega=\omega_{r}+i\gamma$. Note that this regime justifies the use
 of the low frequency Ohm's law, where $|\omega| \ll \min{(\nu_{in},\Omega_{i})}$ is required.
Let us first consider high frequency solutions $\omega_{r} \gg Np,Nn$, of which 
$\omega_{r}=0, ~ \gamma = \mathcal{O}(1)$ is one. 
Balancing the largest terms in the 
dispersion relation (\ref{eq13c}) for this scaling, the quartic and cubic terms, yields 
Equation (\ref{EqNum3}) below. Next we consider intermediate solutions, where $|\omega| \sim N_{p},N_{n}$, 
of which $\omega_{r} = \mathcal{O}(\epsilon), ~ \gamma = \mathcal{O}(\epsilon)$ is one. Then the largest terms
in the dispersion relation (\ref{eq13c}) are the cubic and linear terms , and balancing these yields
Equation (\ref{EqNum4}) below. Finally we consider low-frequency solutions, where $|\omega| \ll N_{p},N_{n}$,
of which $\omega_{r} = 0, ~ \gamma = \mathcal{O}(\epsilon^2)$ is one. Then the largest terms in 
the dispersion relation (\ref{eq13c}) are the linear and constant terms, and balancing these yields
Equation (\ref{EqNum5}) below:}
\bln
\begin{eqnarray}
  \omega \approx -i \left( \nu_{in} + \nu_{ni} \right);
  \label{EqNum3}\\
  \omega^2 \approx \left( \nu_{in} N_n^2 + \nu_{ni} N_p^2 \right) / \left(\nu_{in} + \nu_{ni} \right);
  \label{EqNum4}\\
  \omega \approx -i N_n^2 N_p^2 / \left( \nu_{in} N_n^2 + \nu_{ni} N_p^2 \right).
  \label{EqNum5}
\end{eqnarray}
\eln

The high-frequency solution in Equation (\ref{EqNum3}) is a strongly damped 
inter-penetrating mode in which the two fluid velocities are 180$^\circ$ 
out of phase. It is always stable.

The pair of intermediate-frequency solutions in Equation (\ref{EqNum4})
are a weighted average of the classical single-fluid Rayleigh-Taylor 
growth rates (or Brunt-V\"ais\"al\"a frequencies) of the neutral gas and 
plasma. Due to the preponderance of neutrals over ions (i.e., $n_{n} \gg n_{i}$, hence $\nu_{in} 
\gg \nu_{ni}$; see Fig.\ \ref{fig:incollisions}) these solutions simplify to 
\bln
\begin{equation}
  \omega^2 \approx N_n^2 = \frac{g}{L_n}.
  \label{EqNum6}
\end{equation}
\eln
They are oscillatory in the chromosphere and the I/T where the
neutrals are stably stratified, $g L_n > 0$. On the other hand, in a 
cool, dense, partially neutral prominence suspended within the ionized 
solar corona, $g L_n < 0$, and one of these modes is a growing 
Rayleigh-Taylor instability. A two-fluid (electrically neutral ionized 
plasma plus neutral gas) numerical simulation of such an unstable 
configuration is presented below in \S \ref{rayleightaylor_sims_prom}. 
Due to the strong collisional coupling of the neutral gas and plasma, 
their velocities are essentially equal,
\bln
\begin{equation}
  {\bf V}_n \approx {\bf V}_p.
  \label{EqNum7}
\end{equation}
\eln
This can be deduced readily from Equations (\ref{eq9c}) and (\ref{eq10c}).

A second potentially unstable solution is the low-frequency mode in
Equation (\ref{EqNum5}). It also is a weighted average of the plasma and
neutral contributions, and as $\nu_{in} \gg \nu_{ni}$,  simplifies to
\bln
\begin{equation}
  \omega \approx -i \frac{N_p^2}{\nu_{in}} = -i \frac{g}{\nu_{in} L_p}.
  \label{EqNum8}
\end{equation}
\eln
This is the classical collision-dominated Rayleigh-Taylor instability in 
the I/T \citep[e.g.,][]{ossakow81}, driven by the upward-increasing 
plasma density, $g L_p < 0$. A similar instability should occur in the ALC7 
solar chromosphere where the ionization fraction increases sufficiently 
rapidly with height above the surface (cf.\ Fig.\ \ref{fig:densities}). A 
multi-fluid simulation of this low-frequency mode is presented below in \S 
\ref{rayleightaylor_sims_iono}. In this case, due to the approximate balance 
between gravity and collisional coupling to neutrals on the part of the ions, Equation (\ref{eq10c}), 
and to the very low frequency and the weak collisional 
coupling to ions on the part of the neutrals, Equation (\ref{eq9c}), the 
velocities satisfy
\bln
\begin{equation}
  \left\vert {\bf V}_p \right\vert \gg \left\vert {\bf V}_n \right\vert.
  \label{EqNum9}
\end{equation}
\eln
All of these features represented by Equations (\ref{EqNum6}--\ref{EqNum9}) are evident in the 
simulation results described next.

\subsection{Simulation Study}
\label{rayleightaylor_sims}

We now present a simulation study applicable to both the solar and 
ionospheric case performed within the HiFi spectral-element multi-fluid modeling 
framework \citep{2008PhDT.........1L}. The calculations use an 
implementation of the partial differential equations describing self-consistent, 
nonlinear evolution of a partially ionized, three-fluid  hydrogen mixture of ions, electrons and neutrals                                                               
\citep[][and references therein]{Leake2013}. 
{\color{black} The set of equations solved are (\ref{eq:cont_i}-\ref{energy_eqn}), but neglecting the electron momentum, and with Equation (\ref{gol_ions}), neglecting the last two terms in the low-frequency approximation, and neglecting the plasma and neutral pressure term (third term on the RHS). The electron pressure terms is kept, because, in some cases, when there are small
variations of out-of-plane B-field from the uniform $\mathcal{O}(1)$ guide field, as
is the case for the I/T simulation, it isn't quite so clear that this term can always be ignored. Hence the Ohm's law used is:
\bln
\be
\mb{E}^{p}  \equiv \mb{E}+(\mb{V}_{p}\times\mb{B}) = \left[\frac{1}{k_{ei}}+\frac{1}{k_{en}+k_{in}}\right]\frac{B}{en}\mb{J} 
+ \left[1-\frac{\xi_{n}k_{in}}{k_{en}+k_{in}}\right]\frac{B}{en}\mb{J}\times\hat{\mb{b}}  
 - \frac{\nabla \cdot \mbb{P}_{e}}{en}
\label{gol_ions}
\ee
\eln
}
We point out that using a hydrogen fluid is not truly appropriate 
for the F-layer I/T, which is dominated by oxygen ions. However, 
the significance of our study is that we can capture the essential physics 
of both situations within the framework of a single model, using parameters 
appropriate to the two different environments.{\color{black} 
The boundary conditions are periodic on the sides, and are closed and reflecting on the top and bottom, respectively.}

In the subsections below, we describe the basic parameters and show the key 
figures for the two cases studied. Some mathematical details are relegated 
to the Appendix to streamline the presentation of the essential results.

\begin{figure}[h]
  \begin{center}
    \includegraphics[width=4in]{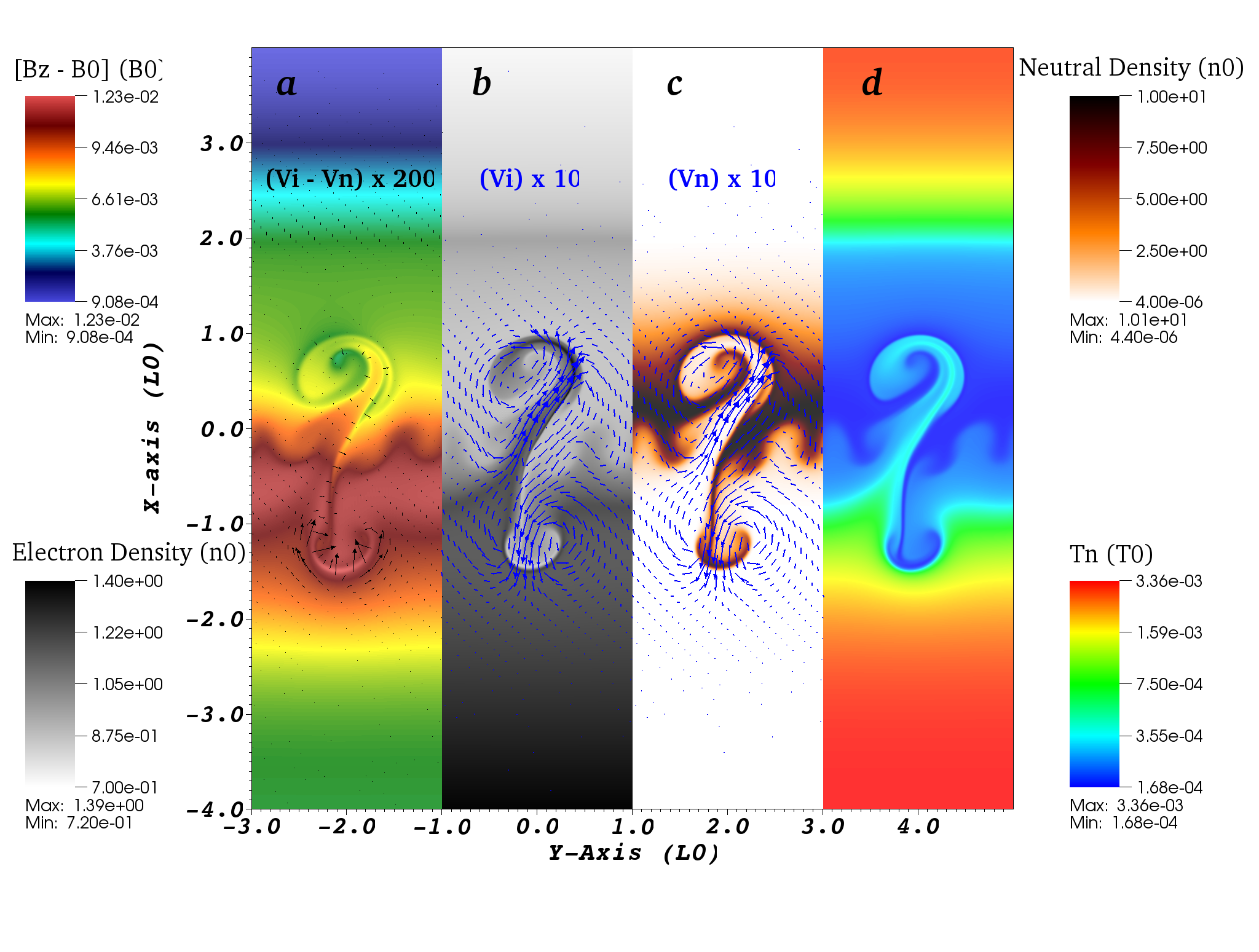}
    \vspace{-20pt}
    \caption{Two-fluid numerical simulation of Rayleigh-Taylor instability in a solar
      prominence: (a) magnetic field $[B_z - B_0]/B_0$; (b) electron or ion density $n_e/n_0 
      = n_i/n_0$; (c) neutral density $n_n/n_0$; (d) neutral temperature $T_n/T_0$. 
      Vector velocities for the neutrals (${\bf V}_n$) and ions (${\bf V}_i$), and 
      their difference (${\bf V}_i - {\bf V}_n$), are shown in panels (c), (b), 
      and (a) respectively.
    \label{fig_sp}}
  \end{center}
\end{figure}

\subsubsection{Solar Prominence}
\label{rayleightaylor_sims_prom}

Our first case is a model for the Rayleigh-Taylor instability in a solar prominence. 
We adopt a two-dimensional slab equilibrium for the prominence structure, 
showing the result of the simulation at a time when the instability is already well-developed 
in Figure \ref{fig_sp}. In the initial equilibrium, the electron (and ion) density $n_e$ of the background 
corona (Fig.\ \ref{fig_sp}b) is exponentially stratified, attaining a value 
$n_0 = 1 \times 10^{15}$ m$^{-3}$ at height $x = 0$ in the Cartesian coordinate 
system scaled to a typical prominence size, $L_0 = 1 \times 10^6$ 
m. The neutral density $n_n (x)$ within the prominence slab (Fig.\ \ref{fig_sp}c) 
is an order of magnitude larger, reaching $10 n_0 = 1 \times 10^{16}$ m$^{-3}$ 
at its central height, $x = 0.5$. The characteristic temperatures of the fluids are 
$T = 2 \times 10^5$ K in the corona and $T = 1 \times 10^4$ K in the prominence; 
the neutral temperature is shown in Figure \ref{fig_sp}d, scaled to a normalization 
temperature $T_0 = 5.8 \times 10^7$ K. The electron and ion temperatures are very 
similar, due to the fast thermal exchange between the fluids. Finally, the deviation 
of the out-of-plane magnetic field $B_z (x)$ from a uniform value $B_0 = 1 \times 
10^{-3}$ T is displayed in Figure \ref{fig_sp}a. As detailed in the Appendix, $B_z$ 
is vertically stratified to balance the initial pressure and gravity forces in 
magnetohydrostatic equilibrium. Because beta is low for both the plasma and neutral 
gas, the associated magnetic-field deviations are relatively small.

This unstable neutral-plasma system is initialized with small-amplitude neutral-density 
perturbations centered at $x = 0$, on the bottom side of the prominence slab. The resulting 
evolved solution at time $t = 4.0 \times 10^2$ s is shown in Figure \ref{fig_sp}. Velocity 
vectors are shown in panels a, b, and c that correspond to the differential ion-neutral 
velocity, ion velocity, and neutral velocity, respectively. We observe that all contours 
have the same shape, and the plasma and neutrals track each other quite closely. The neutral 
gas is unstable and, due to the collisional coupling to the plasma, the plasma follows the 
neutrals. We determined the e-folding growth time $\tau$ of the flow velocity 
and the resulting growth rate is $\tau^{-1} 
\sim 2 \times 10^{-2}$ s$^{-1}$. For solar gravity and the chosen neutral density profile, 
the Brunt-V\"ais\"al\"a frequency, which is the analytic growth rate for the strong-coupling limit, 
see Equation (\ref{EqNum6}), is $N_n \approx 2.6 \times 10^{-2}$ s$^{-1}$, which is in 
very good agreement with the numerically determined growth rate, given the simplicity of the derivation 
compared to the complexity of the simulation model. An evaluation of the collision frequencies 
at $x = 0$ gives $\nu_{in} = 2.2 \times 10^2$ s$^{-1}$ and $\nu_{ni} = 5.2 \times 10^1$ s$^{-1}$. 
Thus, this calculation lies in the strong-coupling regime of \S \ref{rayleightaylor_growth_limit}, 
and Equations (\ref{EqNum6}) and (\ref{EqNum7}) apply.

\begin{figure}[h]
  \begin{center}
    \includegraphics[width=4in]{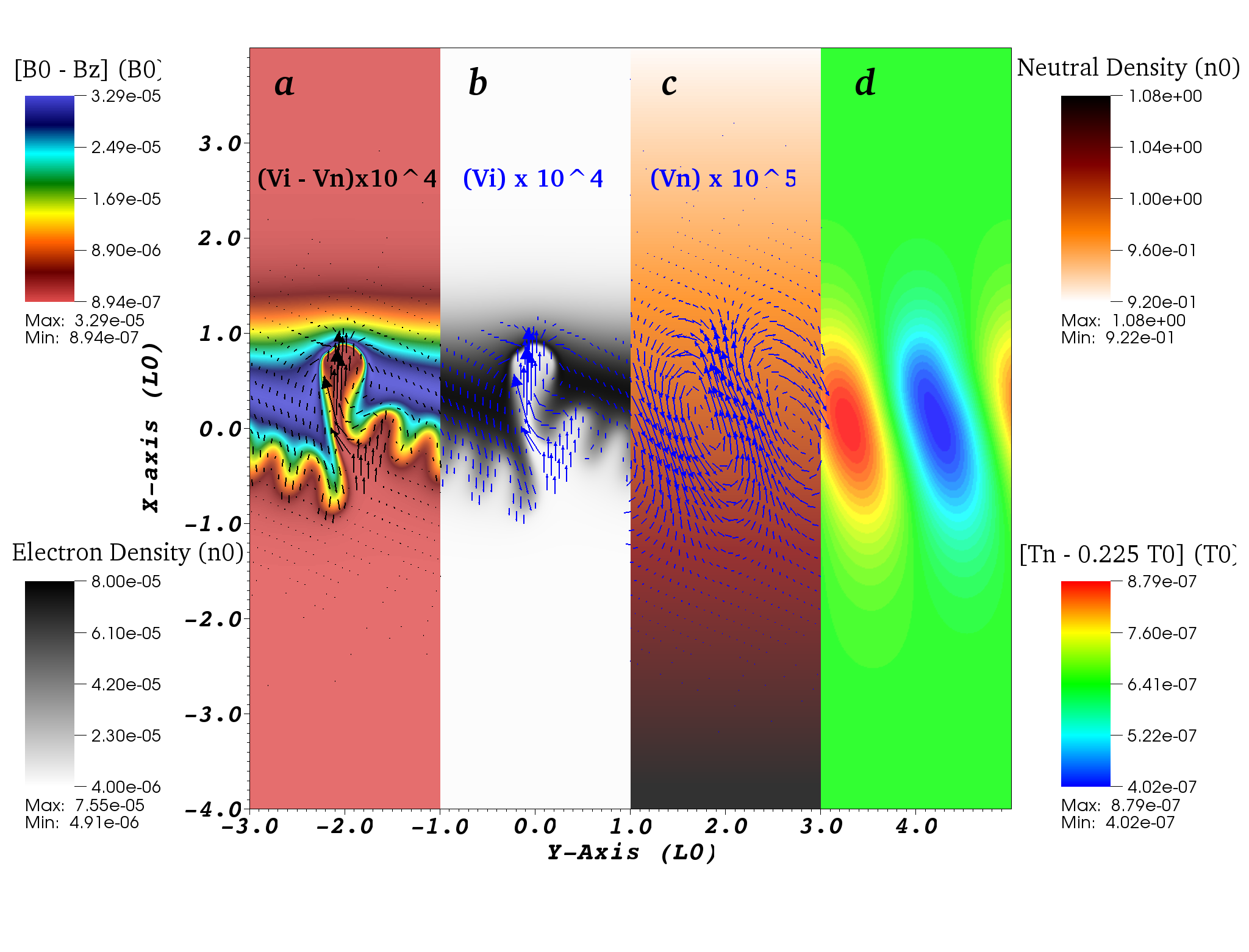}
    \vspace{-20pt}
    \caption{Three-fluid numerical simulation of Rayleigh-Taylor instability in the 
      Earth's I/T: (a) magnetic field $[B_0 - B_z]/B_0$; (b) electron or ion density 
      $n_e/n_0 = n_i/n_0$; (c) neutral density $n_n/n_0$; (d) neutral temperature 
      $[T_n - T_b]/T_0$. Vector velocities for the neutrals (${\bf V}_n$) and ions 
      (${\bf V}_i$), and their difference (${\bf V}_i - {\bf V}_n$), are shown in 
      panels (c), (b), and (a) respectively.
    \label{fig_iono}}
  \end{center}
\end{figure}

\subsubsection{Terrestrial Ionosphere}
\label{rayleightaylor_sims_iono}

Our second case is a model for the Rayleigh-Taylor instability in the I/T. 
The well-developed instability in the simulation initialized with a two-dimensional slab equilibrium is shown in Figure \ref{fig_iono}. In this 
case, the neutral density $n_n (x)$ of the background atmosphere (Figure \ref{fig_iono}c) 
is exponentially stratified, attaining a value $n_0 = 1 \times 10^{16}$ m$^{-3}$ at 
height $x = 0$ in the Cartesian coordinate system scaled to a typical F-layer size, 
$L_0 = 2 \times 10^4$ m. The electron (and ion) density $n_e (x)$ within the I/T 
(Figure \ref{fig_iono}b) is four orders of magnitude smaller, reaching $10^{-4} n_0 
= 1 \times 10^{12}$ m$^{-3}$ at its central height, $x = 0.5$. The temperatures of the 
fluids are all uniform and equal, at $T_b = 1.2 \times 10^3$ K; the deviation of the neutral 
temperature from this value is shown in Figure \ref{fig_iono}d, scaled to a normalization 
temperature $T_0 = 5.3 \times 10^3$ K. Finally, the deviation of the out-of-plane magnetic 
field $B_z (x)$ from a uniform value $B_0 = 3 \times 10^{-5}$ T is displayed in Figure 
\ref{fig_iono}a. Here, $B_z$ is vertically stratified to balance the plasma pressure 
and gravity forces in magnetohydrostatic equilibrium. Because the plasma beta is far 
smaller in the I/T than in the chromosphere and corona, the resulting ionospheric 
magnetic-field deviations are similarly smaller, by comparison.

This unstable neutral-plasma system is initialized with small-amplitude electron-density 
perturbations centered at $x = 0$, on the bottom side of the ionization layer. The resulting 
evolved solution at time $t = 4.6 \times 10^5$ s is shown in Figure \ref{fig_iono}.  Velocity 
vectors again are shown in panels a, b, and c that correspond to the differential ion-neutral 
velocity, ion velocity, and neutral velocity, respectively. In this case, we observe that the 
plasma and neutrals do not track each other very well. The plasma is unstable and develops a 
low-density bubble that rises through the plasma layer, as is observed in the I/T 
(Fig.\ \ref{fig:rayleightaylor1}). The contours of perturbed magnetic field closely resemble 
those of the plasma density. The formation of the bubble and the generation of the ion flows 
are strongly affected by drag exerted by the neutrals. Because the collisional coupling to 
the neutrals by the plasma is very weak, however, the neutrals move only slowly and somewhat 
independently, as shown by the neutral density, velocity, and temperature. 
The numerically calculated growth rate is $\tau^{-1} \sim 1.3 \times 10^{-5}$ 
s$^{-1}$. For terrestrial gravity and the chosen plasma density profile, the Brunt-V\"ais\"al\"a 
frequency is $N_p = 3.7 \times 10^{-2}$ s$^{-1}$, which is much smaller than the ion-neutral 
collision frequency $\nu_{in} = 1.1 \times 10^2$ s$^{-1}$. This calculation also lies in the 
strong-coupling regime of \S \ref{rayleightaylor_growth_limit}, but here Equations (\ref{EqNum8}) 
and (\ref{EqNum9}) are the relevant solutions. The low-frequency growth rate calculated from 
Equation (\ref{EqNum8}) is $\gamma \approx 1.3 \times 10^{-5}$ s$^{-1}$, in excellent agreement 
with the numerically determined rate. We point out that the growth rate for an O$^+$ plasma 
in the real I/T would be about ten times greater than the rate for our simulated H$^+$ 
plasma, with a resulting e-folding time of about 2 hrs, in reasonable agreement with observations.

\subsection{Summary}
These simulations show how a common framework has been used to describe the Rayleigh-Taylor instability in 
both the chromosphere and I/T, even though historically, the phenomena has been approached in two very 
different ways. There are many other common phenomena between the chromosphere and I/T, and much knowledge 
can be gained from applying this universal approach.

\section{Conclusions}
\label{conclusions}

In this paper we have compared the Sun's chromosphere and Earth's ionosphere/thermosphere (I/T). Both are weakly ionized, stratified mixtures of plasma and neutral gas with an increasing ionization fraction with height (altitude). {\color{black} Both have typical plasma 
$\beta$ less than one, and a neutral (or total) $\beta$ which transitions from above to less than one. Thus plasma motions alone are not capable of creating large perturbations in magnetic field, but if the coupling between plasma and neutrals is strong enough, the average motions of neutrals and plasma can potentially create large perturbations in the field. 
For the chromosphere, where the neutral-ion collision frequency is larger than $10^{3}$ Hz, then for phenomena with timescales longer than $10^{-3}$ s, the coupling is strong enough. For the I/T the neutral-ion collision frequency is larger than $10^{-6}$ Hz, so most timescales of interest (e.g. minutes to hours, or 60-3600 s) are too quick for the coupling to be sufficient. This difference is brought about by the much lower ionization level in the I/T compared to the chromosphere, something which creates many important difference between the two atmospheres, such as neutral-plasma collision rates and conductivity. It also affects fluid instabilities, as demonstrated by our discussion of the Brunt-V\"ais\"al\"a frequency. In both environments, this frequency $N_p$ is smaller than the ion-neutral collision frequency, so that the 
Brunt-V\"ais\"al\"a oscillations (or Rayleigh-Taylor 
instabilities) of the plasma are affected by coupling to the neutrals.  In the chromosphere, the oscillations and instabilities 
of the neutrals are similarly strongly affected by coupling to the plasma, since there the neutral Brunt-V\"ais\"al\"a frequency is less than the neutral-ion 
collision frequency. However, 
the opposite is true in the I/T, and so  the oscillations of the stably stratified neutral gas are 
unaffected by the plasma, and neutral motion is essentially undisturbed by the evolution of unstably stratified plasma.}

{\color{black} Both environments exhibit a variation of the magnetization of the ions and electrons with altitude, 
with magnetization being defined by the ratio 
of collision frequency to gyrofrequency. The magnetization is also a measure of how mobile the ions and electrons
are in the presence of neutrals. As the magnetization depends on magnetic field, we found a range of behavior in the chromosphere, but for certain magnetic field cases we found a similar behavior of the magnetization and mobility in both the chromosphere and  the I/T. 
In general, at low heights, the electrons and ions are unmagnetized due to high collision rates with neutrals. 
With increasing altitude, the electrons become mobile first, driving first mainly Pedersen currents, then Hall and Pedersen currents. Higher up still, the ions become magnetized and drive Pedersen currents also. We found that for the 1D I/T model, there are three regions, representing unmagnetized plasma (M1), magnetized electrons and unmagnetized ions (M2), and magnetized plasma (M3). In the middle of the M2 region is a height at which the conductivity became anisotropic (this is where $\Gamma=\xi_{n}k_{e}k_{in}$ become larger than unity, and is also the height at which Pedersen heating becomes important see below). Similar transitions were seen in the chromosphere, with a variation of the altitude of these transitions for different magnetic field models. This anisotropy can also be seen in the relative contribution of current dissipation by perpendicular currents and parallel currents, and is a factor in the location at which the 
field becomes force-free.
}

{\color{black} The large disparity in plasma density between the chromosphere and I/T also creates differences in the modeling and analysis of 
phenomena in the two atmospheres. We showed that the chromosphere is ideal (non-resistive) on length scales much larger than a km, such that the evolution of the magnetic field is dominated by advection by coupled plasma-neutral flows, but resistive below these lengths, where the field becomes decoupled from the average flow. However, because the conductivity is so much lower in the I/T, the Earth's atmosphere is essentially resistive such that the field is always decoupled from the average flow and coupled to electron flow. This generally leads 
to the treatment of I/T phenomena with electrodynamics, where $\mb{E}$ and $\mb{J}$ are considered primary variables and drivers, while in the chromosphere $\mb{B}$ and $\mb{V}$ are the primary variables. An example of this was applied to the phenomena of wind driven dynamos in \S \ref{neutral_wind_dynamos}.
 \citet{Parker_2007}, \citet{Vas_2001}, \citet{Vas_2011}, and \citet{Vas_2012} criticize the E-J paradigm, stating that for long time scales we can remove the dynamical equations for $\partial \mb{E}/\partial t$ and $\partial \mb{J}/\partial t$ and have $\mb{B}$ and $\mb{V}$ be the primary variables (V-B paradigm), but the converse is not true, and the E-J paradigm is not tractable.
Despite these facts, we find that using the E-J paradigm may allow one to arrive at the same result as the 
V-B paradigm, as was achieved in a general sense for wind-driven dynamos and the Rayleigh-Taylor instability in this review. However, in general this may not always be the case. 
}

{\color{black} The two atmospheres of the Earth and the Sun have different 
mechanisms that contribute to the heating. We know more about the heating 
mechanisms in the I/T compared to the chromosphere. In the I/T the dominant heating 
term for the neutral gas  is absorption of UV/EUV radiation, which forms the ionosphere through 
photoionization and raises neutral and 
plasma temperatures to more than one thousand degrees. However, frictional 
heating of the neutral gas by collisions with ions can be as significant as solar 
heating during geomagnetic storms and represents the most variable source of 
energy to quantify in the I/T energy equation. Other processes, such as ion friction 
and the Farley-Buneman instability can also contribute to energy transfer. In the chromosphere, the 
presence of a temperature minimum and temperature gradient reversal is a major challenge for solar physics. 
The observed average chromospheric temperature profile is created by a balance of a few major 
processes. Radiation in the chromosphere is optically thick, due to both spectra and line emission, and 
formed in non-LTE. The downward conduction of energy from the much hotter corona (the heating of
which is also a major open question in solar physics) is important in the 
upper chromosphere, but in the lower chromosphere, the radiation must be balanced by some heating 
mechanism. In fact the chromosphere  requires roughly ten times more heat input than the corona 
to maintain its elevated temperature, due to a much larger density \citep{1990SSRv...54..377N}.

In \S \ref{joule}, we showed how frictional heating provides a mechanism for the conversion of
electromagnetic energy into thermal and kinetic energy of the plasma-neutral gas mix, via the 
term $\mb{E}\cdot\mb{J}$. This frictional heating includes the well known ``Joule heating'' term due to 
ion-electron collisions, and a plasma-neutral frictional term, the latter of which dominates when 
the ions start to become mobile in the presence of neutrals, i.e. when they start to become magnetized.
We showed that when this occurs the currents perpendicular to the magnetic field become dominated
by Pedersen currents, driven by ion motions. As shown in \citet{Goodman2000,Goodman2004a}
the efficiency of Pedersen current heating $\mb{E}_{\perp}\cdot\mb{J}_{\perp} = \eta_{\perp}J_{\perp}^{2}$ increases as $\Gamma k_{e}k_{in}$
starts to become large, which is when the conductivity (and resistivity) tensor becomes anisotropic.
Below this region, the heating is mainly Ohmic Joule heating (electron-ion collisions). This analysis 
applies to a general dissipation mechanism, and to derive the actual heating term one must model 
the generation of such currents in the atmosphere. 
\citet{Goodman2000,Goodman2001}, \citet{Song_2011}, and \citet{Tu_2013} considered the propagation of MHD waves
into the chromosphere. The center of mass flow due to the wave motion has a component perpendicular to 
$\mb{B}$ which drives a center of
mass electric field, which in turn drives the Pedersen current. Future simulations with self-consistent thermodynamics and
resolved non-linear wave interactions and reflections are key to confirming the hypothesis that wave damping by Pedersen current 
dissipation is responsible for the required chromospheric heating.
}

{\color{black} In reviewing the two atmospheres in terms of plasma-neutral coupling, we have brought to light some of the many open questions and issues. We review here some of those issues, and suggest what future studies should address, and what are the main challenges.}

{\color{black} As mentioned, one of the main unanswered questions in the chromosphere is the mechanism which maintains its
elevated temperature. Over the last 60 or so years there have been many proposed mechanisms, and reviews of such 
mechanisms can be found in \citet{1990SSRv...54..377N}. These include dissipation of acoustic, magneto acoustic and Alfv\'{e}n wave which originate at the turbulent convection zone, which is capable of creating a spectrum of waves. Other mechanisms include magnetic reconnection, the Farley-Buneman instability, and low frequency current dissipation. The majority of these proposed mechanism are
only efficient at small scales, which creates one of the main challenges in modeling them. A numerical model must be able to 
sufficiently resolve the scales on which processes such as magnetic reconnection and wave dissipation operate (down to 10m or smaller), but it must also cover the spatial scales of interest, namely the extent of the chromosphere which is $\sim 2$ Mm. The other major problem when modeling proposed heating mechanisms is that to self-consistently model the heating, the model must include the complicated 
radiation in the chromosphere. Recent advances in modeling the chromosphere have included the coupling of the 3D non-LTE radiative physics to the  MHD physics \citep[see e.g.,][and references therein]{Carlsson_2012} on regions of small extent. 
The approach of using such detailed simulations to parameterize the radiation physics in terms of MHD variables (total density, temperature) and model chromospheric proposed heating mechanisms, is a possibly fruitful approach to solving the chromospheric heating problem. However, care must be taken to ensure the correct mechanism are resolved sufficiently, and that on the length and time scales of interest, the correct equations are used, particularly when it comes to the generalized Ohm's law. The problems in understanding the thermodynamic structure of the chromosphere also apply to prominences, which as mentioned are structures of chromospheric material suspended in the corona. The cause of fine structure in prominences is also an open issue, as well as fine structure in the chromosphere such as fibrils, jets, and surges. In addition to explaining the emission in the average, or quiet chromosphere, the increase of emission during flares is also an interesting issue, and has been newly investigated with 
simulations \citep[e.g.][]{Russell_2013}. As with the generic chromospheric heating problem, this issue is an example of the need for well-resolved self-consistent simulations with all of the MHD and radiative physics and with the ability too reproduce high fidelity observations of the atmosphere.
}

The relatively lesser amount of knowledge of the chromosphere is surely due to the fact that we can only remotely sample the plasma in the chromosphere, and this part of the solar atmosphere is optically thick in some lines, which makes for a difficult interpretation of the spectra obtained. The recently launched Interface Region Imaging Spectrograph \cite[IRIS,][]{IRIS} will be valuable in the effort to identify chromospheric heating mechanisms, {\color{black} using relatively high resolution (0.33 arscec) spectra and  line emissions from plasma at temperatures between 5,000 K and 10 MK. These observations will also be used to constrain improved chromospheric simulations which couple the non-LTE radiation that occurs in the chromosphere to the MHD evolution of the plasma \citep[e.g.,][]{Carlsson_2012}. These simulations are not currently able to resolve all chromospheric physics, but are valuable in the ongoing effort to test proposed heating mechanisms.


The Earth's I/T system is better understood than the chromosphere because it is easier to make a wide range of measurements in the former domain. Despite this there are many challenges remaining to I/T science. \citet{Rishbeth_2007} outlined a number of them including: semiannual variations, and the annual asymmetry in both the ionosphere and thermosphere; why the I/T survives at night; day-to-day atmospheric and ionospheric variability, its forcing mechanisms and the I/TÕs apparent predilection for certain time scales; and ionospheric memory and preconditioning. Other challenges include the solar cycle change in the winter anomaly \citep{Torr_and_Torr} and the way that high latitude forcing apparently drives changes in the low latitude I/T \citep{2008JGRA..113.2301W}. 
{\color{black} Understanding these phenomena is difficult because the I/T system is driven from below and above. The Earth's ionosphere is affected by
both tropospheric/stratospheric dynamic through modification of the neutral composition/temperature/winds, and the magnetosphere though high-latitude currents and heating, so one must have a good understanding of these other regions as well as the ability to incorporate these effects self-consistently into an ionosphere model.
Moreover, ionospheric dynamics spans an enormous range of spatial and temporal scales and different physical processes are important in different regions. It is difficult to bring all of this together in a single, coherent model (this is also a generic issue with most geospace and solar regions).

Virtually all ionosphere models assume the magnetic field lines are equipotentials. This reduces the potential equation to 2D and is readily solvable. However, it is 
clear that this is an approximation and should be relaxed. Recently, \cite{Aveiro_and_Hysell_2010, Aveiro_and_Hysell_2012, Aveiro_and_Huba_2013} developed a 3D electrostatic model of the
 ionosphere and applied it to the development of equatorial spread F. Although, in a general sense, the results are similar to the 2D results there are differences which
 could be important. \citet{NRL} have been able to embed a very high resolution grid (.06 degrees over a 60 degree sector) within the context of a global model. They were able to simulate for the first time the onset and evolution of equatorial bubbles (scale sizes 10s km) in a global model. Such coupled simulations are one possible solution to deal with the multi-scale problem of I/T physics. This approach could also be applied to traveling ionospheric disturbances and gravity waves.}

The recently selected NASA I/T missions ICON (Ionospheric Connection) and GOLD (Global-scale Observations of the Limb and Disk) will also address some of the key I/T  issues.  In particular, ICON will obtain the baseline characterization of the internally driven non-linear coupling between the neutral atmospheric drivers of winds, composition changes and the ionospheric responses of plasma densities and ion drifts. During periods of enhanced solar and geomagnetic activity ICON will determine how these parameters deviate from their baseline and will relate them to the strength of the solar wind electrical forcing that is externally applied to the global ionosphere-magnetosphere system.  The GOLD mission will investigate the significance of atmospheric waves and tides propagating from below on the temperature structure of the thermosphere, and it will resolve how the structure of the equatorial ionosphere influences the formation and evolution
of equatorial plasma density irregularities.  

{\color{black} One of the goals of this review paper is to highlight how considering the commonalities of two different atmospheres can
shed light on what can be learnt from one about the other. We have found a lot of commonalities, 
and have been able to talk about the two atmospheres within a common framework, 
but the differences between the two atmospheres create barriers to common studies. The main difference is the plasma density, which affects neutral-ion 
collisions frequencies, conductivity, and resistive vs convective magnetic field nature. 
However, the Rayleigh-Taylor instability is one particular phenomena that we were able
to simulate in both atmospheres using the same equations and model (\S \ref{rayleightaylor}). Rather than search for
common phenomena that exist in both atmospheres, the key to future collaboration is to identify 
common fundamental partially ionized plasma physics problems such as the Rayleigh-Taylor 
instability. Other common fundamental partially ionized plasma physics studies include the
development of two stream instabilities in regions where ions are unmagnetized but electrons are magnetized, and the 
subsequent generation of turbulence (e.g. the Farley-Buneman instability). Also the neutral wind-driven dynamo 
is a common problem that can occur in the chromosphere and I/T.
 The use of the V-B paradigm to describe
I/T phenomena that were previously described in the E-J paradigm, as was done for the neutral wind dynamo 
by \citet{Vas_2012}, will help shed light on the fundamental physics. The use of detailed
simulations based on these fundamental studies is also a possible route to better understanding. Previous examples 
include the work  by \citep[e.g.][]{Tu_2008} on driving by electric fields. Future studies could also involve numerical experiments of a localized neutral wind in a static, uniform background magnetic field, plasma, and neutral gas, using a  three-fluid model (neutrals, electrons, and ions) that retains the displacement current in Amp\`ereÕs law, which would help test the 
magnetohydrodynamic  interpretation of the neutral wind-driven dynamo.
Simpler experiments could use a two-fluid model (neutrals and charge-balanced plasma) to determine whether charge separation in the electrodynamic description truly is required to generate the dynamo or, instead, is transient and merely incidental to the process.}  
}

\appendix

\section{Appendix}
\label{appendix}

The two-dimensional simulation results shown in Figures \ref{fig_sp} and \ref{fig_iono} 
were performed with the HiFi spectral-element multi-fluid model \citep{2008PhDT.........1L}. 
Effective grid sizes of 480$\times$1920 and 180$\times$720 were used in the solar-prominence 
and ionosphere cases, respectively, along the horizontal ($y$) and vertical ($x$) directions. 
Periodic conditions were applied at the side boundaries ($y$), while closed, reflecting, 
free-slip, perfect-conductor conditions were applied at the top and bottom boundaries ($x$), 
which were placed sufficiently far from the unstable layer to have negligible effect on the 
Rayleigh-Taylor evolution. Details of the plasma and neutral profiles and parameters used 
in the simulations are given below.

\subsection{Prominence}

Normalization constants for this case are number density $n_0 = 1 \times 10^{15}$ 
m$^{-3}$, length scale $L_0 = 1 \times 10^6$ m, and magnetic field $B_0 = 1 \times 
10^{-3}$ T. Using these in a hydrogen plasma, normalization values for the time 
$t_0 = 1.45$ s and temperature $T_0 = 5.76 \times 10^7$ K can be derived.  The ion
inertial scale is so much smaller than any scale of interest that, in this case, 
its value has been set explicitly to zero, $d_i = (c/\omega_{pi0})/L_0 = 0$. {\color{black} This is equivalent 
to neglecting the Hall term in the Ohm's law. Using the ratio of Hall term to Pedersen in the center of mass Ohm's law (\ref{gol_CM}), 
and the simplifications used in \S \ref{current_diss}, this is equivalent to $\xi_{n}^{2}k_{in} \gg 1$, which is valid for
these simulations.}

The electron and ion density profiles are given by atmospheric stratification, 
\bln
\begin{equation}
n_i(x) = n_e(x) = n_0 \exp \left( -\frac{x}{x_0} \right).
\end{equation}
\eln
The scale height $x_0$ is set by the solar gravitational acceleration, 
$g_S = 2.74 \times 10^2$ m s$^{-2}$, and the assumed background temperature 
of the corona, $T_b = 3.5 \times 10^{-3} T_0 = 2.02 \times 10^5$ K, scaled 
to the normalization length $L_0$; its value is $x_0 = 12.2$. The density 
profile of the neutral fluid that constitutes the prominence is given by 
a prescribed function of $x$ plus a very low uniform background value,
\bln
\begin{equation}
  n_n(x) = n_{n0} \sech^2 \left( 2x-1 \right) + n_{nb}.
\end{equation}
\eln
The peak neutral number density enhancement is taken to be $n_{n0} = 1 
\times 10^{16}$ m$^{-3} = 10 n_0$, while $n_{nb} = 3.5 \times 10^{-7} 
n_{n0}$, corresponding to the neutral fraction obtained in the HiFi 
ionization/recombination equilibrium at the background temperature 
$T_b$. We chose an artificially low value of $T_b$ (compared to a 
typical coronal temperature of about $2 \times 10^6$ K) in order 
to prevent the background neutral density $n_{nb}$ from being far 
smaller still.

The electron, ion, and neutral temperatures are all assumed to be equal 
to each other initially.  The temperature profile is given by a prescribed 
function $f(x)$, 
\bln
\begin{equation}
  T(x) = T_b f(x) = T_b \frac{\cosh^2(x-0.5)}{\cosh^2(x-0.5) + \lambda},
\end{equation}
\eln
which approaches $T_b$ away from the prominence and attains a minimum 
value $T_p = T_b / \left( 1 + \lambda \right)$ within the prominence.  
To obtain a temperature approximately corresponding to that observed 
on the Sun, with an associated low ionization fraction $(n_{e0}/(n_{e0} 
+ n_{n0})) = 0.091$, we set the parameter $\lambda = 20$. The resulting 
prominence temperature $T_p = 9.60 \times 10^3$ K.

The magnetic field is initialized to lie in the out-of-plane direction 
$\hat{\bf e}_z$, so that it is perpendicular to both gravity in the 
$-\hat{\bf e}_x$ direction and the instability wavenumber $k$ in the 
$\hat{\bf e}_y$ direction.  It is given by 
\bln
\begin{equation}
  \Bvec = B_0 \hat{\mb{e}}_z \left[ 1 + \beta\left\{   \frac{n_e(x)}{n_0} 
  \left[ 1 - f(x)  \right] - \frac{n_n(x)}{2n_0}f(x) - \frac{1}{x_0} 
  \frac{n_{n0}}{2n_0} \left[ \tanh \left( 2x-1 \right) - 1 \right] 
  \right\} \right]^{1/2},
\end{equation} 
\eln
where $\beta = 1.4 \times 10^{-2}$ is the plasma beta evaluated using the 
background plasma pressure at $x = 0$ and $B_0$. The magnetic field profile 
so constructed accommodates (1) the plasma pressure change from the isothermal 
hydrostatic profile (2) the neutral pressure, and (3) the gravitational force 
exerted on the bulk of the neutral fluid (neglecting the small background 
contribution $n_{nb}$) throughout the atmosphere.

{\color{black} This initial condition is not an exact solution to the multi-fluid model, 
including ionization and recombination, with no flow. It is close enough to 
the solution, however, that any flows created by pressure gradients driven by ionization/recombination 
of the initial condition are small compared to the
flows initiated by the instability.}
The instability is initiated by introducing a small neutral density 
perturbation localized in $x$ on the bottom side of the prominence, 
\bln
\begin{equation}
  \Delta n_n(x,y) = \delta n_n(x)
  \exp[-4x^2]\frac{1}{5}\sum_{j=1}^5\sin[j \pi y],
\end{equation}
\eln
where we chose $\delta = 10^{-2}$, and $y$ is the normalized distance 
along the gravitational equipotential surface.

\subsection{Ionosphere}

The normalization constants are number density $n_0 = 1 \times 10^{16}$m$^{-3}$,
length scale $L_0 = 2 \times 10^4$ m, and magnetic field $B_0 = 3 \times 
10^{-5}$ T.  Using these in a hydrogen plasma, normalization values 
for the time $t_0 = 3.06$ s, temperature $T_0 = 5.19 \times 10^3$ K, 
and ion inertial scale length $d_i = (c/\omega_{pi0})/L_0 = 1.14 \times 
10^{-4}$ can be derived.

The neutral density profile is given by atmospheric stratification, 
\bln
\begin{equation}
n_n(x) = n_{0} \exp \left( -\frac{x}{x_0} \right).
\end{equation}
\eln
The scale height $x_0$ is set by Earth's gravitational acceleration, 
$g_E = 9.81$ m s$^{-2}$, and the assumed background temperature of the 
ionosphere, $T_b = 0.225 T_0 = 1.17 \times 10^3$ K, scaled to the 
normalization length $L_0$; its value is $x_0 = 49.1$. The electron/ion 
density profile of the plasma is given by a prescribed function of $x$ 
plus a low uniform background value,
\bln
\begin{equation}
  n_e(x) = n_i(x) = n_{e0} \sech^2 \left( 2x-1 \right) + n_{eb}.
\end{equation}
\eln
The peak electron number density is taken to be $n_{e0} = 1 \times 10^{12}$ 
m$^{-3} = 10^{-4} n_0$, while $n_{eb} = 0.05 n_{e0}$. The electron, ion, 
and neutral temperatures are all assumed to be equal and uniform initially, 
at $T_b = 1.17 \times 10^3$ K.

As before, the magnetic field is initialized to lie in the out-of-plane 
direction.  It is given by 
\bln
\begin{equation}
  \Bvec = B_0 \hat{\bf e}_z \left[ 1 - \beta \left\{ \frac{2n_e(x)}{n_0} 
  + \frac{1}{2x_0} \left[ \tanh \left( 2x - 1 \right) - 1 \right] 
  \right\} \right]^{1/2},
\end{equation} 
\eln
where $\beta = 0.450$ is the neutral beta calculated using the background 
neutral pressure at $x = 0$ and $B_0$. The magnetic field profile so constructed 
accommodates (1) the plasma pressure and (2) the gravitational force exerted 
on the bulk of the plasma (neglecting the small background contribution $n_{eb}$) 
throughout the atmosphere.

The instability is initiated with a small electron density perturbation localized 
in $x$ on the bottom side of the ionization layer,
\bln
\begin{equation}
\Delta n_e(x,y) = -\delta n_e(x)
\exp[-4x^2]\frac{1}{5}\sum_{j=1}^5\sin[j \pi y],
\end{equation}
\eln
where again we chose $\delta = 10^{-2}$, and $y$ is the normalized distance 
along the gravitational equipotential surface.

\begin{acknowledgements}
This work was funded by NASA's ``Living with a Star'' Targeted Research and Technology program "Plasma-Neutral Gas Coupling in the Chromosphere and Ionosphere" . 
Numerical simulations were performed using a grant of computer time from the DoD High Performance Computing Program. NCAR is sponsored by the National
Science Foundation.
\end{acknowledgements}

\bibliographystyle{aps-nameyear}      

\providecommand{\noopsort}[1]{}\providecommand{\singleletter}[1]{#1}%

\end{document}